# Quantum Image Processing: the truth, the whole truth, and nothing but the truth about its problems on internal image-representation and outcomes recovering

Mario Mastriani
ORCID Id: 0000-0002-5627-3935

*Qubit Reset LLC, 2875 NE 191, suite 801, Aventura, FL 33180, USA*

mmastri@qubitreset.com

**Abstract** – In this paper, three tecniques of internal image-representation in a quantum computer are compared: Flexible Representation of Quantum Images (FRQI), Novel Enhanced Quantum Representation of digital images (NEQR), and Quantum Boolean Image Processing (QBIP). All conspicuous technical items are considered in this comparison for a complete analysis: i) performance as Classical-to-Quantum (Cl2Qu) interface, ii) characteristics of the employed qubits, iii) sparsity of the used internal registers, iv) number and size of the required registers, v) quality in the outcomes recovering, vi) number of required gates and its consequent accumulated noise, vi) decoherence, and vii) fidelity. These analyses and demonstrations are automatically extended to all variants of FRQI and NEQR. This study demonstrated the practical infeasibility in the implementation of FRQI and NEQR on a physical quantum computer (QPU), while QBIP has proven to be extremely successful on: a) the four main quantum simulators on the cloud, b) two QPUs, and c) optical circuits from three labs. Moreover, QBIP also demonstrated its economy regarding the required resources needed for its proper function and its great robustness (immunity to noise), among other advantages, in fact, without any exceptions.

## 1. Introduction

Since its beginnings, three decades ago, Quantum Image Processing (QImP) was not immune to controversies, however, its mere mention was attractive from the beginning and, as time passed by, more and more researchers have shown their interest in it. In 2011, the first apparently practical technique of internal representation of an image in a QPU appears, named FRQI [1], in the form of a normalized state which captures information about colors and their corresponding positions in the images. All the experiment contributed in that work were implemented on a high-level interpreter (HLI) [2] and programmed by the same author, who applied it to the compression of a 256×256 gray image of Lena with *8 bits-per-pixel*. Two years later, the second apparently practical technique of internal representation of an image in a QPU appears, known as NEQR [3], with the argument that it improves FRQI when using a qubit sequence to store the gray-scale value of each pixel in the image for the first time, instead of the probability amplitude of a qubit, as in FRQI. According to its authors, the comparison of the performances between FRQI and NEQR reveal that NEQR can achieve a quadratic speedup in quantum image preparation, increase the compression ratio of quantum images by approximately 1.5X, and retrieve digital images from quantum images accurately. In this case, the authors, curiously, do not even mention the platform on which they perform the experiment, or the technical characteristics of the parameters of such an implementation, that is, it is almost a *gedankenexperiment*. This is incomprehensible from the practical point of view, since it is the first time in literature that a technique is presented by opposition with a previous one without clarifying the parameters of the experiment from which the success of the comparison results.

On the other hand, in 2015, a new technique for the internal image-representation in a QPU appears, called QBIP [4], which exclusively works with the first bitplane of each color channel of the image allowing us to reduce the storage problems of the previous techniques, facilitate the design and operation of the Cl2Qu interface, automatically recover the outcomes with maximum fidelity and less noise, and dramatically reduce the size and amount of internal registers, as well as, the number of gates, among many other advantages. Two years later, a paper showing the unfeasibility in the recovery of the outcomes by FRQI appears, raising a disproportionate controversy [5], and becoming the precedent of the present work. Moreover, today we have unrestricted access to the main QPUs via cloud services [6-9] allowing any interested researcher of QImP to confirm the results of this pioneering work, which only tried to show the inexcusable truth. In fact, QImP is one of the most attractive and promising tools within the Quantum Technology toolbox, as long as, the internal image-representation techniques used really work.

Finally, this work will demonstrate what works and what does not work as an internal image-representation technique inside QImP, subjecting the result to the unavoidable scrutiny of the sovereign scientific method. This will also demonstrate the weakness of the argument presented by several authors of QImP, which results from the implementation of their works using FRQI [1] and NEQR [3] on an HLI [2] in a classic machine with the argument that QPUs are not yet well developed, which is completely inaccurate for two reasons:

1. QBIP now works on a QPU, as it did at the time of its creation on an optical circuit, instead

2. FRQI (and its variants), as well as, NEQR (and its variants) did not work at the time of its creation on an optical circuit, neither do they now on a QPU nor will they do so in the future, due to problems that will be demonstrated in Sections 3, 4, and 6.

Showing up next, the comparative elements between FRQI, NEQR, and QBIP are outlined in Section 2. In Sections 3, 4, and 5, we present the pros and cons for FRQI, NEQR, and QBIP, respectively. In Section 6, the implementations on quantum platforms for the three techniques are presented. Section 7 includes the discussions about the experimental results, while, Section 8 provides the conclusions.

## 2. Comparative elements between FRQI, NEQR, and QBIP

This section introduces the necessary tools to evaluate the performance of every technique for the internal representation of an image inside a quantum computer. It goes from its generation in the classical world (digital), going through its insertion in the quantum world, until its recovery via quantum measurement, once the quantum image has been submitted to the treatment by an internal quantum algorithm.

### *2.1 Typical scheme of QImP*

A typical scheme of QImP necessarily implies an architecture like the one presented in Fig. 1. Evidently, this scheme does not differ much from that used by quantum computing for any other task. Figure 1 shows that the classical image (digital) must be introduced into the quantum computer for further processing, which is the responsibility of a quantum algorithm allocated inside a quantum computer.

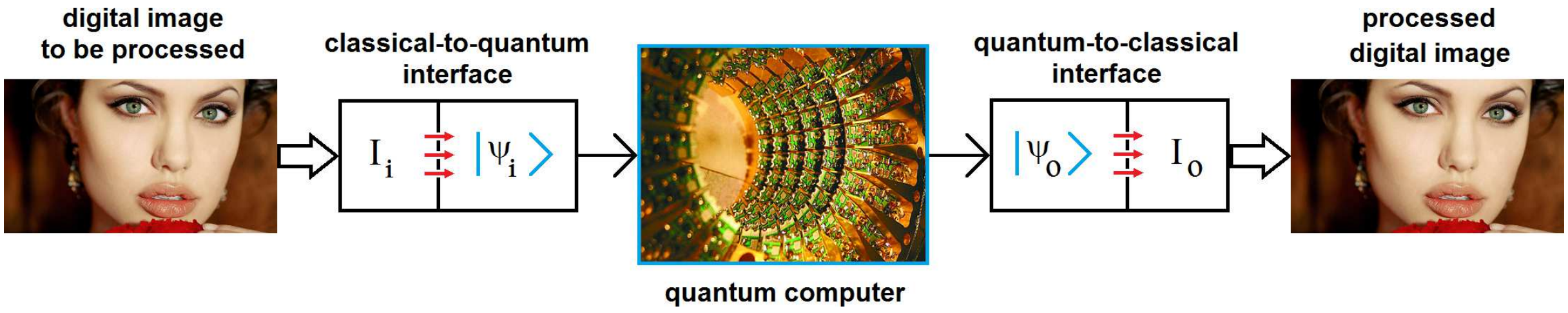

**Figure 1.** Scheme of Quantum Image Processing (QImP).

Let us suppose that the aforementioned processing consists in the filtering of the noise that the digital image brings from its classical origin, e.g., a camera, the channel through which it was received, etc. Consequently, the quantum algorithm will have the function of filtering said noise. Clearly, the problem is to introduce a quantum version of the original classical image into the quantum computer. There are two possibilities to achieve this:

a) by hand, i.e., preparing qubit by qubit in an artisanal way, or
b) automatically, using a Classical-to-Quantum interface (Cl2Qu).

Now, suppose that the camera takes color pictures at a resolution of 1920x1080 pixels (i.e., a common camera), so, each digital image will be composed of 1920x1080x3x8 bits (approximately 50 Million bits), which should be converted into approximately 50 Million of qubits inside the quantum computer. If we analyze this real case in order, option (a) is not viable, since no laboratory in the world is able to prepare 50 Million qubits, let alone there are no humans who would do this artisanal imitation work, i.e., convert each one of the 50 Million of bits in its quantum counterpart. This is obviously unthinkable. Regarding (b), we find the following options according to the QImP literature, i.e., via:

I. FRQI (and its variants) [1],
II. NEQR (and its variants) [3], and a third option,
III. which are constituted by a family of proven Cl2Qu interface, and that are used by QBIP [4].

Next, we will postulate the essential and inexcusable conditions that must be met by any Cl2Qu interface in order to be used on a QPU like Rigetti [7] or IBM Q Experience [6]. Therefore, with this information, we will evaluate throughout this paper, the performance as Cl2Qu interface of the three techniques suggested above. What is clear even from this precise moment is that any effort to implement a QImP scheme is absolutely unfeasible without a real and practical Cl2Qu interface.

### *2.2 Conditions to be met by a Cl2Qu interface*

All Cl2Qu interface must meet, *sine qua non* conditions such as the following:

a) ***Traceability:*** every component of the image must not lose the necessary traceability throughout the entire process of Fig.1, which will also be seen as an inexcusable link between the value of the pixel bit and its location in the image at all times. What is the reason for such a mandatory condition? The reason been is that no physical quantum computer (QPU) allows intermediate instances of quantum measurement or purges of qubits, known as qubit reset gate (obviously with implementations of this gate in a time less than coherence time, see Rigetti [7]). In other words, if the outcomes are not what we expected, then there is no way to go back on the same exact process to identify the source of the problem. This is clearly in tune with the fact that the quantum world and its implementation in gates that respond to the different Algebras (Clifford, Pauli, etc.) represent an eclectic environment.

b) ***Homotopical relation:*** every bit of the original image (digital, i.e., classical) must maintain a homotopical relation of type $bit \rightarrow [\,Cl2Qu\,] \rightarrow |bit\rangle \equiv qubit$ with its counterpart within the physical quantum computer (QPU), for the same reason mentioned above.

c) ***Automatic intake:*** every bit of the original image (digital, i.e., classical) must be converted to its quantum counterpart in an automatic way. The reason is more than obvious, let us go back to the example of a 1920x1080x3x8 bit image: who can and/or wants to prepare 50 Million of qubits by hand?

d) ***Facilitation in the design of the Quantum-to-Classical (Qu2Cl) interface:*** this condition is essential for a correct recovery of outcomes, obviously, considering the problem of quantum measurement [10] and complementarity [11] that will be explained later. In other words, if the three previous conditions are met, then the Qu2Cl interface (i.e., which recovers the outcomes) is extremely simple. In fact, a measurement exclusively on the *z* axis of the Bloch's sphere (Fig.2) is precisely how the physical quantum computers (QPUs) currently in service work. See Rigetti [7] and IBM Q Experience [6].

These conditions are only recognized in those techniques of QImP implemented in a laboratory on an optical circuit or on a QPU. These can never be detected in such implementations of QImP made in an HLI [2] on a classic computer.

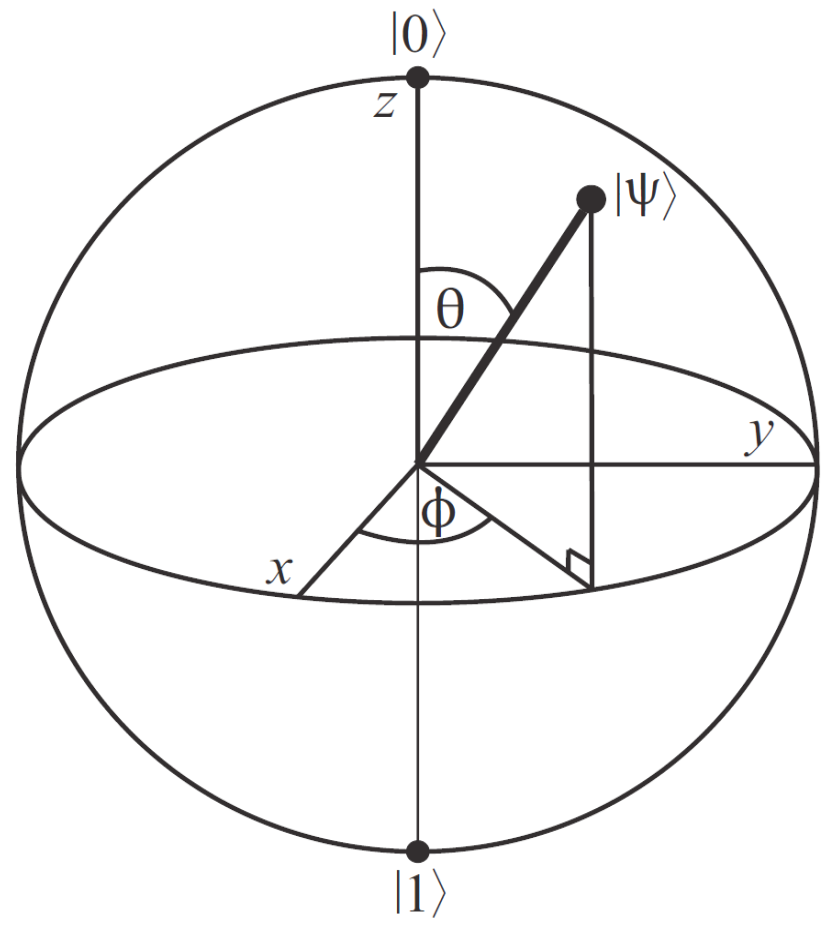

**Figure 2.** Bloch's sphere.

### *2.3 Characteristics of the employed qubits*

From Quantum Information Processing [12-14], we know that if we resort to the superposition principle, then every arbitrary qubit can be expressed as follows:

$$|\psi\rangle = \alpha|0\rangle + \beta|1\rangle, \tag{1}$$

where, $\{|0\rangle, |1\rangle\}$ are the Computational Basis States (CBS), being

$$|0\rangle = \begin{bmatrix} 1 \\ 0 \end{bmatrix} \equiv \textit{north pole} \tag{2}$$

and

$$|1\rangle = \begin{bmatrix} 0 \\ 1 \end{bmatrix} \equiv \textit{south pole} \tag{3}$$

both poles of the Bloch's sphere of Fig.2. Besides, $\alpha$ and $\beta$ must meet the following conditions:

$$\alpha = cos\frac{\theta}{2},\ \ \beta = e^{i\phi}\, sin\frac{\theta}{2},\ \ |\alpha|^2 + |\beta|^2 = 1 \quad / \alpha \wedge \beta \in \mathbb{C} \tag{4}$$

with $0 \le \theta \le \pi$, $0 \le \phi < 2\pi$ on Bloch's sphere of Fig.2. Then, we can rewrite Eq.(1) as:

$$|\psi\rangle = cos\frac{\theta}{2}|0\rangle + e^{i\phi}\, sin\frac{\theta}{2}|1\rangle \cdot \tag{5}$$

Evidently, CBS form an orthogonal base, which is ideal to face the different stages in which a qubit is involved in its trajectory along the QPU. Besides, the CBS have another fundamental attribute to work in the quantum world, which is that they only have projection on a single axis of the Bloch's sphere. Why is this attribute important? This attribute is important because the QPU, such as IBM Q Experience [6], only yields outcomes (exclusively at the end of the quantum algorithm, i.e., not in intermediate instances) in terms of $\alpha$ and $\beta$, but not them directly. That is, if the technique of internal representation of the image generates projections on more than one axis (i.e., another apart from the $z$ axis), these additional projections are lost, giving a completely different noisy result to that which existed inside the QPU. In other words, what has been recovered does not reflect the physical reality inside the quantum machine since the phase information related to the angle $\phi$ is lost. It is clear that the *orthogonality* between the qubits that conform the employed basis is not enough to obtain appropriate outcomes at the output of a QPU. As we can see throughout this work FRQI does not fulfill this premise, while NEQR and QBIP do.

### *2.4 Sparsity, number and size of the required registers*

Another issue, no less important, is the inappropriate nature of the registers required to express the internal representation of the image. The most sensitive of these issues is undoubtedly sparsity, which is evidenced by the presence of Kronecker's product in FRQI [1] and NEQR [3]. The use of this dimensionally expansive operation represents, in itself, the inappropriate use of a large amount of the big sparse matrix inside both techniques. This will be seen in depth in Sections 3 and 4, respectively, however, we can anticipate that this fully impacts on the increase of:

- the computational cost,
- the storage,
- the excessive use of quantum gates, and the noise introduced by these in the total process,
- the complexity in the implementation of the experiment and in an even greater complexity of the quantum algorithms used after the internal image representation technique,
- the size and number of registers used inside the QPU,
- among many others.

It is clear that QBIP [4] is not a victim of any of these problems.

### *2.5 Quantum measurement and the complementarity problem*

The quantum measurement [10] problem has to do with the complementarity [11] present between the uncertainties (when trying to measure along the three axes of the Bloch's sphere, i.e., $x$, $y$, and $z$) or trade-off of the three inequalities of Eq.(6),

$$\Delta x\, \Delta y \leq 1/4\pi, \quad \Delta x\, \Delta z \leq 1/4\pi, \quad \Delta z\, \Delta y \leq 1/4\pi, \tag{6}$$

where $\pi \cong 3.141592$. Now, if we measure the state of a qubit along the $z$ axis of the Bloch's sphere with total accuracy, then, the uncertainty when measuring in the other two axes will be total. This manifestation of the Heisenberg's Uncertainty Principle [12] is unfortunately ignored by the authors of the papers that use FRQI [1], since it is erroneously believed that this technique eludes the complementarity problem, however, the reality shows us that FRQI has essential projections on the three axis of the Bloch's sphere, i.e., it is an obvious victim of complementarity, which will be demonstrated in Sections 3 and 6, while, NEQR [3] has an even greater problem when it comes to recovering its outcomes, which is known as *entanglement coupling* and which we will explain and demonstrate in Sections 4 and 6. We must take into account that both the quantum simulators with cloud service of [6-9] as well as the QPUs of IBM Q [6] and Rigetti [7] (with accesss from the cloud too) only measure along the $z$-axis. In fact, if we take a look at the main equations of FRQI [1], we can clearly see that this technique makes extensive use of the principle of superposition, which is averaged by a power of 2. The final result, $|I\rangle$ is far from being a CBS $\{|0\rangle,|1\rangle\}$, i.e., it is far from Eqs(2) and (3), therefore, the outcome of the quantum algorithm will be completely affected and altered by quantum measurement, which is what we can clearly see if we make simulations with [6-9] like those of Section 6. Obviously, this alteration of the result is not seen in the codes made with an HLI [2] on a classic computer, which shows the theoretical reality of FRQI, but which is far from the physical reality, being the latter what ultimately really matters. This problem can only be successfully over-come if working exclusively with CBS as is in the case of QBIP [4], and apparently in NEQR [3]. In all other cases, quantum measurement [10] inexorably alters the result. A simple demonstration to show that the CBS are independent of this problem can be seen from a projective measurement of both CBS based on their density matrices,

$$\hat{M}_0 = |0\rangle\langle 0| = \begin{bmatrix} 1 \\ 0 \end{bmatrix}\begin{bmatrix} 1 & 0 \end{bmatrix} = \begin{bmatrix} 1 & 0 \\ 0 & 0 \end{bmatrix}, \tag{7}$$

and

$$\hat{M}_1 = |1\rangle\langle 1| = \begin{bmatrix} 0 \\ 1 \end{bmatrix}\begin{bmatrix} 0 & 1 \end{bmatrix} = \begin{bmatrix} 0 & 0 \\ 0 & 1 \end{bmatrix}. \tag{8}$$

Then, the CBS post-measurement will be,

$$|0_0\rangle_{pm} = \frac{\hat{M}_0|0\rangle}{\sqrt{\langle 0|\hat{M}_0^\dagger \hat{M}_0|0\rangle}} = \frac{\begin{bmatrix} 1 & 0 \\ 0 & 0 \end{bmatrix}\begin{bmatrix} 1 \\ 0 \end{bmatrix}}{\sqrt{\begin{bmatrix} 1 & 0 \end{bmatrix}\begin{bmatrix} 1 & 0 \\ 0 & 0 \end{bmatrix}\begin{bmatrix} 1 & 0 \\ 0 & 0 \end{bmatrix}\begin{bmatrix} 1 \\ 0 \end{bmatrix}}} = \frac{\begin{bmatrix} 1 \\ 0 \end{bmatrix}}{\sqrt{1}} = \begin{bmatrix} 1 \\ 0 \end{bmatrix} = |0\rangle \tag{9}$$

and

$$|1_1\rangle_{pm} = \frac{\hat{M}_1|1\rangle}{\sqrt{\langle 1|\hat{M}_1^\dagger \hat{M}_1|1\rangle}} = \frac{\begin{bmatrix} 0 & 0 \\ 0 & 1 \end{bmatrix}\begin{bmatrix} 0 \\ 1 \end{bmatrix}}{\sqrt{\begin{bmatrix} 0 & 1 \end{bmatrix}\begin{bmatrix} 0 & 0 \\ 0 & 1 \end{bmatrix}\begin{bmatrix} 0 & 0 \\ 0 & 1 \end{bmatrix}\begin{bmatrix} 0 \\ 1 \end{bmatrix}}} = \frac{\begin{bmatrix} 0 \\ 1 \end{bmatrix}}{\sqrt{1}} = \begin{bmatrix} 0 \\ 1 \end{bmatrix} = |1\rangle \tag{10}$$

Evidently, they do not suffer alterations by the measurement process. It is precisely with this type of qubits, the CBS, with which QBIP works [4], and apparently so does NEQR [3]. Instead, any other type of qubit will be altered by the quantum measurement process, for example,

$$|\psi_0\rangle_{pm} = \frac{\hat{M}_0|\psi\rangle}{\sqrt{\langle \psi|\hat{M}_0^\dagger \hat{M}_0|\psi\rangle}} = \frac{\begin{bmatrix} 1 & 0 \\ 0 & 0 \end{bmatrix}\begin{bmatrix} \alpha \\ \beta \end{bmatrix}}{\sqrt{\begin{bmatrix} \alpha & \beta \end{bmatrix}\begin{bmatrix} 1 & 0 \\ 0 & 0 \end{bmatrix}\begin{bmatrix} 1 & 0 \\ 0 & 0 \end{bmatrix}\begin{bmatrix} \alpha \\ \beta \end{bmatrix}}} = \frac{\begin{bmatrix} \alpha \\ 0 \end{bmatrix}}{\sqrt{\alpha^2}} = \frac{\alpha}{|\alpha|}|0\rangle \tag{11}$$

and,

$$|\psi_1\rangle_{pm} = \frac{\hat{M}_1|\psi\rangle}{\sqrt{\langle \psi|\hat{M}_1^\dagger \hat{M}_1|\psi\rangle}} = \frac{\begin{bmatrix} 0 & 0 \\ 0 & 1 \end{bmatrix}\begin{bmatrix} \alpha \\ \beta \end{bmatrix}}{\sqrt{\begin{bmatrix} \alpha & \beta \end{bmatrix}\begin{bmatrix} 0 & 0 \\ 0 & 1 \end{bmatrix}\begin{bmatrix} 0 & 0 \\ 0 & 1 \end{bmatrix}\begin{bmatrix} \alpha \\ \beta \end{bmatrix}}} = \frac{\begin{bmatrix} 0 \\ \beta \end{bmatrix}}{\sqrt{\beta^2}} = \frac{\beta}{|\beta|}|1\rangle \tag{12}$$

That can be a serious problem for all the techniques of internal representation of an image that work with generic qubits, which was mentioned in [5] and that as we will see in Section 3 for the case of FRQI is even more serious. In fact, as we will see in the implementations of Section 6, this problem will cause the FRQI outcomes to completely lose color tones, collapsing into strict black and white pixels, exclusively, which is dramatically far from that reported by papers which implement it on an HLI [2].

### *2.6 Gate noise, decoherence and fidelity*

Quantum gates, particularly those optically implemented [15], inherently generate a series of noises, specifically, three types:

a. bit flip,
b. phase flip (or phase damping), and
c. bit-phase flip

Undoubtedly, the influence of these noises on total process performance is in direct proportion to the number of gates used by the aforementioned process, which is composed by the interfaces (Cl2Qu, and Qu2Cl), the internal image representation technique, and the quantum algorithm. This influence is extremely high in both FRQI [1] and NEQR [3], since as we will see in Sections 3 and 4, respectively, they require a large number of gates for their respective implementations. Besides, decoherence is also present at the moment of recovering the outcomes, which is strongly conditioned by the quantum measurement method [16] chosen. Consequently, we can observe a degradation of the fidelity due to the use of gates that manipulate the phase, as in the case of the *H* (Hadamard), *CNOT*, and Toffoli's gates, which triggers the intervention of the aforementioned noises. But, what is fidelity? It is a metric widely used in Quantum Communications [17, 18], which allows us to evaluate the quality of the reconstruction when rebuilding a state after going through a quantum process. For example, let us suppose a quantum process whose incoming state is $|\psi_{in}\rangle$, and whose outgoing state is $|\psi_{out}\rangle$, then, their respective density matrices will be,

$$\rho_{in} = |\psi_{in}\rangle\langle\psi_{in}| \text{ and } \rho_{out} = |\psi_{out}\rangle\langle\psi_{out}| \tag{13}$$

Fidelity quantifies a transformation performance between $|\psi_{in}\rangle$ and $|\psi_{out}\rangle$ states as,

$$f = Tr\left[\left(\sqrt{\rho_{out}}\,\rho_{in}\sqrt{\rho_{out}}\right)^{1/2}\right]. \tag{14}$$

This last metric is the most important one used in Quantum Teleportation [15], which is reasonable, given that, fidelity $f$ is a metric that shows us how similar the initial and final states in a quantum process are [19]. It has widely been used to characterize the performance of various quantum information tasks. Fidelity $f$ is bounded by $0 \leq f \leq 1$, where the unit fidelity ($f = 1$) implies that the initial and final states are equivalent [20, 21]. It gives us a protocol for its application on any gate used in a quantum algorithm implemented on a QPU. Let us consider a simple example, the fidelity of an *H* (Hadamard) gate. For this case, we resort to Eq.(15) where the multiplication of two Hadamard matrices with each other results in,

$$|\psi_{out}\rangle = H\,H\,|\psi_{in}\rangle = \left(\begin{bmatrix}1 & 1\\ 1 & -1\end{bmatrix}\frac{1}{\sqrt{2}}\right)\left(\begin{bmatrix}1 & 1\\ 1 & -1\end{bmatrix}\frac{1}{\sqrt{2}}\right)|\psi_{in}\rangle = \begin{bmatrix}1 & 0\\ 0 & 1\end{bmatrix}|\psi_{in}\rangle = |\psi_{in}\rangle. \tag{15}$$

The fidelity measured in Eq.(14) will indicate the noisy nature of gate *H* of Eq.(15) which will depend on how far fidelity is from 1. It is evident that for this elementary example both density matrices of Eq.(13) should be equal, given the identity of Eq.(15), with a fidelity like that represented in Eq.(14) clearly equal to 1, however, laboratory tests on an optical circuit [15] gives us a very different value to 1. Let us imagine then a technique of internal representation of the image that makes excessive use of this type of gates, and we have not even reached the quantum algorithm yet. In other words, excessive use of this type of gates by the internal image representation technique generates phase noise regardless of the quantum algorithm used, in fact, it unnecessarily feeds said quantum algorithm also with noise. Gate noise and measurement on more than one axis of the Bloch's sphere represent an explosive cocktail when correctly recovering the outcomes resulting from a QImP process.

We can do this with all the gates used by the process since said gates are reversible and unitary. If we do not work with an optical circuit, then we can request this data from the QPU manufacturers [6, 7], which are made collectively obvious by decoherence shown in the results. What we cannot do is ignore the noise of the gates, decoherence, and the consequences of the indiscriminate use of those gates, among others. Besides, so many gates inefficiently consume the coherence time of all the qubits involved in relation to the timeline of the entire process, i.e., from input to output, starting with the preparation of the qubits (or their intake thanks to a Cl2Qu interface), and continuing with the huge process involved behind FRQI and NEQR, the quantum algorithm and the Qu2Cl interface. Worse would be the case if the aforementioned quantum algorithm or the Cl2Qu interface included entangled particles [12-14, 22-24], i.e., a spurious entanglement process at the exit of NEQR that we will see in detail in Sections 4 and 6, and that is known as *entanglement coupling*. That is, FRQI and NEQR move away from all the recommendations of the QPU manufacturers [6, 7], as well as, any reasonable laboratory practice on an optical circuit.

It is extremely curious that all the QImP papers that use FRQI and NEQR implemented on a classic machine and coded in an HLI [2] do not even mention this problem. For this reason it is obvious why this sensitive topic of the quantum world is absent in those works, losing their scientific validity. In fact, there is no QImP paper that mentions any of these topics.

### *2.7 Special Gate Requirements*

Regularly, both the techniques for the internal representation of an image as well as the quantum algorithms that derive from these have special requirements of quantum gates and commands that cannot be implemented on a QPU. In this section, we will explain those special requirements.

***Anti-control (or zero control) condition on a qubit being OFF:*** There is no QPU, such as IBM Q [6], or Rigetti [7] (in fact, not even their simulators or any other quantum simulator like Quantum Programming Studio [8]) that directly accepts this condition. Only Quirk simulator does [9]. As we will see in this work, NEQR, quantum comparator, quantum counter, and practically all the quantum algorithms used in the papers of QImP require this type of control. Figure 3 shows both types of conditions, i.e., condition on a qubit being OFF and ON.

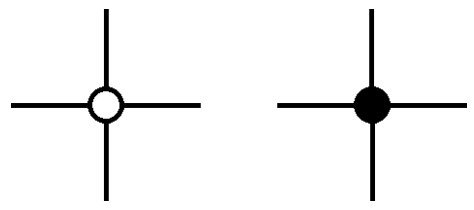

**Figure 3.** Anti-control (or zero control) condition on a qubit being OFF on the left, and control condition on a qubit being ON on the right.

However, in certain cases, we can construct an anti-control condition using a control condition, as shown in Fig.4 where both version are equivalent.

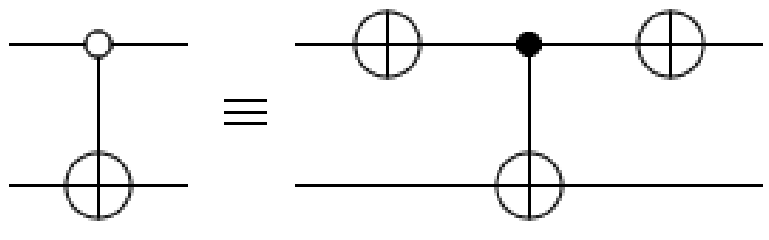

**Figure 4.** Building an anti-control condition from a control condition.

NEQR, quantum comparator and quantum counter, among many others, require this type of control in a more general context like that of Fig.5.,These configurations can only be implemented on a QPU at the cost of an excessive number of additional $X$ gates which introduce the noises saw in Subsection 2.6, which undoubtedly complicates the implementation of NEQR and its associated quantum algorithms.

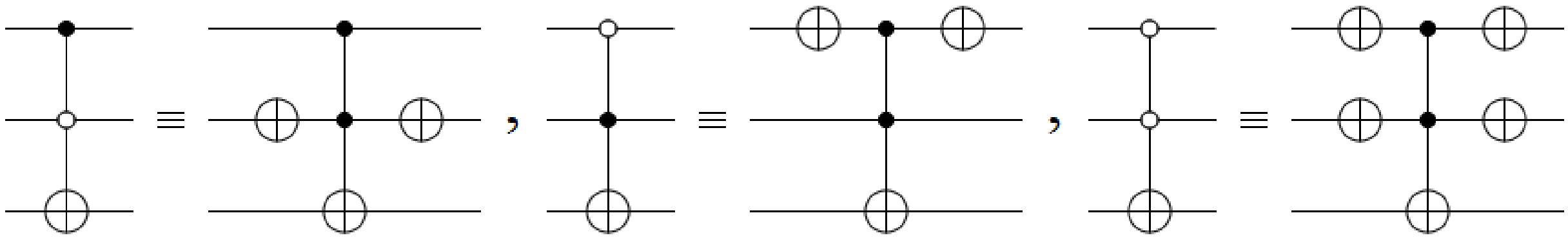

**Figure 5.** This type of control cases cannot be directly implemented on a QPU, were it not for their equivalences, which inappropriately increase the number of gates involved.

***if-then-else statement:*** There is a general consensus in both companies [6, 7] about the difficulties in the practical implementation of this statement. To date, none have succeeded. On the other hand, said companies have advanced so much that they do not foresee implementing this statement in the medium term, to be possible. It is important to clarify that, unfortunately, there is no QImP paper that does not use this statement. But from the response of both manufacturers, we see that we cannot implement an "*if-then-else*" statement even for a CBS. A similar difficulty is found when trying to implement this statement in an optical circuit [15], where the result of an apparent "*if-then-else*" is an episodic presence of certain quantum gates, in which researchers emulate the operational presence of the statement by hand positioning the optional gates derived from it, or by means of a non-optical electronic circuit relative to the aforementioned configuration. That is to say, in the optical circuit, an action of *putting or taking out* gates is used to emulate the presence of an "*if-then-else*" statement.

Notwithstanding the above, it is a matter of time for the physical implementation of this statement to take place, but, what type of qubits will this be for? Definitely, this statement will be strongly conditioned by the type of qubit used by the internal image representation technique, since it is known that it is impossible to clone [25] or compare [26] two generic qubits (as is the case required by FRQI [1]), with only one exception: if both qubits are CBS $\{|0\rangle,|1\rangle\}$, as in the case of QBIP [4].

Summing-up, a generic qubit cannot be cloned, copied or compared, but, why is this so?

*No-Cloning Theorem:* This theorem states that [25]: *There is no unitary operator U acting on* $\mathrm{H}_4^{A\cup B} = \mathrm{H}_2^A \otimes \mathrm{H}_2^B$ *in such a way that for all the normalized states* $|\psi\rangle_A$ *and* $|e\rangle_B$ *in* **H** *comply with* $U\left(|\psi\rangle_A |e\rangle_B\right) = e^{i\alpha(\psi,e)} |\psi\rangle_A |\psi\rangle_B$ *for some real number* $\alpha$ *depending on* $\psi$ *and* $e$.

To prove this theorem, an arbitrary pair of states $|\phi\rangle_A$ and $|\psi\rangle_A$ are selected in a Hilbert's space **H**, then, since *U* is a unitary operator,

$$\begin{aligned}\langle\phi|\psi\rangle\langle e|e\rangle &\equiv \langle\phi|_A \langle e|_B |\psi\rangle_A |e\rangle_B = \langle\phi|_A \langle e|_B U^{\dagger} U |\psi\rangle_A |e\rangle_B \\ &= e^{-i(\alpha(\phi,e)-\alpha(\psi,e))} \langle\phi|_A \langle\phi|_B |\psi\rangle_A |\psi\rangle_B = e^{-i(\alpha(\phi,e)-\alpha(\psi,e))} \langle\phi|\psi\rangle^2\end{aligned} \tag{16}$$

Since the quantum state $|e\rangle$ is assumed to be normalized, i.e., $\langle e|e\rangle = 1$, then we get

$$\left|\langle\phi|\psi\rangle\right|^2 = \left|\langle\phi|\psi\rangle\right|, \tag{17}$$

This implies that either $\left|\langle\phi|\psi\rangle\right| = 1$ or $\left|\langle\phi|\psi\rangle\right| = 0$. Hence by the Cauchy–Schwarz inequality either $\phi = e^{i\beta}\psi$ or $\phi$ is orthogonal to $\psi$. However, this cannot be the case for two arbitrary states. Therefore, a single universal *U* cannot clone a general quantum state. This proves the No-Cloning Theorem [25] with one exception, since if both states $|\phi\rangle$ and $|\psi\rangle$ are CBS $\{|0\rangle,|1\rangle\}$, then, Eq.(17) is fulfilled without problems, which is the case that supports QBIP [4],

$$\left(\left|\langle 0|0\rangle\right|^2 = \left|\begin{bmatrix}1 & 0\end{bmatrix}\begin{bmatrix}1\\0\end{bmatrix}\right|^2 = |1|^2 = 1\right) = \left(\left|\langle 0|0\rangle\right| = \left|\begin{bmatrix}1 & 0\end{bmatrix}\begin{bmatrix}1\\0\end{bmatrix}\right| = |1| = 1\right), \tag{18a}$$

$$\left(\left|\langle 1|1\rangle\right|^2 = \left|\begin{bmatrix}0 & 1\end{bmatrix}\begin{bmatrix}0\\1\end{bmatrix}\right|^2 = |1|^2 = 1\right) = \left(\left|\langle 1|1\rangle\right| = \left|\begin{bmatrix}0 & 1\end{bmatrix}\begin{bmatrix}0\\1\end{bmatrix}\right| = |1| = 1\right), \tag{18b}$$

Which justifies why an internal image representation technique such as QBIP [4] can clone, copy and compare its qubits without any problem.

***What can and what cannot be done on a QPU?*** Quantum computer manufacturers have established to date what gates and statements can and cannot be implemented on their QPUs. These manufacturers defined a series of problematic or teratological gates, whose implementations were doubtful and which are represented in Table I.

**TABLE I:** Possibility to implement problematic gates on QPUs

| Gate | IBM Q Experience [6] | Rigetti [7] |
|---|---|---|
| barrier | Yes | No |
| qubit reset \|0> | No | Yes (but, under some conditions) |
| if-then-else statement | No | No |
| quantum measurement in 3 axis | No | No |

Given that QImP papers require these gates directly or indirectly, we will describe the possibility of their physical implementation according to the manufacturers of quantum computers: IBM Q [6] and Rigetti [7]. On the other hand, Table I is ordered from lower to higher complexity of implementation, from top to bottom, respectively.

*Barrier:* This gate allows us to interrupt a process of simplification (synthesis or optimization) of gates by the same quantum platform, which would alter the original quantum algorithm. This is consistent with what we have said about these devices, which have the natural tendency to use the least number of gates possible. This *barrier* gate, which can include all the qubits involved, interrupts this tendency in a compulsory way forcing the processor to involve all the present gates in the prefixed layout. For now, it is only used by IBM Q [6]. The Rigetti's QPU [7] does not make those optimizations *per se*.

*Qubit reset gate* [|0>]*:* This gate is essential in QImP since it is required in several instances of FRQI, NEQR and the quantum algorithms, since it is used for the purpose of cleaning the registers (with ancillas) in order to be reused. Let us recall the excessive use of qubits and gates by FRQI, NEQR as well as the quantum algorithm that is coupled to their outputs. IBM Q [6] directly does not have this tool in its toolbox. Maybe, IBM Q can incorporate it in the future. Although Rigetti [7] only allows us to use it at the beginning of each algorithm in order to clean the registers, it does not allow us to do it anywhere else today. There is a possibility that Rigetti [7] could implement a more ductile version of this gate in the future with a shorter time than decoherence time.

*If-then-else statement:* Everything related to this statement was previously developed in this same subsection.

*Quantum measurement with total precision on the three axis of Bloch's sphere:* This is definitively and unequivocally impossible as it was explained above. The reality is that IBM Q [6] only gives us a version of Eqs.(11) and (12), so does Rigetti [7]. In other words, we completely lose the notion of the projections of outcomes on the plane established by the $x$ and $y$ axes. The papers of QImP that use

FRQI and NEQR need to measure the projections on the 3 axes of the Bloch's sphere with total precision. Therefore, this indicates that we are facing a serious problem since decoherence itself generates spurious projections on axes not originally involved, degrading the correct values in the axes that do matter. Besides, the quantum measurement can only be implemented at the end of the entire process in a QPU [6, 7] and not in intermediate instances as simulators [6-9] allow or as indicated by several QImP papers.

Another very common problem in the papers of QImP is the use of an indiscriminate quantum Arithmetic, in particular, an excessive use of the quantum addition. It is perfectly possible to do a quantum adder for CBS words using XOR (for the addition) and AND (for the carry). However, a complete Arithmetic like a generic adder of qubits "+" is not possible. For example, if quantum Arithmetic was freely implemented, then the following could be done:

$$(I+Z)/2 = 1/2\begin{bmatrix}1 & 0\\ 0 & 1\end{bmatrix} + 1/2\begin{bmatrix}1 & 0\\ 0 & -1\end{bmatrix} = \begin{bmatrix}1 & 0\\ 0 & 0\end{bmatrix}. \tag{19}$$

This gate is no unitary, and no reversible, in fact, it is a qubit reset gate [|0>]. Specifically, none of the free access QPUs in the cloud allow us to define our own gates as a result of examples such as Eq.(19), where the innocent sum of two Pauli's gates gives a new gate unacceptable by the QPU. Therefore, we must be very careful when including algorithms that include this type of practice in our papers. That is, QImP papers employ gates that may never be physically implemented. This predicament does not justify the response of several authors of QImP that say: "when quantum computers are developed our algorithms can be implemented". What we are showing here has to do with the expert opinion of professionals working in the main companies in the area, which admit that several tools required by the papers of QImP will be of dubious implementation in the future as well as impossible on their QPUs.

### *2.8 Entanglement coupling*

This is a spurious effect that occurs when we resort to an indiscriminate use of *H* (Hadamard) gates combined with *CNOT* gates, the result of which can be so devastating that it nullifies the original functionality of the quantum algorithm in question. In fact, as we will see in Section 4, this devastating effect generates at the output of NEQR [3] an undesirable entanglement between the qubits that represent the values of each pixel and the qubits that represent their respective positions, whereby different things like values of pixels and their respective positions change together to similar states, constituting something without any sense.

### *2.9 Final considerations*

In this section, we have seen all the elements that every technique of internal representation of an image must take into account. If this technique does not meet all these conditions, any traceability analysis is impossible (going back to our steps, from the outcomes to the inputs) in case something does not work out as expected, in an environment as ecliptic as that of Quantum Mechanics [27]. In regards to the QImP literature, everything said so far leads us to a clear preliminary conclusion: there is a great deal of confusion in a large part of the QImP community about what can and cannot be done on a QPU [6, 7] or in an optical circuit [15]. It was difficult to understand, then, why the resistance to review these issues in a laboratory, despite the fact that, for years, the vast majority of experts in Quantum Mechanics [27], who do not work in QImP, have had doubts about the fidelity of the results obtained by part of the authors of QImP, who work on classic computers, and code their techniques and quantum algorithms in an HLI [2].

Summing-up, the fundamental reason for the doubt is that the community of authors of QImP obtains results with classic computers programmed in an HLI [2], that the rest of the scientists of Quantum Mechanics [27] have never been able to obtain with QPUs [6, 7] or optical circuits [15] in their respective areas. It is clear that QImP cannot be an island within Quantum Mechanics [27], therefore, the conclusion is evident, something must be wrong.

## 3. Flexible Representation of Quantum Images (FRQI)

### *3.1 Scheme of operation*

In this section, we will analyze the most outstanding features of FRQI [1] and its behavior against the aspects established in the previous section, only those that affect FRQI the most, therefore, we will begin directly with its way of representing an image, which is established by Eq.(20)

$$\left|I(\theta)\right\rangle = \frac{1}{2^n}\sum_{i=0}^{2^{2n-1}}\left(cos\,\theta_i\left|0\right\rangle + sin\,\theta_i\left|1\right\rangle\right)\otimes\left|i\right\rangle, \tag{20}$$

$$\theta_i \in \left[0,\frac{\pi}{2}\right], \quad i = 0,1,\ldots,2^{2n}-1, \tag{21}$$

where the angles $\theta_i$ encode the colors of the pixels of the corresponding tile, while $\left|i\right\rangle$ represents the location of the elements of said tile by a horizontal rafter. The FRQI state is a normalized state, i.e., $\left\|I(\theta)\right\| = 1$ as given by

$$\left\|I(\theta)\right\| = \frac{1}{2^n}\sqrt{\sum_{i=0}^{2^{2n}-1}\left(\cos^2\theta_i + \sin^2\theta_i\right)} = 1 \tag{22}$$

Figure 6 shows an example for an image of 2-by-2 pixels and its corresponding codification.

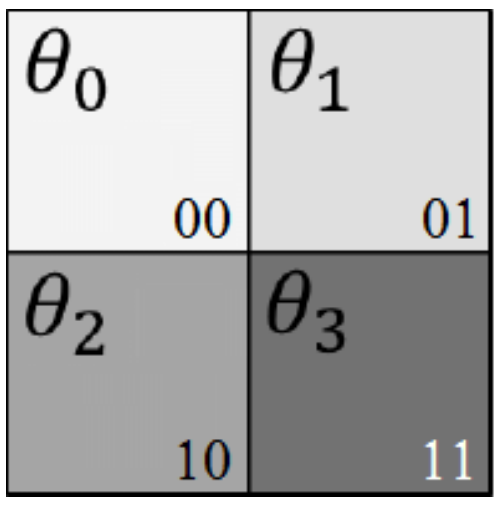

$$|I(1)\rangle = \frac{1}{2}[(\cos\theta_0\,|0\rangle + \sin\theta_0|1\rangle)\otimes|00\rangle + (\cos\theta_1|0\rangle + \sin\theta_1|1\rangle)\otimes|01\rangle + (\cos\theta_2\,|0\rangle + \sin\theta_2|1\rangle)\otimes|10\rangle + (\cos\theta_3|0\rangle + \sin\theta_3|1\rangle)\otimes|11\rangle]$$

**Figure 6.** FRQI for and image of 2-by-2 pixel.

Basically, as we have already mentioned, FRQI attempts to encode the colors of a classical image through angles. This is done from identity and Hadamard's matrices (in fact, poorly defined on page 66 of the original paper [1]) which are applied on a battery of ancillas of type $\left|0\right\rangle$ and the intervention of the location of the pixels $\left|i\right\rangle$ to arrive at a state $\left|H\right\rangle$. From here, when we apply the matrices of rotation on this state, finally $\left|I(\theta)\right\rangle$ is reached. With this technique, it is even difficult to imagine the human intervention at the time of preparing the qubits by hand. This technique could not be more counterintuitive and more difficult to implement in the world of quantum computers [6, 7].

### *3.2 Cl2Qu interface and data intake*

In all the equations that make up this extensive, contrived and counterintuitive procedure, we have not only lost the traceability but the very existence of the original image (classic, i.e., digital). The classic image to be introduced or represented internally in the quantum machine is not present anywhere. On the other hand, the procedure as a whole, as well as each of its constituent elements in particular, do not respect any of the conditions to be met by a Cl2Qu interface and that are necessary to be implemented on a QPU. See Subsection 2.2.

### *3.3 Characteristics of the employed qubits*

It is clear that the final state of this technique $|I(\theta)\rangle$ has a projection on more than one axis of the Bloch's sphere. Therefore, the use of *if-then-else* statement as well as more general quantum comparators [28] is absolutely forbidden. We should remember from Subsection 2.7 (What can and what cannot be done on a physical quantum computer?) that any comparison statement between qubits can only be applied on a CBS, exclusively.

### *3.4 Sparsity, number and size of the required registers*

*3.4.1 Number of gates and computational cost*

Consistent with what was said in the previous subsection, FRQI requires an excessive amount of gates to try to express the simplest and insignificant thing, as in fact happens with Eq.(20) for a 2-by-2 pixel image of Fig.6. This triggers two major problems:

a) *Consumption of the coherence time:* every gate has a time of implementation. The sum of all of them constitutes the execution time of the quantum algorithm in the QPU. If the gates are too many, there is a risk that the execution time of the quantum algorithm will be greater than the coherence time of its constituent states (the weakest link in the chain), therefore, the quantum algorithm will not be fully executed, that is, we will not get the correct outcomes.

b) *Computational cost:* unlike QBIP, in FRQI we must introduce all the 24 bitplanes of the original image in the quantum machine, since the very first moment. QBIP instead works with only 3 of the 24 bitplanes of the original image, since it only interacts with the Most-Significant-Bit (MSB) of each color [4, 29]. The saving of resources of the QPU is evident when using QBIP instead of FRQI.

*3.4.2 Number, dimensionality and size of registers*

From Eq.(20) and Fig.6, we can see the intervention of the Kronecker's product inside FRQI. Subsection 2.4 (Sparsity, number and size of the required registers) warns us about these types of practices, which require the use of an immense amount of large registers (sparsity) by the QPU. Besides, we must remember that the difficulty of implementing the technique of internal representation of the image, in this case FRQI, unequivocally implies the difficulty of implementing the quantum algorithm that follows it, given that FRQI is the one that makes the intake of the mentioned quantum algorithm. In this way, we cannot speak of the same version of the quantum algorithm to be implemented for FRQI as for QBIP for a same problem to be solved. In other words, the version to be implemented of the algorithm is dramatically more complex in FRQI than in QBIP. This directly impacts everything said above, in particular, the computational cost, as well as the storage.

*3.4.3 Storage*

The previous subsection speaks by itself, however, we still need to consider the interfaces. In the same way that we talk about an excessive requirement of resources for calculations, we must also mention a similar requirement in storage. Here also, the dramatic difference between FRQI and QBIP is conserved as explained in Subsection 3.5. It is curious that the authors of QImP omit in their papers the complete processing and storage requirements of FRQI: Cl2Qu interface (if any) + FRQI + Quantum Algorithm + Qu2Cl interface. We must always keep in mind when making a comparison of storage between FRQI and QBIP that the latter exclusivelly works with 3 of the 24 bitplanes of the original image, then, we should not assign to QBIP a storage that it does not have.

### *3.5 Quantum measurement and the complementarity problem*

In this subsection, we will describe the difficulties related to the recovery of outcomes in FRQI thanks to the problem of quantum measurement and complementarity. As we can imagine, this is because $|I(\theta)\rangle$ of the Eq.(20) has projections on more than one axis of the Bloch's sphere. Consequently, the single procedure of measuring along a single axis of the Bloch's sphere, for a technique that is expressed on more than one axis, introduces noise. If we return to the example of Subsection 2.1, where we hypothetically considered a quantum algorithm for filtering an image, we will find what is indicated in [5], that is, the process of recovery of outcomes for FRQI introduces more noise than that which supposedly eliminated the quantum algorithm used. Moreover, we have not even taken into account decoherence, which is completely related to the criteria used in the papers of QImP with implementations exclusively in an HLI [2], which is completely different from the physical reality.

Let us see this in detail, FRQI [1] captures information about colors and the corresponding positions of those colors, where $\otimes$ is the Kronecker's product, $|0\rangle$, $|1\rangle$, are the $2\times1$ computational basis states (CBS), $i = 0, 1, \ldots, 2^{2n}-1$ are the $2^{2n}\times1$ CBS, and $\theta = \left(\theta_0, \theta_1, \ldots, \theta_{2^{2n}-1}\right)$ is the vector of angles encoding colors. There are two parts in the FRQI representation of an image; $cos\,\theta_i\,|0\rangle + sin\,\theta_i\,|1\rangle$ which encode the information about colors, and $|i\rangle$ is the corresponding position in the image, respectively. An example of a $2\times2$ image is shown in Fig. 6.

Now, according to Eq.(21),

$$cos\,\theta_i = \left|cos\,\theta_i\right| \geq 0, \quad and \quad sin\,\theta_i = \left|sin\,\theta_i\right| \geq 0, \quad \forall i, \tag{23}$$

given that for that range in the angle values of $\theta_i$, $cos\,\theta_i$ and $sin\,\theta_i$ are always positive.

If we call $|f_i\rangle$ to $cos\,\theta_i\,|0\rangle + sin\,\theta_i\,|1\rangle$, we will have,

$$|f_i\rangle = cos\,\theta_i\,|0\rangle + sin\,\theta_i\,|1\rangle = cos\,\theta_i \begin{bmatrix}1\\0\end{bmatrix} + sin\,\theta_i \begin{bmatrix}0\\1\end{bmatrix} = \begin{bmatrix}cos\,\theta_i\\ sin\,\theta_i\end{bmatrix}, \tag{24}$$

where the respectives density matrices for each CBS $|0\rangle$, and $|1\rangle$ will be,

$$M_{|0\rangle} = |0\rangle\langle 0| = \begin{bmatrix}1\\0\end{bmatrix}\begin{bmatrix}1 & 0\end{bmatrix} = \begin{bmatrix}1 & 0\\0 & 0\end{bmatrix}, \text{ and} \tag{25}$$

$$M_{|1\rangle} = |1\rangle\langle 1| = \begin{bmatrix}0\\1\end{bmatrix}\begin{bmatrix}0 & 1\end{bmatrix} = \begin{bmatrix}0 & 0\\0 & 1\end{bmatrix}. \tag{26}$$

Therefore, applying projective measurement on $|f_i\rangle$, the impact of quantum measurement [16] on FRQI is evidenced,

$$\begin{aligned}\left|f_{i,|0\rangle}\right\rangle_{pm} &= \frac{M_{|0\rangle}\,|f_i\rangle}{\sqrt{\langle f_i|\,M_{|0\rangle}M_{|0\rangle}\,|f_i\rangle}} \\ &= \frac{\begin{bmatrix}1 & 0\\0 & 0\end{bmatrix}\begin{bmatrix}cos\,\theta_i\\ sin\,\theta_i\end{bmatrix}}{\sqrt{\begin{bmatrix}cos\,\theta_i & sin\,\theta_i\end{bmatrix}\begin{bmatrix}1 & 0\\0 & 0\end{bmatrix}\begin{bmatrix}1 & 0\\0 & 0\end{bmatrix}\begin{bmatrix}cos\,\theta_i\\ sin\,\theta_i\end{bmatrix}}} = \frac{\begin{bmatrix}cos\,\theta_i\\ 0\end{bmatrix}}{\sqrt{cos^2\,\theta_i}} = \frac{cos\,\theta_i}{\left|cos\,\theta_i\right|}\begin{bmatrix}1\\0\end{bmatrix} = |0\rangle,\end{aligned} \tag{27}$$

$$\left| f_{i,|1\rangle} \right\rangle_{pm} = \frac{M_{|1\rangle} | f_i \rangle}{\sqrt{\langle f_i | M_{|1\rangle} M_{|1\rangle} | f_i \rangle}}$$

$$= \frac{\begin{bmatrix} 0 & 0 \\ 0 & 1 \end{bmatrix} \begin{bmatrix} cos\,\theta_i \\ sin\,\theta_i \end{bmatrix}}{\sqrt{\begin{bmatrix} cos\,\theta_i & sin\,\theta_i \end{bmatrix} \begin{bmatrix} 0 & 0 \\ 0 & 1 \end{bmatrix} \begin{bmatrix} 0 & 0 \\ 0 & 1 \end{bmatrix} \begin{bmatrix} cos\,\theta_i \\ sin\,\theta_i \end{bmatrix}}} = \frac{\begin{bmatrix} 0 \\ sin\,\theta_i \end{bmatrix}}{\sqrt{sin^2\,\theta_i}} = \frac{sin\,\theta_i}{|sin\,\theta_i|} \begin{bmatrix} 0 \\ 1 \end{bmatrix} = |1\rangle. \tag{28}$$

In other words, Eqs.(27) and (28) highlight the effect of quantum measurement [16] on FRQI, which had already been set out in an earlier paper [5]. Although the result of positive-operator valued measure (POVM) has a module practically equal to one in all cases, in FRQI, which encodes the colors by means of states such as those of Eq.(24), Eqs.(27) and (28) have values identical to the case of CBSs. Therefore, we must to explore the impact of quantum measurement on the color information of the image coded in $|f_i\rangle$, which represents a factor of central importance in FRQI. This will be done in Subsection 6.1. Finally, in addition to not complying with the minimum conditions agreed in Section 2 for a Cl2Qu interface, FRQI does not meet any other conditions, in fact, and precisely as seen Subsection 6.1, it cannot keep the quality of the original image due to two reasons: a) decoherence for accumulating so many gates per qubit, and b) the high level of noise introduced into the outcomes during the measurement. This is what we have seen throughout these years every time we try to implement FRQI on a platform with cloud services such as IBM [6] and Rigetti [7], although the same can be seen on an optical circuit [15].

### *3.6 Gate noise, decoherence and fidelity*

It is evident that the complete FRQI procedure that goes from the ground states $|0\rangle$ to $|I(\theta)\rangle$ implies an immense amount of quantum gates. As we have seen in Subsection 2.6, this is directly proportional to the noise introduced as a whole by said gates in the obtained results. This decoherence weakens the fidelity of each block impacting the final quality of the outcomes. As we have indicated previously, no QImP paper mentions a single word about this. Obviously, neither their simulations in an HLI [2] show this problem.

### *3.7 Special Gate Requirements*

This technique does not require special gates or anti-control (or zero control) condition.

### *3.8 Entanglement coupling*

So far, we have not been able to detect the entanglement coupling in this technique, however, our fine analysis of this problem could never take place in the presence of the main FRQI problem, as explained in Subsection 3.5 and as it will be demonstrated experimentally in Section 6, which masks all other problems.

### *3.9 Final considerations*

If we consider the idea of implementing FRQI in the simulators [6-9], or on the QPUs of IBM Q [6] and Rigetti [7], we will quickly find the first difficulty that distances us from physical reality and that consists in the manual preparation of the qubits of FRQI, in the absence of a Cl2Qu interface. The very idea of introducing the image in the FRQI procedure to make a test on a simulator and even worse, on a quantum machine is impractical. There is not even in the study phase a single interface in literature that can automatically feed FRQI in the future. Even the manual work of preparing qubits in a laboratory by human beings for this task is absolutely not intuitive. To this, we must add the absolute loss of traceability from the pixel to be treated until the eventual recovered outcome. When the tracea-bility of the pixels is lost within the set of steps involving FRQI, we also lose any possibility of going backwards to analyze errors and thus individualize the block (or blocks) responsible for them.

Summing-up, we obtained some fundamental conclusions for an eventual practical implementation of FRQI:

a) Since FRQI consists in coding the color of the pixels through angles, it is difficult to imagine a Cl2Qu interface that manages to do so even in an inefficient way.
b) To this, we must add that traceability is lost even more because the location of the pixels is represented by a horizontal rafter of the considered mosaic, that is, for every row all the columns are traversed generating a vector of locations as a result of the aforementioned horizontal rafter.
c) Alternative paths to those indicated in [1] to obtain $|I(\theta)\rangle$ are not valid regarding the practical impossibility of implementing simple Arithmetic operations to be able to operate with arbitrary qubits or between gates, as in the case of the implementation of Pauli's Algebra on a QPU.
d) The simulations of the authors of QImP will eventually work because in their programs in an HLI [2] the physical realities mentioned here are not present.
e) The quantum algorithms used to solve the same problem are more complicated in FRQI than in QBIP as a consequence of the way that FRQI has of representing the image internally.
f) There is nothing more intuitive than the homotopic representation that QBIP uses to analyze the different phases in which the image is involved inside the QPU and that FRQI does not use.
g) Unfortunately, it is believed that the natural juxtaposition of quantum gates will always lead inexorably to the desired result. This implies a complete lack of knowledge regarding the impact of decoherence on the states involved, particularly if these are needlessly more than what is necessary.
h) We should consider that in an environment as complex as the quantum world with non-adiabatic machines subjected to decoherence and, therefore, with a fidelity significantly lower than 1 (due to the noise of the gates), the original recommendations are to try by all means the following:

   I. not to lose the traceability of the image inside the quantum machine by not losing the homotopic representation of it, something that FRQI does not do and QBIP does.
   II. to use the least number of gates (which are sources of noise) so as not to degrade the fidelity unnecessarily, being this a metric which all of us, who work in Quantum Information Processing [12-14], must pay close attention to.

## 4. Novel Enhanced Quantum Representation of digital images (NEQR)

### *4.1 Scheme of operation*

To improve FRQI, the newly proposed NEQR model uses two entangled qubit sequences to store the gray-scale and position information; and the whole image in the superposition of the two-qubit sequences. Suppose the gray range of image is $2^q$, binary sequence $C_{YX}^0 C_{YX}^1 \dots C_{YX}^{q-2} C_{YX}^{q-1}$ encodes the gray-scale value $f(Y, X)$ of the corresponding pixel $(Y, X)$ as in:

$$f(Y,X) = C_{YX}^0 C_{YX}^1 \dots C_{YX}^{q-2} C_{YX}^{q-1}, \quad C_{YX}^k \in [0,1], \quad f(Y,X) \in \left[0, 2^q - 1\right]. \tag{29}$$

The representative expression of a quantum image for a $2^n \times 2^n$ frame can be written as in:

$$|I\rangle = \frac{1}{2^n} \sum_{Y=0}^{2^{2n-1}} \sum_{X=0}^{2^{2n-1}} |f(Y,X)\rangle |Y X\rangle = \frac{1}{2^n} \sum_{Y=0}^{2^{2n-1}} \sum_{X=0}^{2^{2n-1}} \bigotimes_{i=0}^{q-1} \left|C_{YX}^i\right\rangle |Y X\rangle, \tag{30}$$

where, according to their authors [3], both $\left|C_{YX}^i\right\rangle$ and $\left|Y X\right\rangle$ are strictly CBS $\{|0\rangle, |1\rangle\}$, otherwise the technique would not work.

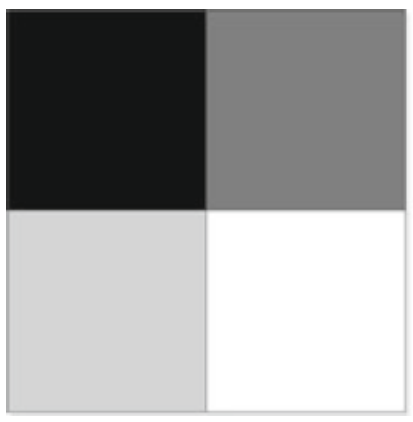

$$|I\rangle = \frac{1}{2}(|0\rangle \otimes |00\rangle + |100\rangle \otimes |01\rangle + |200\rangle \otimes |10\rangle + |255\rangle \otimes |11\rangle)$$
$$= \frac{1}{2}(|00000000\rangle \otimes |00\rangle + |01100100\rangle \otimes |01\rangle + |11001000\rangle \otimes |10\rangle + |11111111\rangle \otimes |11\rangle)$$

**Figure 7.** A 2×2 example image and its representative expression in NEQR

In Fig. 7 we have a 2×2 NEQR image example. According to this technique [3], making a series of transforms we must go from the representation of a pixel of the digital image of Eq.(29) to its quantum counterpart of Eq.(30). The example in Fig. 7 is represented, in detail, in the following equation,

$$\begin{aligned} |I\rangle &= \frac{1}{2}\begin{pmatrix} \left|C_{00}^0 C_{00}^1 C_{00}^2 C_{00}^3 C_{00}^4 C_{00}^5 C_{00}^6 C_{00}^7\right\rangle|00\rangle + \left|C_{01}^0 C_{01}^1 C_{01}^2 C_{01}^3 C_{01}^4 C_{01}^5 C_{01}^6 C_{01}^7\right\rangle|01\rangle \\ +\left|C_{10}^0 C_{10}^1 C_{10}^2 C_{10}^3 C_{10}^4 C_{10}^5 C_{10}^6 C_{10}^7\right\rangle|10\rangle + \left|C_{11}^0 C_{11}^1 C_{11}^2 C_{11}^3 C_{11}^4 C_{11}^5 C_{11}^6 C_{11}^7\right\rangle|11\rangle \end{pmatrix} \\ &= \frac{1}{2}\left(|00000000\rangle|00\rangle + |01100100\rangle|01\rangle + |11001000\rangle|10\rangle + |11111111\rangle|11\rangle\right), \end{aligned} \tag{31}$$

where the equivalent image $|I\rangle$ is represented by 4 CBS sequences $C_{YX}^0 C_{YX}^1 \ldots C_{YX}^{q-2} C_{YX}^{q-1}$ projected on the respective coordinates also represented by CBS, i.e., $|0\rangle$, and $|1\rangle$. Therefore, we only have to test this technique. We begin with the quantum circuit of NEQR [3] (its Fig. 7) on Quirk platform [9] in Fig. 8, where, the letters *B*, *P* and *D* at the top of the figure mean: *Bloch's sphere*, *probability of* $|1\rangle$, and *density matrix*, respectively. In this technique, the qubits are prepared according to the quantum transformation of $\Omega_{YX}$ to set a grayscale for the pixel,

$$\Omega_{YX}|0\rangle^{\otimes q} = \bigotimes_{i=0}^{q-1}\left(\Omega_{YX}^i|0\rangle\right) = \bigotimes_{i=0}^{q-1}\left|0 \oplus C_{YX}^i\right\rangle = \bigotimes_{i=0}^{q-1}\left|C_{YX}^i\right\rangle = \left|f(Y,X)\right\rangle. \tag{32}$$

This group of transformations involves 14 2-*CNOT* gates, 2 Hadamard's gates, and 8 identity matrices for the preparation of one pixel on a grayscale, however, the preparation of the qubits is by hand.

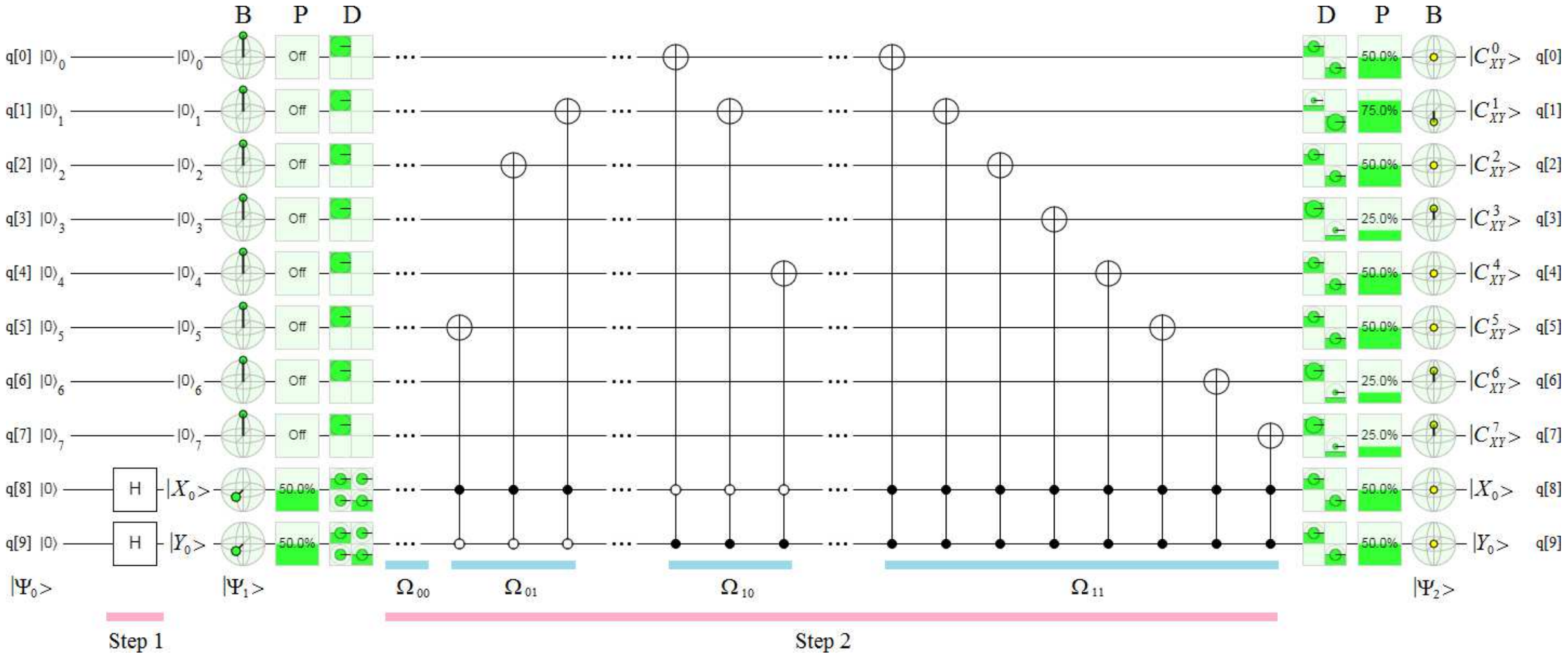

**Figure 8.** Implementation of NEQR on Quirk [9].

From left to right of Fig.8, NEQR consists of 2 steps. Step 1 starts with 10 CBS ground state $|0\rangle$, and 2 Hadamard's gates on the qubits q[8] and q[9] with which the procedure represents the coordinates of the pixel in the tile $(|Y_0\rangle,|X_0\rangle)$. If we analyze the qubits q[8] and q[9] of $|\Psi_2\rangle$, we see that they are entangled states, since both the Bloch's spheres and the density matrices indicate an entanglement between them, which is incorrect because the coordinates or location of the pixel would be always represented by CBS, i.e., $\{|0\rangle,|1\rangle\}$, see Fig. 7, and Eq.(31). But the worst is yet to come. None of the $|C_{YX}^i\rangle$ at the output of NEQR are CBS, as the first 8 elements of the vector $|\Psi_1\rangle$, i.e., none of the elements of the vector $|\Psi_2\rangle$ represent orthogonal qubits. Therefore, NEQR does not feed the quantum algorithm to its output with orthogonal qubits as mentioned in all papers that use NEQR [3]. In fact, NEQR gives three types of outputs completely different to CBS's:

$$wavefunction: |C_{YX}^{0,2,4,5}\rangle = \tfrac{1}{\sqrt{2}}|00\rangle + \tfrac{1}{\sqrt{2}}e^{i\varphi}|11\rangle \rightarrow statevector: |C_{YX}^{0,2,4,5}\rangle = (50\%|00\rangle, 50\%|11\rangle) \neq \text{CBS} \quad (33a)$$

$$wavefunction: |C_{YX}^{1}\rangle = \tfrac{\sqrt{3}}{2}|00\rangle + \tfrac{1}{2}e^{i\varphi}|11\rangle \rightarrow statevector: |C_{YX}^{1}\rangle = (75\%|00\rangle, 25\%|11\rangle) \neq \text{CBS} \quad (33b)$$

$$wavefunction: |C_{YX}^{3,6,7}\rangle = \tfrac{1}{2}|00\rangle + \tfrac{\sqrt{3}}{2}e^{i\varphi}|11\rangle \rightarrow statevector: |C_{YX}^{3,6,7}\rangle = (25\%|00\rangle, 75\%|11\rangle) \neq \text{CBS} \quad (33c)$$

This outcomes has nothing to do with CBS: $\{|0\rangle,|1\rangle\}$. Instead, they are maximally entangled states like q[0]=q[2]=q[4]=q[5] and $|Y_0\rangle = |X_0\rangle = |C_{YX}^{0,2,4,5}\rangle$, and not maximally entangled states like q[1] and q[3]= q[6]=q[7]. NEQR does not meet any of the required conditions of Section 2 for a Cl2Qu interface because its outcomes do not belong to a base, in fact, it does not even meet the working definition of NEQR [3], which are clear in Eqs.(29) and (31); and Fig. 7. Therefore, taking into account everything seen so far, we can draw some preliminary conclusions:

1. NEQR does not work as a Cl2Qu interface not even doing the data intake by hand,

2. The orthogonality dies at the output of NEQR and at the input of the quantum algorithm. Then, NEQR cannot sustain the orthogonality, let alone, the quantum algorithm because it is not its function,

3. These results completely impact the outcomes, since the $|C_{YX}^i\rangle$'s are not orthogonal having projections in more than one axis. Now, considering that the measurements in the QPU only measure with absolute precision in a single axis, then in addition to the noise of gates we will have measurement noise, given that the single act of measuring will introduce even more lack of precision in the absence of orthogonality and uniprojection,

4. Let us imagine what would happen if instead of analyzing NEQR we analyzed FRQI. It is evident that the results are even more catastrophic, and

5. As it is evident, all these results would worsen dramatically if Quirk [9] considered the noise, since even on the most permissive platform, NEQR did not pass the test.

6. In the presence of the loss of orthogonality and uniprojection, we must say goodbye to the use of *if-then-else* statement and quantum comparators [28] for NEQR [3].

7. Clearly, these results do not coincide with those obtained in an HLI [2] on a classic computer by the NEQR users-authors.

Another version of NEQR also obtained from [3] (its Fig. 9), which we implement in Fig. 9 on the Quirk platform [9] again, confirms the results of Fig. 8. There is nothing more to add, the scientific evidence speaks for itself.

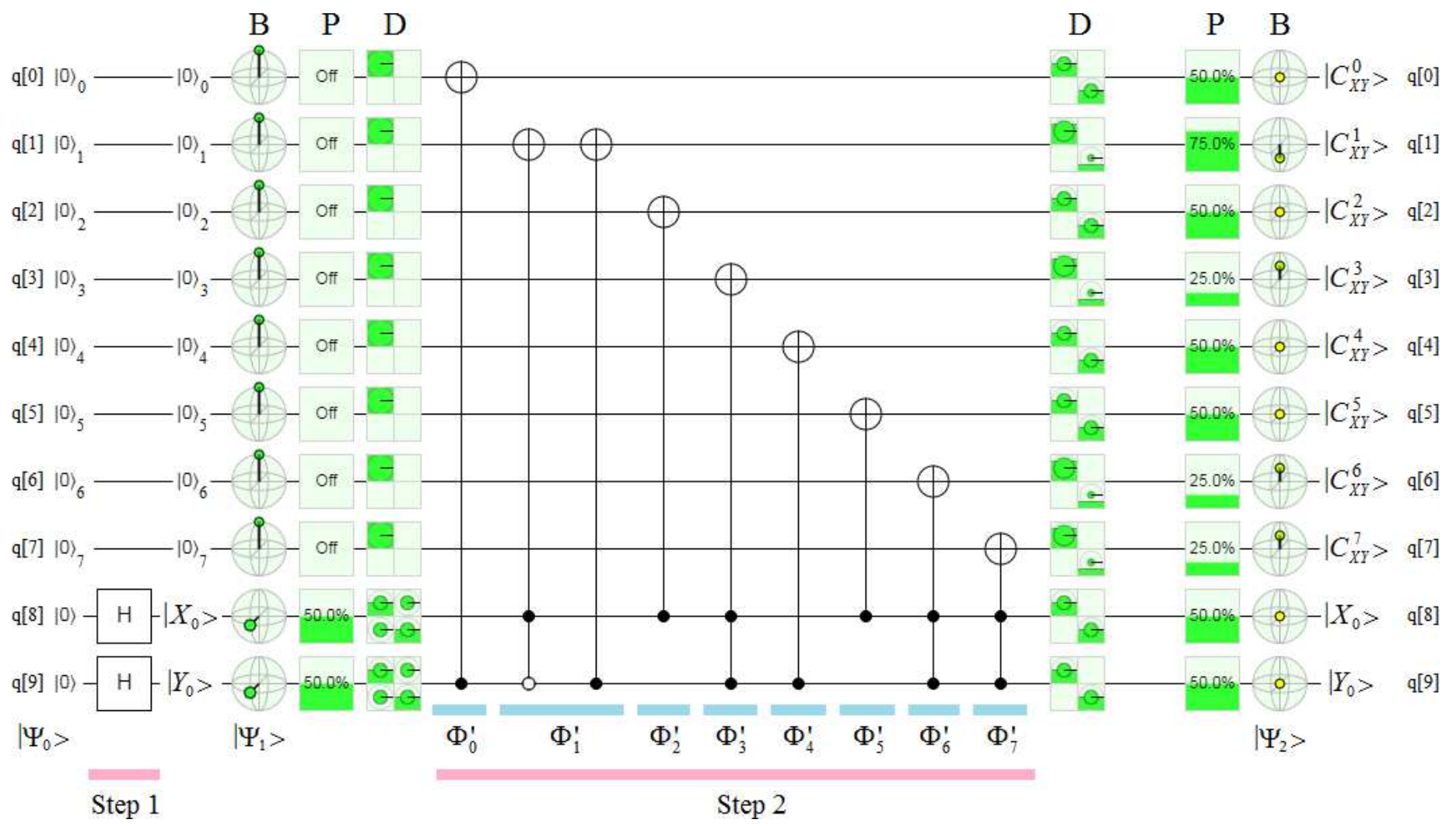

**Figure 9.** Implementation of an alternative version of NEQR on Quirk [9].

In Section 6, we will see the devastating effect of the seven items mentioned above.

### *4.2 Cl2Qu interface and data intake*

This technique [3] is defined from Eq.(30) and Fig.7, however, the complete procedure can be seen in Fig.10, which is structured in a series of consecutive steps.

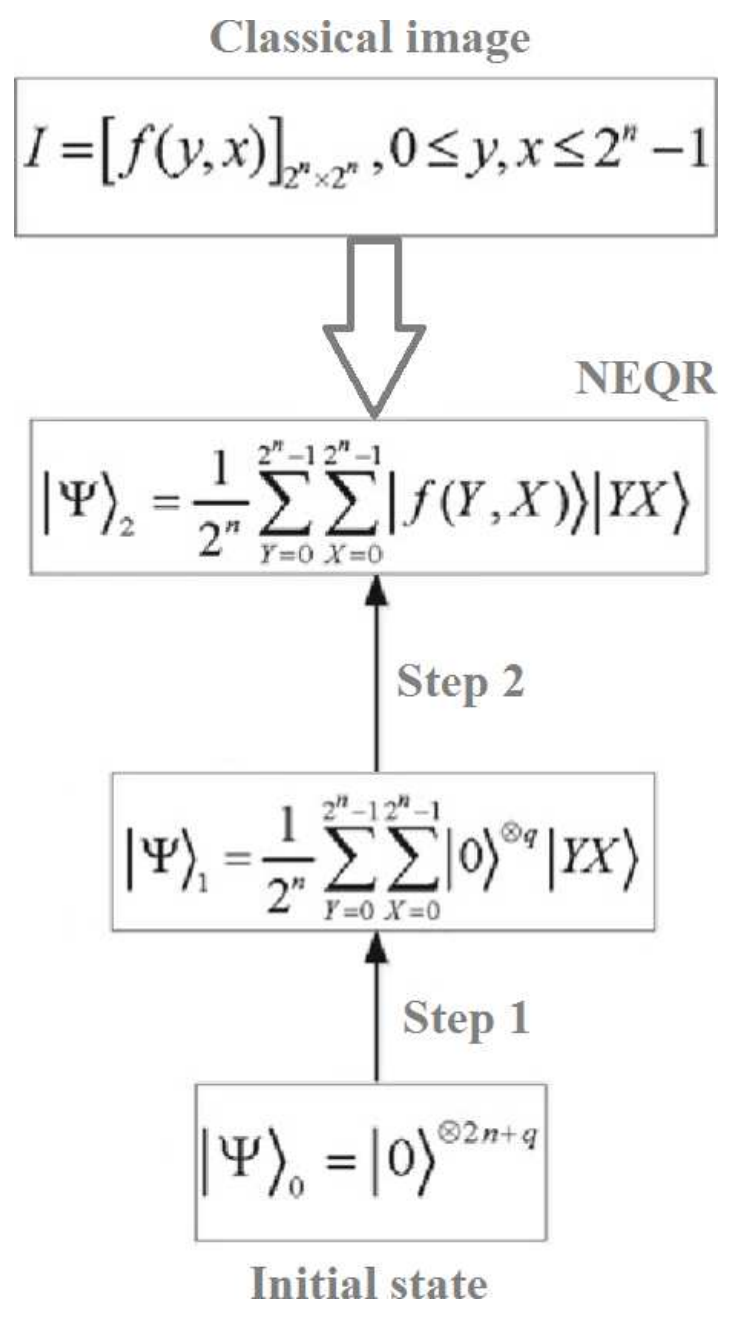

**Figure 10.** Complete scheme of NEQR.

The main problem consists in the intervention of the classical image in the process. It is difficult to understand how the classical image can be incorporated to the NEQR without a Cl2Qu interface, because, the practical problem treated so far consisting in a classic image of 1920-by-1080-by-3-by-8 bits (we are talking about 50 Million of bits, approximately) to be introduced by hand in a QPU seems to be quite a challenge. However, there is a more serious problem: papers that use NEQR, treat this as Cl2Qu interface, which is clearly incorrect.

Figure 11 clarifies the situation exposed in the last subsection, i.e., the qubits are prepared by hand according to the quantum transformation of $\Omega_{YX}$ to set the grayscale for the pixel, as we can see in Eq.(34),

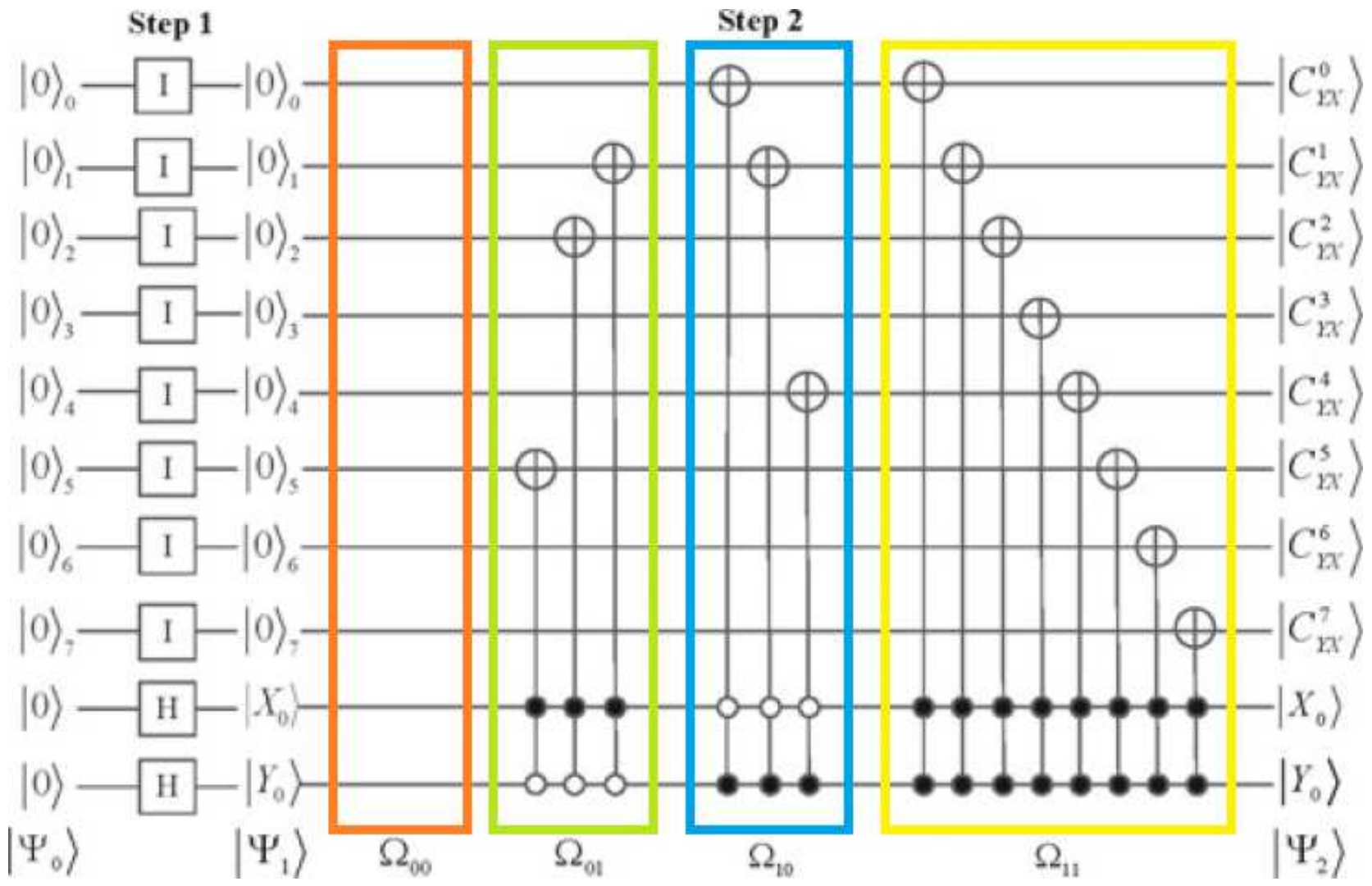

**Figure 11.** Quantum circuit for NEQR preparation of the image in Fig.10. In this case, 14 2−*CNOT* gates are needed to construct only one pixel of the quantum image.

$$\Omega_{YX}\left|0\right\rangle^{\otimes q}=\bigotimes_{i=0}^{q-1}\left(\Omega_{YX}^{i}\left|0\right\rangle\right)=\bigotimes_{i=0}^{q-1}\left|0\oplus C_{YX}^{i}\right\rangle=\bigotimes_{i=0}^{q-1}\left|C_{YX}^{i}\right\rangle=\left|f\left(Y,X\right)\right\rangle \quad (34)$$

This group of transformations involves 14 2-*CNOT* gates, 2 Hadamard's gates (*H*), and 8 identity matrices (*I*) for the preparation of one pixel in a grayscale, however, the preparation of the qubits is by hand. In other words, NEQR does not meet any of the conditions required by a Cl2Qu interface as established in Section 2 by the manufacturers of QPUs. To make matters worse, NEQR's outcomes are not CBS as indicated by Eq.(33) unlike what has been established by their authors [3] and users. All this must be added to the entanglement coupling problem that we will see in Subsection 4.8.

### *4.3 Characteristics of the employed qubits*

The intervention of so many *H* (Hadamard) and *CNOT* gates in an inappropriate way within NEQR, to encode a single pixel in 8 levels of gray, added to those of the corresponding quantum algorithm cause a race against the coherence time. The noise of so many gates conspires against the orthogonality of the constituent elements of NEQR, which affects its performance and that of the quantum algorithm. This loss of performance is expressed as the decoherence that corrupts the orthogonality of the intervening states. This is clearly seen in the implementations of Figures 8 and 9 on the Quirk [9] platform, where the outcomes are represented in Eq.(33). The loss of orthogonality by decoherence also conspires, undoubtedly, with the possibility of using *if-then-else* statement and quantum comparators [28].

### *4.4 Sparsity, number and size of the required registers*

#### *4.4.1 Number of gates and computational cost*

Figure 11 shows an excessive amount of gates to express the most simple and insignificant element of the image, i.e., a single pixel. Precisely in this case, NEQR requires 2 Hadamard's gates (*H*) and 14 2-*CNOT* gates with control, i.e., conditions on a qubit being ON; and with anti-control, or zero control, i.e., conditions on a qubit being OFF. The anti-control condition will not allow us to implement it on the QPUs of IBM Q [6], Rigetti [7] and Quantum Programming Studio [8] (QPS) in a simple way, as explained in Subsection 2.7, due to the requirement of too many *X* gates for it. From Fig.11, we ignore the presence of 8 identity matrices (*I*).

*4.4.2 Number, dimensionality and size of registers*

The presence of Kronecker's product in Eq.(30), Fig.10, and Eq.(32) are responsible for the circuital overexpression of Fig.11. The expansion in number, dimension and size of the registers (which are the host of the data inside QPU) to represent a single pixel impacts the race against coherence time and storage. We must keep in mind that all these difficulties are in exchange for nothing, since the data intake is at hand and still needs to perform a similar analysis with the quantum algorithm that is coupled to the output of NEQR.

*4.4.3 Storage*

The authors who use NEQR do not mention the storage of the different instances of the complete process, that is, the cumulative, since the Cl2Qu interface and the quantum algorithm to be used are still missing. The complexity and, therefore, storage required by NEQR for a single pixel is directly proportional to the complexity of the quantum algorithm at its output, hence, the storage required by them. The cost of introducing almost 50 million qubits by hand should not be underestimated. We must remember that the chain is always broken by the weakest link.

### 4.5 Quantum measurement and the complementarity problem

According to all the authors who have been using NEQR, unlike FRQI, NEQR works with orthogonal outputs, which would simplify any recovery of outcomes at the exit of the quantum algorithm when performing a quantum measurement. If the outputs of NEQR, i.e., $\left|C_{YX}^{i}\right\rangle$ are CBS: $\{|0\rangle,|1\rangle\}$, there will be no problem, but if they are not, everything would collapse. This is precisely what happens given the results of the implementation on Quirk [9] of Fig. 8 and Eq.(33).

### 4.6 Gate noise, decoherence and fidelity

NEQR is a clear victim of the three types of gate noises introduced in Subsection 2.6. The reason is simple, and it is the excessive amount of *H* (Hadamard) and 2-*CNOT* gates required to be able to express a single mosaic like the one of Fig. 7. Therefore, we only have to imagine the inadmissible amount of noise that would be generated if NEQR could be processed by a QPU for a standard image, let us say, a classic image of 1920-by-1080-by-3-by-8 bits (we are talking about 50 Million of bits, approximately), in other words, we would be in the presence of a real noise generator.

### 4.7 Special Gate Requirements

Figures 8, 9, and 11 show the need to use anti-control (or zero control) condition on a qubit being OFF for any version of NEQR implementation [3]. This requirement extends to its associated quantum algorithms: quantum comparator [28], quantum counter, among others. As we know, this condition cannot be implemented without using an excessive amount of additional *X* gates [12-14], in the manner required by NEQR and its associated quantum algorithms, on the QPUs of IBM Q [6], and Rigetti [7], among many others, as well as in its simulators, including Quantum Programming Studio (QPS) [8].

### 4.8 Entanglement coupling

In this subsection, it will be demonstrated that there is an undesirable coupling due to the entanglement between the outcomes of NEQR. This is a very common trap in which several quantum algorithms that indiscriminately use *H* (Hadamard) and *CNOT* gates between their threads fall, entangling their outputs in a catastrophic way. In NEQR, this problem is even more dramatic because this undesirable coupling entangles the representation elements of each pixel, i.e., the $\left|C_{YX}^{i}\right\rangle$'s, which should be CBS $\{|0\rangle,|1\rangle\}$ and from Eqs.(33) we see that they are not, with the positions of said pixels, i.e., the pairs $\left|Y\,X\right\rangle$. As will be seen in the implementations of Section 6, this coupling causes the morphological destruction of the treated image.

The following experiment is very simple, since if the NEQR outputs are really entangled, then, they should serve as a vehicle for teleporting an arbitrary qubit when forming an EPR channel. On the other hand, if teleportation were not successful, then we would conclude that the coupling does not exist. For this experiment we will use the Quirk simulator [9] given the difficulties in implementing NEQR on QPUs [6, 7] and almost all simulators with free access in the cloud [6-8].

In Fig. 12, we can see the NEQR technique implemented on Quirk [9], and at its output a quantum teleportation module with a classical disambiguation channel (a CNOT gate between q[11] and q[12], and another vertical line between q[10] and q[12] as a control on a *Z* Pauli's gate). In qubit q[10], we prepare the state to be teleported, whose value will be: *wavefunction* = 0.9238 |0> + 0.3826 |1>, and *statevector*: 0.854 for |0>, and 0.146 for |1>. This last value can be seen in q[10], in the *Probability of |1>* module of Quirk [9] with a red point below. We assume that the outcomes q[0] and q[9] of NEQR are entangled, then, we introduce them respectively from the output of NEQR, into the inputs q[11] and q[12] of the quantum teleportation protocol as an EPR pair. Quantum teleportation can only take place iff q[0] (i.e., $\left|C_{YX}^{0}\right\rangle$) and q[9] (i.e., $\left|Y_{0}\right\rangle$) of NEQR constitute an entangled pair. Figure 12 shows that after strictly applying the quantum teleportation protocol in Quirk [9], we obtain that the outcome $\left|C_{YX}^{0}\right\rangle$, and the coordinate ($\left|Y_{0}\right\rangle$) are entangled with dire consequences on the treated image. The most conclusive evidence of this entanglement has to do with the fact that quantum teleportation was completely successful, since we were able to recover the teleported qubit at q[12] at the lower-right end of Fig. 12, as it is evidenced by the *Probability of |1>* module of Quirk with a red point below. The first of these consequences is that, as we mentioned in this paper, Eqs.(33), $\left|C_{YX}^{0}\right\rangle$ is not a CBS, i.e., it does not belong to { $\left|0\right\rangle, \left|1\right\rangle$ } as this technique requires within its definition but it is an element like this $1/\sqrt{2}\left(\left|00\right\rangle+\left|11\right\rangle\right)$, while the second dire consequence can be seen later in the implementations of Section 6. Next, another version of the quantum teleportation protocol confirms the same results.

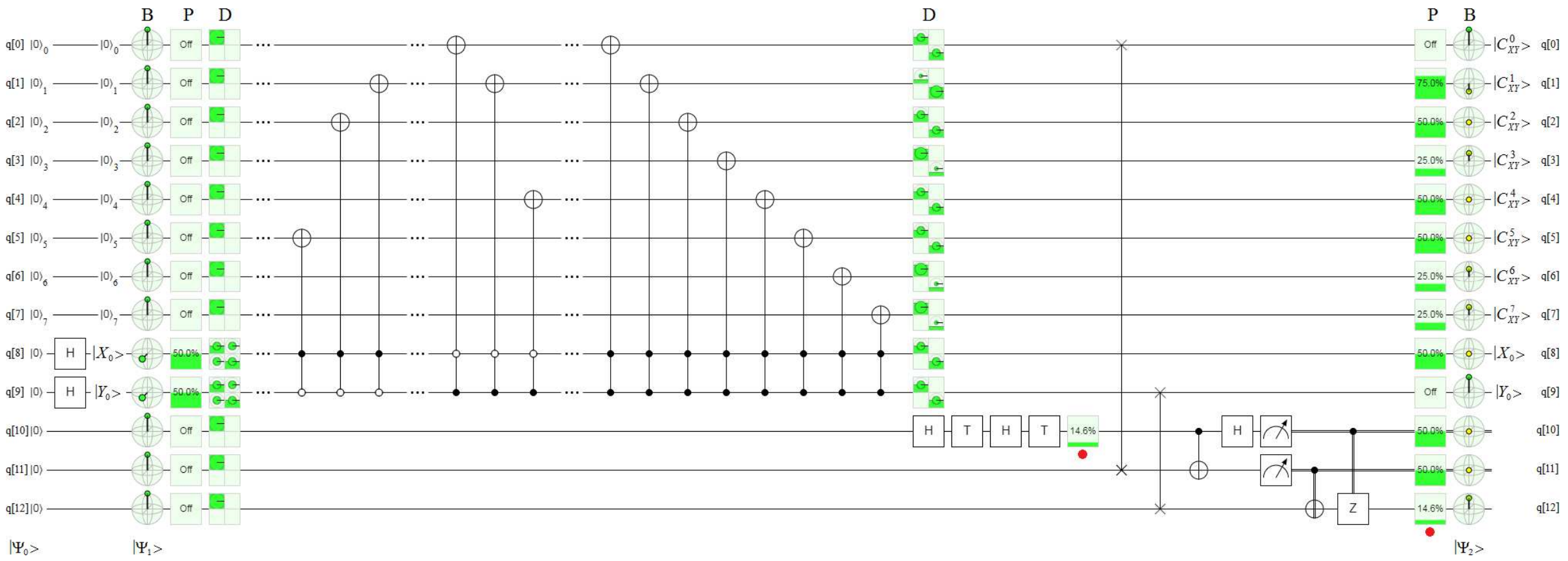

**Figure 12.** Implementation of NEQR on Quirk [9] with a teleportation module at its output. The red points under the *Probability of* |1> in qubits q[10] and q[12] shows the success of the teleportation.

The evidence is clear and blunt, this entanglement coupling is responsible for the disastrous results obtained in Section 6 for this technique of internal image representation. As expected, similar results were obtained with the implementation of the alternative version of NEQR on Quirk [9] of Fig. 9. This shows us that this undesirable effect is an intrinsic problem of the technique itself, which invalidates the operation of any quantum algorithm that follows the output of NEQR.

### *4.9 Final considerations*

In this section, the two most serious and insurmountable problems of NEQR have been revealed:

I. neither the qubits that represent the elements of each pixel, i.e., the $\left|C_{YX}^{i}\right\rangle$'s, nor the coordinates of the pixel in the tile $(|Y_0\rangle,|X_0\rangle)$ are CBS $\{|0\rangle,|1\rangle\}$ as its authors [3] and users affirm, and as if this was not catastrophic enough,

II. some of its outcomes have an undesirable entanglement coupling, which will inevitably destroy the structure of the treated image.

There is nothing more to add, the scientific evidence speaks for itself.

## 5. Quantum Boolean Image Processing (QBIP)

This technique consists of a pair of interfaces, i.e., Cl2Qu: $bit \rightarrow [\,Cl2Qu\,] \rightarrow |bit\rangle \equiv qubit$, and Qu2Cl: $qubit \equiv |bit\rangle \rightarrow [\,Qu2Cl\,] \rightarrow bit$, with a quantum algorithm between them. QBIP strictly respects the configuration stablish in Fig. 1, where the three mentioned elements work strictly and exclusively with CBS $\{|0\rangle,|1\rangle\}$, not being altered by the quantum measurement [16].

### *5.1 Scheme of operation*

If we have a Lena's color image like that of Fig.13,

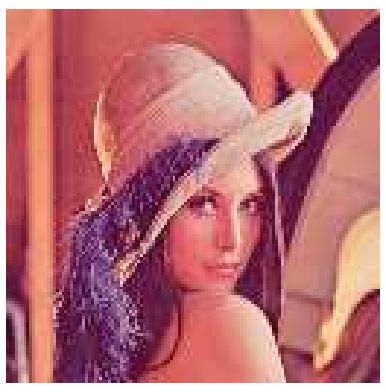

**Figure 13.** Lena's color image of 128-by-128-by-3-by-8 bits.

and we descompose it in its 24 bitplanes, then, we will obtain Fig.14, which shows a simplified detail of this procedure for a tile of only 4-by-4 pixels of Fig. 13 (for the purpose of not complicating the drawing), which is carried out by a *bit-slicer*() function implemented on a classic computer.

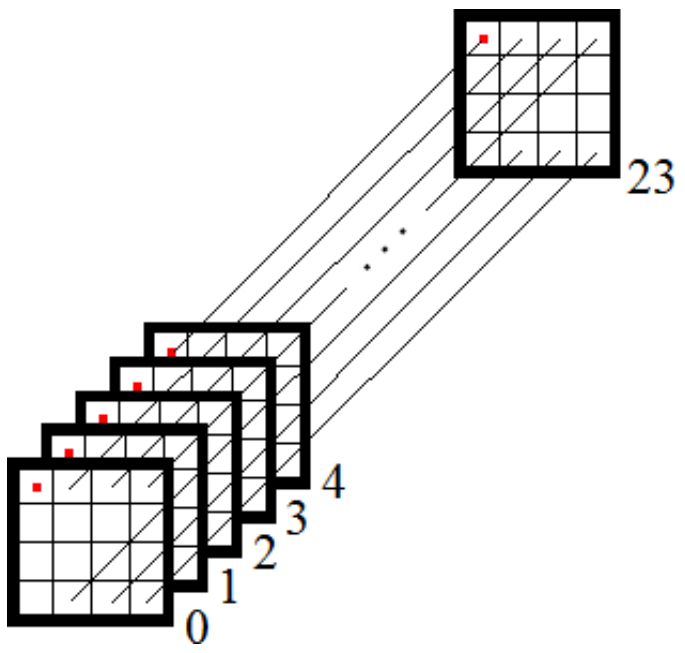

**Figure 14.** Slicing of 4-by-4 pixels of the original image in its 24 bitplanes.

The action of the mentioned *bit-slicer*() function has a counterpart, it is the *bit-reassembler*() function, which allows us to reconstruct the image from its bitplanes. The joint action of both functions can be seen in Fig.15. However, in order not to complicate said figure, we only represent the 8 bitplanes of the red channel.

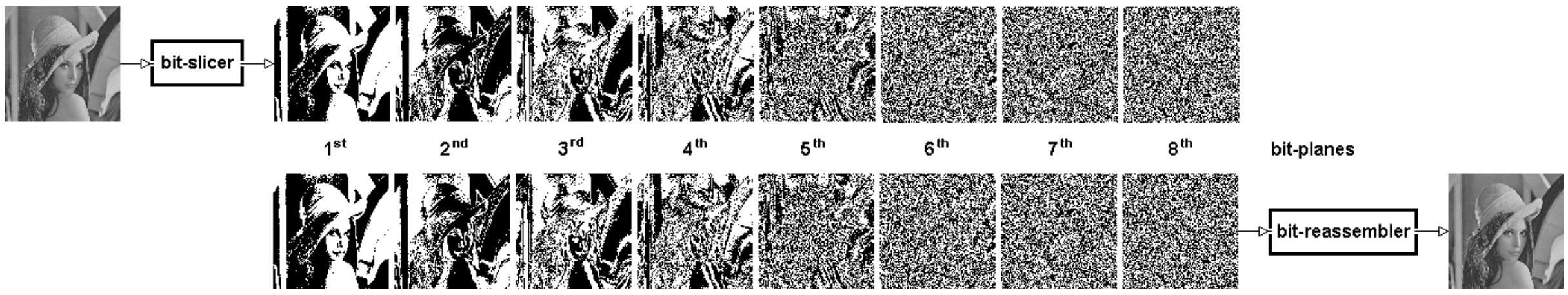

**Figure 15.** The combined action of the *bit-slicer*() and *bit-reassembler*() functions on Lena's red channel.

QBIP only works with the most significant bit (MSB) of each color channel of the original image, as we can see in Fig. 16 (where only it is represented the MSB of the red channel), which for an image of $R\times C\times CoC\times BpPpC$ bits (where, $R$ is the number of rows, $C$ the number of columns, $CoC$ means *channels-of-color*, which they always are 3, and $BpPpC$ is the number of *bit-per-pixel-per-channel*, which are generally 8) we can only work with $R\times C\times CoC$ bits, dramatically lowering the storage respect to FRQI and NEQR at least 8 times.

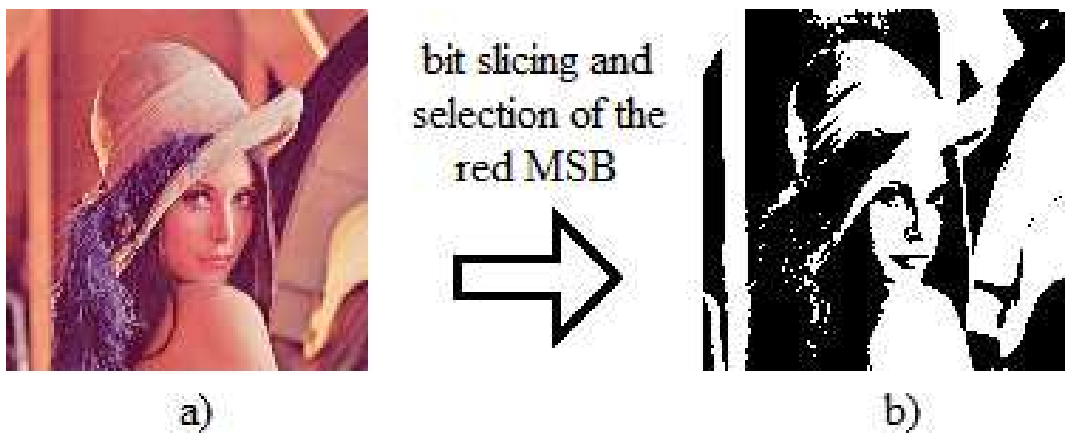

**Figure 16.** a) Lena's color image of 128-by-128-by-3-by-8 bits. b) Lena's MSB bitplane of the red channel with 128-by-128-by-1 bits.

It is precisely and exclusively on the mentioned MSBs (of the respective three color channels) that the quantum algorithm acts.

### *5.2 Cl2Qu interface and data intake*

For this work several configurations can be used, which range from the most efficient and simple to the most inefficient and complicated, which proves the ductility of QBIP in the intake of the image and in the recovery of the outcomes.

*Efficient and simple:* QBIP is extremely robust due to the fact that it exclusively works with CBS, in fact, if we want to map a bit (0 or 1) into a qubit (|0> or |1>) all we need is a classically-controlled NOT quantum gate acting on a |0> input. Then, if the control is 0, the output remains |0> and if the control is 1 the output is changed to |1>. That is, only one gate is necessary.

*Inefficient and complicated:* Even another Cl2Qu interface, based on the first part of the superdense coding protocol [30], which is ridiculously complicated for this simple task, can be used with success in QBIP. This Cl2Qu interface can be seen in Fig. 17, while the Qu2Cl interface is seen in its last part, that is, the quantum measurement [16] along the *z*-axis of the Bloch's sphere, which will be explained in detail later. Although Fig. 17 is not the most efficient option of Cl2Qu interface since one EPR pair must be prepared for each pair of bits to be transformed into their respective CBS, it is surely the most pedagogical one, where: $\left|\beta_{b_1b_2}\right\rangle = 1/\sqrt{2}\left(\left|b_1b_2\right\rangle + \left(-1\right)^{b_2}\left|\bar{b}_1\bar{b}_2\right\rangle\right)$, while $X$ and $Z$ are Pauli's matrices. In order to simplify the explanation of its operation, we start with a reduced version of only 2 bits, in a completely homotopic way: $\{b_1, b_2\} \rightarrow Cl2Qu \rightarrow \{\left|b_1\right\rangle, \left|b_2\right\rangle\}$. However, both the interface and its explanation are absolutely transferable to any number of bits.

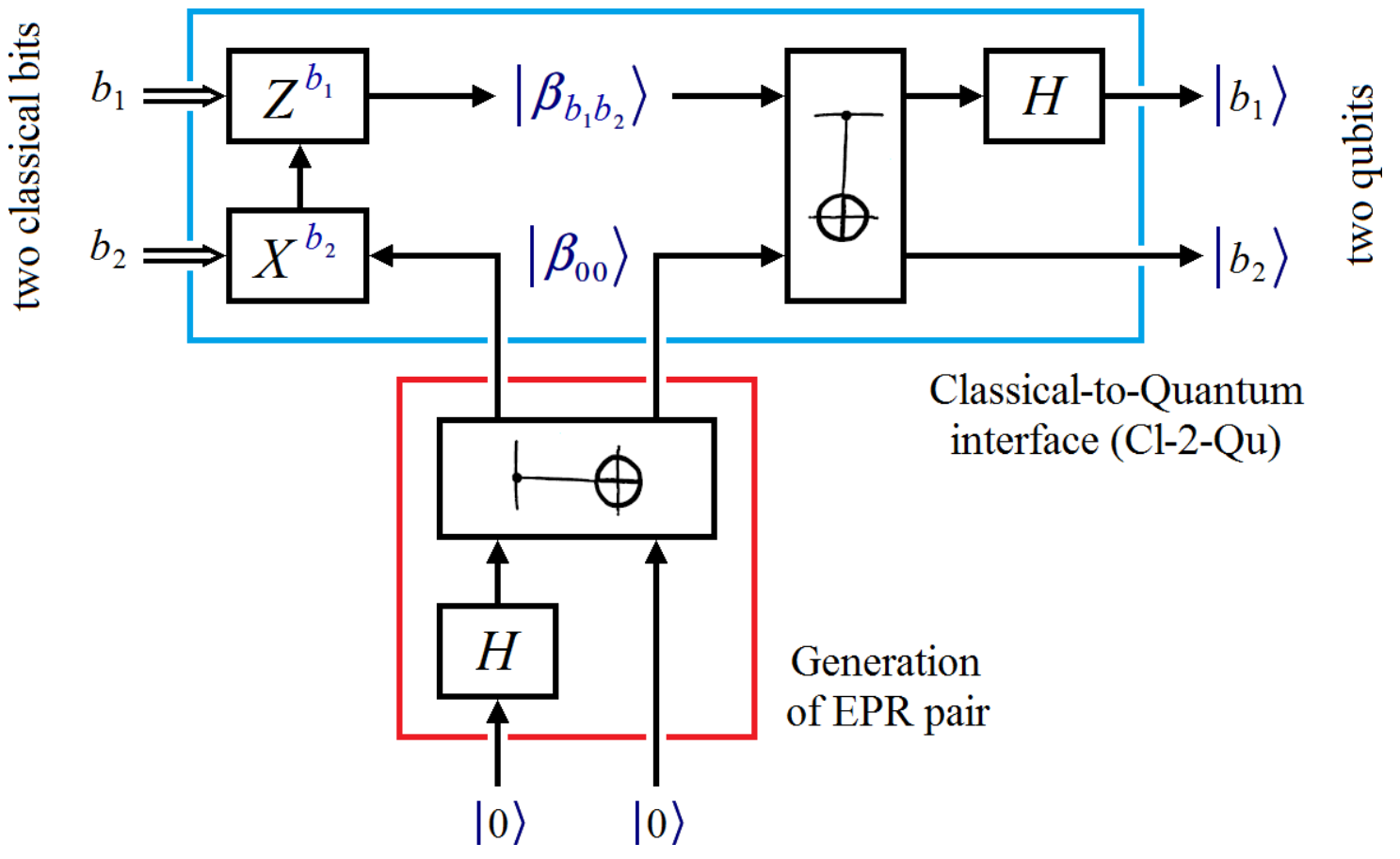

**Figure 17.** Example of a possible pair of Cl2Qu and Qu2Cl interfaces for QBIP extracted from the superdense coding protocol [30], where, $|b_1\rangle$ and $|b_2\rangle$ are the CBS versions of the classical bits $b_1$ and $b_2$, respectively.

We begin with the creation of an EPR pair (two $|\beta_{00}\rangle$ entangled states) from two ancilla ground states $|0\rangle$. Then, we have two possibilities according to $b_2$, i.e., 0 or 1. If $b_2 = 0$, then, at the output of the Pauli's gate $X$, we will have $|\beta_{00}\rangle$ again. See Fig.17. But, if $b_2 = 1$, then, at the output of the Pauli's gate $X$, we will have,

$$(X \otimes I)|\beta_{00}\rangle = \left(\begin{bmatrix} 0 & 1 \\ 1 & 0 \end{bmatrix} \otimes \begin{bmatrix} 1 & 0 \\ 0 & 1 \end{bmatrix}\right) \begin{bmatrix} 1/\sqrt{2} \\ 0 \\ 0 \\ 1/\sqrt{2} \end{bmatrix} = \begin{bmatrix} 0 & 0 & 1 & 0 \\ 0 & 0 & 0 & 1 \\ 1 & 0 & 0 & 0 \\ 0 & 1 & 0 & 0 \end{bmatrix} \begin{bmatrix} 1/\sqrt{2} \\ 0 \\ 0 \\ 1/\sqrt{2} \end{bmatrix} = \begin{bmatrix} 0 \\ 1/\sqrt{2} \\ 1/\sqrt{2} \\ 0 \end{bmatrix}. \tag{35}$$

If, at the output of the Pauli's gate $Z$, $b_2 = 0$ and $b_1 = 0$, we will obtain $|\beta_{00}\rangle$. But if $b_2 = 0$ and $b_1 = 1$, we will have,

$$(Z \otimes I)|\beta_{00}\rangle = \left(\begin{bmatrix} 1 & 0 \\ 0 & -1 \end{bmatrix} \otimes \begin{bmatrix} 1 & 0 \\ 0 & 1 \end{bmatrix}\right) \begin{bmatrix} 1/\sqrt{2} \\ 0 \\ 0 \\ 1/\sqrt{2} \end{bmatrix} = \begin{bmatrix} 1 & 0 & 0 & 0 \\ 0 & 1 & 0 & 0 \\ 0 & 0 & -1 & 0 \\ 0 & 0 & 0 & -1 \end{bmatrix} \begin{bmatrix} 1/\sqrt{2} \\ 0 \\ 0 \\ 1/\sqrt{2} \end{bmatrix} = \begin{bmatrix} 1/\sqrt{2} \\ 0 \\ 0 \\ -1/\sqrt{2} \end{bmatrix} \tag{36}$$

Now, if $b_2 = 1$ and $b_1 = 1$, we will use the result of Eq.(35) to obtain the state at the output of the Pauli's gate $Z$,

$$(Z \otimes I)\begin{bmatrix} 0 \\ 1/\sqrt{2} \\ 1/\sqrt{2} \\ 0 \end{bmatrix} = \left(\begin{bmatrix} 1 & 0 \\ 0 & -1 \end{bmatrix} \otimes \begin{bmatrix} 1 & 0 \\ 0 & 1 \end{bmatrix}\right) \begin{bmatrix} 0 \\ 1/\sqrt{2} \\ 1/\sqrt{2} \\ 0 \end{bmatrix} = \begin{bmatrix} 1 & 0 & 0 & 0 \\ 0 & 1 & 0 & 0 \\ 0 & 0 & -1 & 0 \\ 0 & 0 & 0 & -1 \end{bmatrix} \begin{bmatrix} 0 \\ 1/\sqrt{2} \\ 1/\sqrt{2} \\ 0 \end{bmatrix} = \begin{bmatrix} 0 \\ 1/\sqrt{2} \\ -1/\sqrt{2} \\ 0 \end{bmatrix} \tag{37}$$

Obviously, the result at the output of the Pauli's gate $Z$ is the same as at the input of the *CNOT* gate in the upper right sector of Fig.17. The four obtained results, according to the values of $b_2$ and $b_1$, are attached to the entrance of the aforementioned *CNOT* gate under the generic name of $|\beta_{b_1b_2}\rangle$. At the entrance of said *CNOT* gate the other entangled state $|\beta_{00}\rangle$ will also be incorporated**.**

Now, if we apply *CNOT* gate to these results, then, for $b_2 = 0$ and $b_1 = 0$, we will obtain,

$$\begin{bmatrix}1&0&0&0\\0&1&0&0\\0&0&0&1\\0&0&1&0\end{bmatrix}\begin{bmatrix}1/\sqrt{2}\\0\\0\\1/\sqrt{2}\end{bmatrix}=\begin{bmatrix}1/\sqrt{2}\\0\\1/\sqrt{2}\\0\end{bmatrix} \tag{38}$$

For $b_2 = 0$ and $b_1 = 1$,

$$\begin{bmatrix}1&0&0&0\\0&1&0&0\\0&0&0&1\\0&0&1&0\end{bmatrix}\begin{bmatrix}1/\sqrt{2}\\0\\0\\-1/\sqrt{2}\end{bmatrix}=\begin{bmatrix}1/\sqrt{2}\\0\\-1/\sqrt{2}\\0\end{bmatrix} \tag{39}$$

For $b_2 = 1$ and $b_1 = 0$,

$$\begin{bmatrix}1&0&0&0\\0&1&0&0\\0&0&0&1\\0&0&1&0\end{bmatrix}\begin{bmatrix}0\\1/\sqrt{2}\\1/\sqrt{2}\\0\end{bmatrix}=\begin{bmatrix}0\\1/\sqrt{2}\\0\\1/\sqrt{2}\end{bmatrix} \tag{40}$$

For $b_2 = 1$ and $b_1 = 1$,

$$\begin{bmatrix}1&0&0&0\\0&1&0&0\\0&0&0&1\\0&0&1&0\end{bmatrix}\begin{bmatrix}0\\1/\sqrt{2}\\-1/\sqrt{2}\\0\end{bmatrix}=\begin{bmatrix}0\\1/\sqrt{2}\\0\\-1/\sqrt{2}\end{bmatrix} \tag{41}$$

Finally, we will apply the Hadamard's gate ($H$) to the last set of equations according to Fig.17 [30], thus, for $b_2 = 0$ and $b_1 = 0$, we will have,

$$\begin{aligned}(H\otimes I)\begin{bmatrix}1/\sqrt{2}\\0\\1/\sqrt{2}\\0\end{bmatrix}&=\left(\begin{bmatrix}1/\sqrt{2}&1/\sqrt{2}\\1/\sqrt{2}&-1/\sqrt{2}\end{bmatrix}\otimes\begin{bmatrix}1&0\\0&1\end{bmatrix}\right)\begin{bmatrix}1/\sqrt{2}\\0\\1/\sqrt{2}\\0\end{bmatrix}=\begin{bmatrix}1/\sqrt{2}&0&1/\sqrt{2}&0\\0&1/\sqrt{2}&0&1/\sqrt{2}\\1/\sqrt{2}&0&-1/\sqrt{2}&0\\0&1/\sqrt{2}&0&-1/\sqrt{2}\end{bmatrix}\begin{bmatrix}1/\sqrt{2}\\0\\1/\sqrt{2}\\0\end{bmatrix}\\&=\begin{bmatrix}1\\0\\0\\0\end{bmatrix}=\begin{bmatrix}1\\0\end{bmatrix}\otimes\begin{bmatrix}1\\0\end{bmatrix}=|0\rangle\otimes|0\rangle=|00\rangle\end{aligned} \tag{42}$$

for $b_2 = 0$ and $b_1 = 1$,

$$\begin{aligned}(H\otimes I)\begin{bmatrix}1/\sqrt{2}\\0\\-1/\sqrt{2}\\0\end{bmatrix}&=\left(\begin{bmatrix}1/\sqrt{2}&1/\sqrt{2}\\1/\sqrt{2}&-1/\sqrt{2}\end{bmatrix}\otimes\begin{bmatrix}1&0\\0&1\end{bmatrix}\right)\begin{bmatrix}1/\sqrt{2}\\0\\-1/\sqrt{2}\\0\end{bmatrix}=\begin{bmatrix}1/\sqrt{2}&0&1/\sqrt{2}&0\\0&1/\sqrt{2}&0&1/\sqrt{2}\\1/\sqrt{2}&0&-1/\sqrt{2}&0\\0&1/\sqrt{2}&0&-1/\sqrt{2}\end{bmatrix}\begin{bmatrix}1/\sqrt{2}\\0\\-1/\sqrt{2}\\0\end{bmatrix}\\&=\begin{bmatrix}0\\0\\1\\0\end{bmatrix}=\begin{bmatrix}0\\1\end{bmatrix}\otimes\begin{bmatrix}1\\0\end{bmatrix}=|1\rangle\otimes|0\rangle=|10\rangle\end{aligned} \tag{43}$$

for $b_2 = 1$ and $b_1 = 0$,

$$
(H \otimes I)\begin{bmatrix} 0 \\ 1/\sqrt{2} \\ 0 \\ 1/\sqrt{2} \end{bmatrix} = \left( \begin{bmatrix} 1/\sqrt{2} & 1/\sqrt{2} \\ 1/\sqrt{2} & -1/\sqrt{2} \end{bmatrix} \otimes \begin{bmatrix} 1 & 0 \\ 0 & 1 \end{bmatrix} \right) \begin{bmatrix} 0 \\ 1/\sqrt{2} \\ 0 \\ 1/\sqrt{2} \end{bmatrix} = \begin{bmatrix} 1/\sqrt{2} & 0 & 1/\sqrt{2} & 0 \\ 0 & 1/\sqrt{2} & 0 & 1/\sqrt{2} \\ 1/\sqrt{2} & 0 & -1/\sqrt{2} & 0 \\ 0 & 1/\sqrt{2} & 0 & -1/\sqrt{2} \end{bmatrix} \begin{bmatrix} 0 \\ 1/\sqrt{2} \\ 0 \\ 1/\sqrt{2} \end{bmatrix}
$$

$$
= \begin{bmatrix} 0 \\ 1 \\ 0 \\ 0 \end{bmatrix} = \begin{bmatrix} 1 \\ 0 \end{bmatrix} \otimes \begin{bmatrix} 0 \\ 1 \end{bmatrix} = |0\rangle \otimes |1\rangle = |01\rangle \tag{44}
$$

for $b_2 = 1$ and $b_1 = 1$,

$$
(H \otimes I)\begin{bmatrix} 0 \\ 1/\sqrt{2} \\ 0 \\ -1/\sqrt{2} \end{bmatrix} = \left( \begin{bmatrix} 1/\sqrt{2} & 1/\sqrt{2} \\ 1/\sqrt{2} & -1/\sqrt{2} \end{bmatrix} \otimes \begin{bmatrix} 1 & 0 \\ 0 & 1 \end{bmatrix} \right) \begin{bmatrix} 0 \\ 1/\sqrt{2} \\ 0 \\ -1/\sqrt{2} \end{bmatrix} = \begin{bmatrix} 1/\sqrt{2} & 0 & 1/\sqrt{2} & 0 \\ 0 & 1/\sqrt{2} & 0 & 1/\sqrt{2} \\ 1/\sqrt{2} & 0 & -1/\sqrt{2} & 0 \\ 0 & 1/\sqrt{2} & 0 & -1/\sqrt{2} \end{bmatrix} \begin{bmatrix} 0 \\ 1/\sqrt{2} \\ 0 \\ -1/\sqrt{2} \end{bmatrix}
$$

$$
= \begin{bmatrix} 0 \\ 0 \\ 0 \\ 1 \end{bmatrix} = \begin{bmatrix} 0 \\ 1 \end{bmatrix} \otimes \begin{bmatrix} 0 \\ 1 \end{bmatrix} = |1\rangle \otimes |1\rangle = |11\rangle \tag{45}
$$

Here Cl2Qu interface finalizes. Basically, a Cl2Qu interface is a block that performs a transfer of type $\{b_1, b_2\} \rightarrow \{|b_1\rangle, |b_2\rangle\}$ for CBS, i.e., $\{0,1\} \rightarrow \{|0\rangle, |1\rangle\}$. On the other hand, since QBIP works with said CBS, then the Quantum-to-Classical (Qu2Cl) interface is reduced to that of Fig.18, that is, a simple quantum measurement [16] procedure on the $z$ axis of the Bloch's sphere.

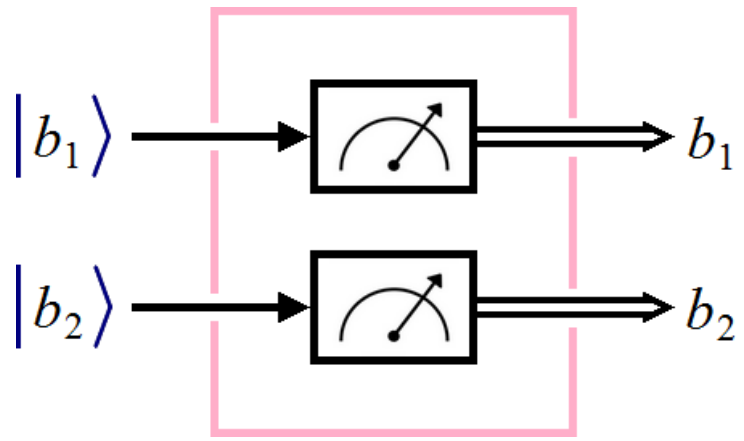

**Figure 18.** Quantum-to-Classical (Qu2Cl) interface.

This demonstrates that when working exclusively with CBSs, virtually anything can be used as a Cl2Qu interface.

### *5.3 Characteristics of the employed qubits*

As we have mentioned in this section, QBIP works with CBSs $\{|0\rangle, |1\rangle\}$, with all the advantages that this implies (and no disadvantage) and that have been explained so far.

### *5.4 Sparsity, number and size of the required registers*

Since QBIP is a criterion, which does not imply quantum circuits by itself, the only storage required is complete and exclusive responsibility of the Cl2Qu interface, which temporarily occupies the registers of the QPU with the 3 MSBs belonging to the three color channels (R, G, B), i.e., with an occupancy rate of said registers of 3/24 100% with respect to FRQI and NEQR.

The computational cost of QBIP is then that of the chosen Cl2Qu interface. The only case in which all bitplanes of the original image should be used is when we must invert the colors of such image. In Section 6, we show just two of the countless examples of successful implementation of QBIP on a quantum platform with free access in the cloud. We chose Quirk [9] for being the most pedagogical one, with an excellent drag-and-drop graphic interface, however, similar implementations made with the other platforms such as IBM Q [6], Rigetti [7], and Quantum Programming Studio [8] yielded the same results.

### *5.5 Quantum measurement and the complementarity problem*

As a direct consequence of Subsection 5.3, and considering that only CBS can be measured without any alteration [26], it is clear that QBIP is not a victim of these serious problems.

### *5.6 Gate noise, decoherence and fidelity*

As we have mentioned, QBIP is a criterion, which does not imply quantum circuits by itself, therefore, this technique does not incorporate gates that: conspire against coherence time (when implemented on superconducting platforms such as IBM Q [6], or Rigetti [7]) or introduce delays or some of the noises mentioned in Subsection 2.6.

### *5.7 Special Gate Requirements*

It does not require them.

### *5.8 Entanglement coupling*

This serious problem of NEQR is totally and absolutely foreign to QBIP.

### *5.9 Final considerations*

QBIP has proved to have the following comparative advantages in relation to FRQI and NEQR:

- by working strictly and exclusively with CBSs $\{|0\rangle,|1\rangle\}$, their qubits can be exactly measured, copied and compared, therefore, both Cl2Qu and Qu2Cl interface are simple to implement, with low computational cost and a minimum temporary storage,
- since it is a criterion, the only quantum gates associated with QBIP are used for their interfaces, that is, it is a minimum number, with the lowest noise from gates of the three techniques analyzed in this work,
- it has no special gate requirements, and finally
- this is not a victim of entanglement coupling.

In other words, the comparative advantages, supported by verifiable evidence, are overwhelming.

## 6. Implementations on quantum platforms

In this section, we will present two sets of implementations:

a) the first set consists of a strict comparison of performance between FRQI, NEQR and QBIP on the Rigetti platform [7], specifically on its Quantum Virtual Machine (QVM), and which consists of taking an image of the classical world, representing it with the technique, and then returning it again to the classical world to evaluate the impact that the technique had on the morphological integrity of the image, i.e., without a quantum algorithm, only the technique,

b) the second set consists of exclusive QBIP implementations for several quantum algorithms, since it is the only technique with which efficient implementations can be made on IBM Q [6], Rigetti [7], Quantum Programming Studio [8], Quirk [ 9], among many others.

***6.1 Comparison of performance between FRQI, NEQR and QBIP:*** What are the consequences of the results of Fig. 12 of the Subsection 4.8 in the case of NEQR, and of the disturbing action of the quantum measurement suspected in Subsection 3.5 in the case of FRQI when they are applied on an image? To answer this question, first, we will consider the red channel (8 bit-per-pixel, 128-by-128 pixels) of Lena's image of Fig. 13.

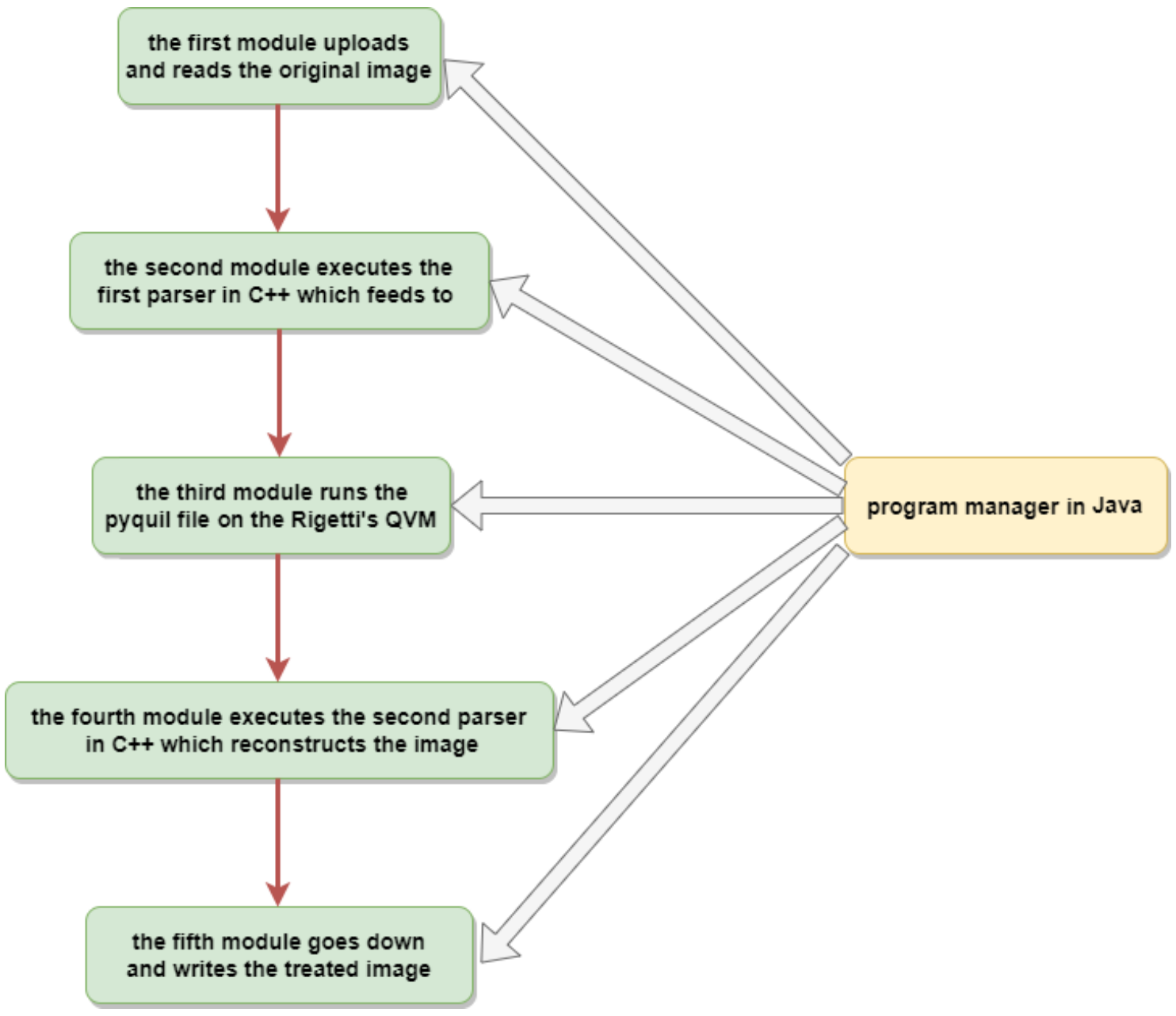

**Figure 19.** Program manager and the different modules designed for this experiment.

Second, we are going to prepare a series of modules (programs) which will interact with each other in a coordinated way thanks to the intervention of a program manager in Java, and whose final objective is to generate an automatic intake in the Rigetti's *quantum virtual machine* (QVM). For this purpose, we design the following modules according to Fig. 19:

*Program manager in Java*

- the first module uploads and reads the original image,
- the second module executes the first parser in C++ which feeds to
- the third module, which runs the pyquil file on the Rigetti's QVM,
- the fourth module executes the second parser in C++, which reconstructs the image, and
- the fifth module goes down and writes the treated image.

Figure 20 shows, from left to right:

- the original image to be treated,
- the FRQI version after applying this technique and the posterior quantum measurement of their respective outcomes. It is evident that the quantum measurement eliminates the gray levels, converting each pixel to a strict CBS $\{|0\rangle, |1\rangle\}$, i.e., destroying the original image,
- the NEQR version, with three types of outcomes (and all its possible combinations), as we can see in Eq.(33) of this paper, destroys the morphology of the image as a result of the devastating effect of entanglement coupling. Finally,
- the QBIP version, where we have applied the technique to the 8 bitplanes of the image, i.e., not just to the first bitplane (MSB) as it is usual in this technique. The outcomes clearly show the total absence of entanglement between them, i.e., entanglement coupling, as well as, the complete immunity to alterations related to quantum measurement.

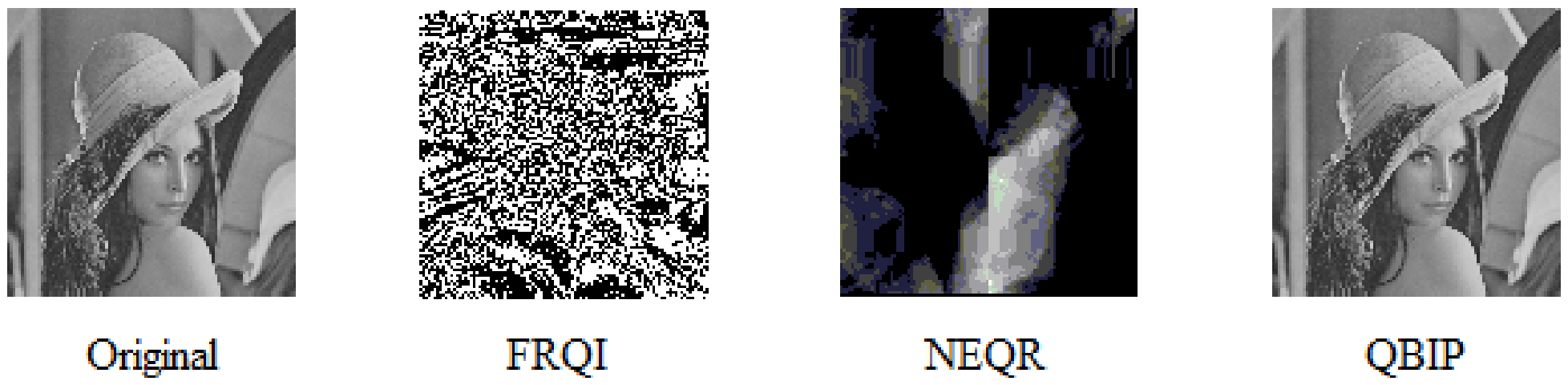

**Figure 20.** Results of the first experiment. From left to right, we have: the original image, the FRQI version after quantum measurement, the NEQR version in terms of the entanglement between some of their outcomes and their respective allocations, and the QBIP version after its treatment and quantum measurement.

Finally, we show another important result of the experiment for the three techniques: the elapsed time each one takes,

- FRQI → 2389 minutes,
- NEQR → 714 minutes, and
- QBIP → 68 minutes, taking into account all the bitplanes of the image + a Classical-to-Quantum interface like that of Subsection 5.2 of this paper + a simple quantum measurement module as a Quantum-to-Classical interface, see Fig. 18.

These results were reached coding in pyquil 2.6 on the QVM of Rigetti [7] on an Intel Core i7-4702 MQ CPU @ 2.20 GHZ/2.20 GHZ with 8.00 GB RAM on a 64-bits operating system,Windows 7 Ultimate. Similar results can be reached implementing FRQI on a 5-qubits IBM Q QPU, where we can appreciate, in detail, the disastrous effect of quantum measurement on the outcomes of this technique.

***6.2 Implementations of several quantum algorithms in QBIP:*** Given its simplicity, there is practically no QImP algorithm that cannot be implemented with QBIP. Recently the following quantum algorithms have been implemented under the QBIP technique, namely:

- Quantum Boolean Wavelet Transform (QBWT),
- Quantum Boolean Orthogonalization Process (QBOP) [31], based on its classical version [32-34],
- Quantum Boolean Uncorrelation Process (QBUP),
- Quantum Boolean Rule of Three (QBRo3), based on its classical version [35], for image denoising, compression and super-resolution,
- Quantum Boolean Inverter (QBInv), which completely inverts the colors of an image. For more information about the classic color inversion of an image, we can see [29, 36],
- Quantum Boolean Cosine Transform (QBCT),
- Quantum Boolean Convolution and Deconvolution (QBCD),
- Quantum Boolean Image Denoising (QBID) [4], and
- Quantum Boolean Edge Detection (QBED) [37].

Table II summarizes the platforms used for all implementations (which will be only two for a matter of space) where the superscripts mean:

1. QPS means: Quantum Programming Studio [8],
2. QPU means: Quantum Processing Unit, i.e., the physical quantum computer,
3. QPS invokes to IBM's [6] and Rigetti's [7] QPUs,
4. DD means: drag and drop,
5. In this subsection, we will implement QBWT and its reverse, i.e., iQBWT, however, if in the middle, between QBWT and iQBWT we need to filter the high-frequency subbands thanks to a technique that uses an *if-then-else* statement like QBID, then, we cannot implement it on any QPUs, and
6. Both QBID and QBED need an *if-then-else* statement for their implementation, therefore, this cannot be implemented on any QPUs.

**TABLE II:** Is it possible to implement the following quantum algorithms on platforms with cloud service?

| Quantum Algorithm | IBM Q [6] | | Rigetti [7] | | QPS[1] [8] | | Quirk [9] |
|---|---|---|---|---|---|---|---|
| | Simulator (DD[4] & qasm) | QPU[2] (DD[4], qasm & qiskit) | Simulator (pyquil) | QPU (quil) | Simulator (DD[4] & qasm) | QPU[3] | Simulator (DD[4]) |
| QBWT[5] | Yes | Yes | Yes | Yes | Yes | Yes | Yes |
| QBOP | Yes | Yes | Yes | Yes | Yes | Yes | Yes |
| QBUP | Yes | Yes | Yes | Yes | Yes | Yes | Yes |
| QBRo3 | Yes | Yes | Yes | Yes | Yes | Yes | Yes |
| QBInv | Yes | Yes | Yes | Yes | Yes | Yes | Yes |
| QBCT | Yes | Yes | Yes | Yes | Yes | Yes | Yes |
| QBCD | Yes | Yes | Yes | Yes | Yes | Yes | Yes |
| QBID | Yes | No[6] | Yes | No[6] | Yes | No[6] | Yes |
| QBED | Yes | No[6] | Yes | No[6] | Yes | No[6] | Yes |

Now, we will proceed to show the implementations of Quantum Boolean Wavelet Transform (QBWT, with its reverse version iQBWT) and Quantum Boolean Orthogonalization Process (QBOP) on the 4 mentioned platforms [6-9].

### *6.2.1 Quantum Boolean Wavelet Transform (QBWT) and its inverse (iQBWT)*

For this experiment, we use Lena's color image of 128-by-128-by-3-by-8 bits of Fig.13, which is digitized inside a classical computer, where, we apply a slicer procedure [1, 8] on the image, and thus we obtain the 24 bitplanes of it through a procedure like those of Figures 14 and 15. Now, we select, the MSB of the red channel, with 128-by-128-by-1 bits, which is shown in Fig.16.

Still in the classical computer, we select the first tile of 2-by-2 bits, beginning in the upper-left margin of the frame, in such a way as to obtain a configuration like that of the Fig.21.

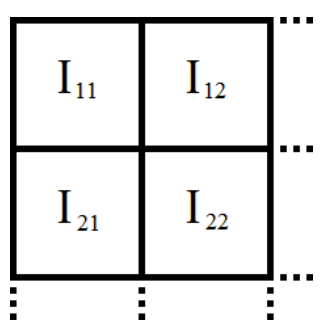

**Figure 21.** Tile of 2-by-2 bits, beginning in the upper-left margin of the frame of Fig.13.

From this point, we need to choose one of the two Cl2Qu interfaces of Subsection 5.2. Obviously, the natural choice corresponds to the simple and efficient version of said interfaces, however, and for the sole purpose of showing the technical feasibility of the complex and inefficient version, we show its use in Fig. 22 for instructional purposes. We implement the Cl2Qu interface of Fig. 22 on the chosen platform from a classical computer. Already within the platform, we apply QBWT,

$$|J_{11}\rangle = |I_{11}\rangle \quad (46a)$$

$$|J_{12}\rangle = |I_{12}\rangle \oplus |I_{11}\rangle \quad (46b)$$

$$|J_{21}\rangle = |I_{21}\rangle \oplus |I_{11}\rangle \quad (46c)$$

$$|J_{22}\rangle = |I_{22}\rangle \oplus |I_{11}\rangle \quad (46d)$$

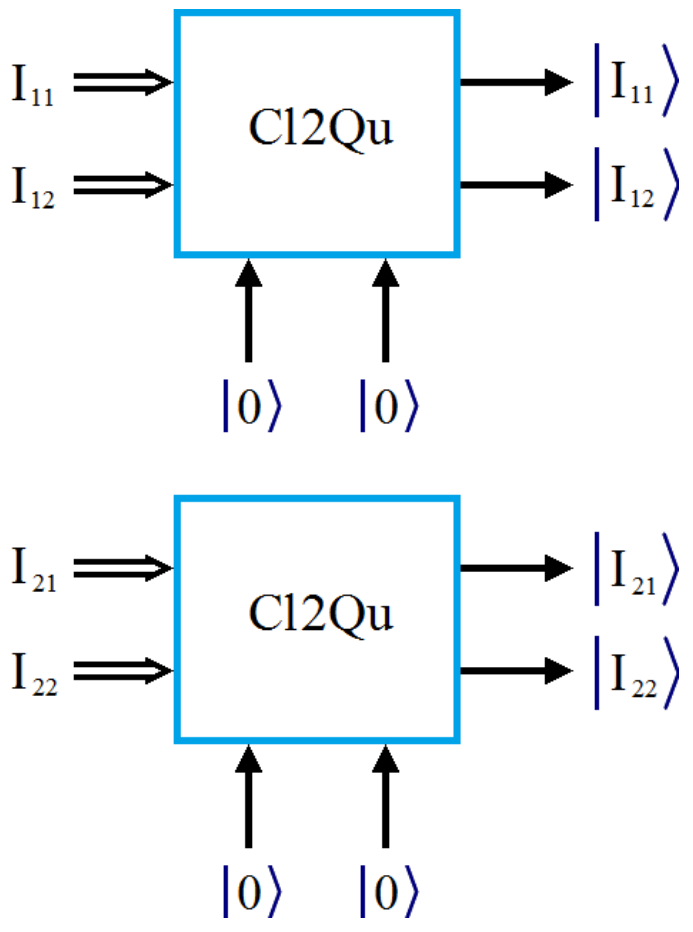

**Figure 22.** Cl2Qu interface for the intake of the 4 bits of Fig.21.

where $\oplus$ means *XOR* gate. Equation (46) will be implemented inside the different platforms thanks to *CNOT* gates, exclusively. Before starting with the 4 mentioned platforms, and in order to compare the outcomes of these with the expected theoretical results, we clarify the real values of the mentioned classical mosaic of Fig.21 in relation to MSB of Fig.16,

$$\begin{bmatrix} I_{11} & I_{12} \\ I_{21} & I_{22} \end{bmatrix} = \begin{bmatrix} 1 & 0 \\ 1 & 1 \end{bmatrix} \rightarrow Cl2Qu \rightarrow \begin{bmatrix} |I_{11}\rangle & |I_{12}\rangle \\ |I_{21}\rangle & |I_{22}\rangle \end{bmatrix} = \begin{bmatrix} |1\rangle & |0\rangle \\ |1\rangle & |1\rangle \end{bmatrix} \tag{47}$$

After applying Eq.(46) to the quantum version of Eq.(47) mosaic, already within the platform, we must obtain a new mosaic inside such platform,

$$\begin{bmatrix} |J_{11}\rangle & |J_{12}\rangle \\ |J_{21}\rangle & |J_{22}\rangle \end{bmatrix} = \begin{bmatrix} |1\rangle & |1\rangle \\ |0\rangle & |0\rangle \end{bmatrix} \tag{48}$$

**Outcomes on Quirk [9]:** It is evident that on Quirk the whole process takes place inside a classical computer, since this is a strict simulator that does not interact with the environment [9]. That is why in these cases we use a single example tile like Eq.(47) as proof of concept. Figure 23 shows, on the left side, the qubits of the right side of Eq.(47), and, on the right side, the bits/qubits of Eq.(48), in a correct correspondence. Why do we say bits/qubits? Let us remember that after a Qu2Cl interface we always obtain bits, however, from Eqs.(9) and (10) of Subsection 2.5, we know that the quantum measurement does not alter the CBS. This represents an unmistakable example of the superior advantage represented by working with CBS. Besides, if Quirk [9] interacted with the environment, then thanks to the Qu2Cl interface we would obtain an image like that of Fig.24 by QBWT action. As in fact, we will see with the other platforms.

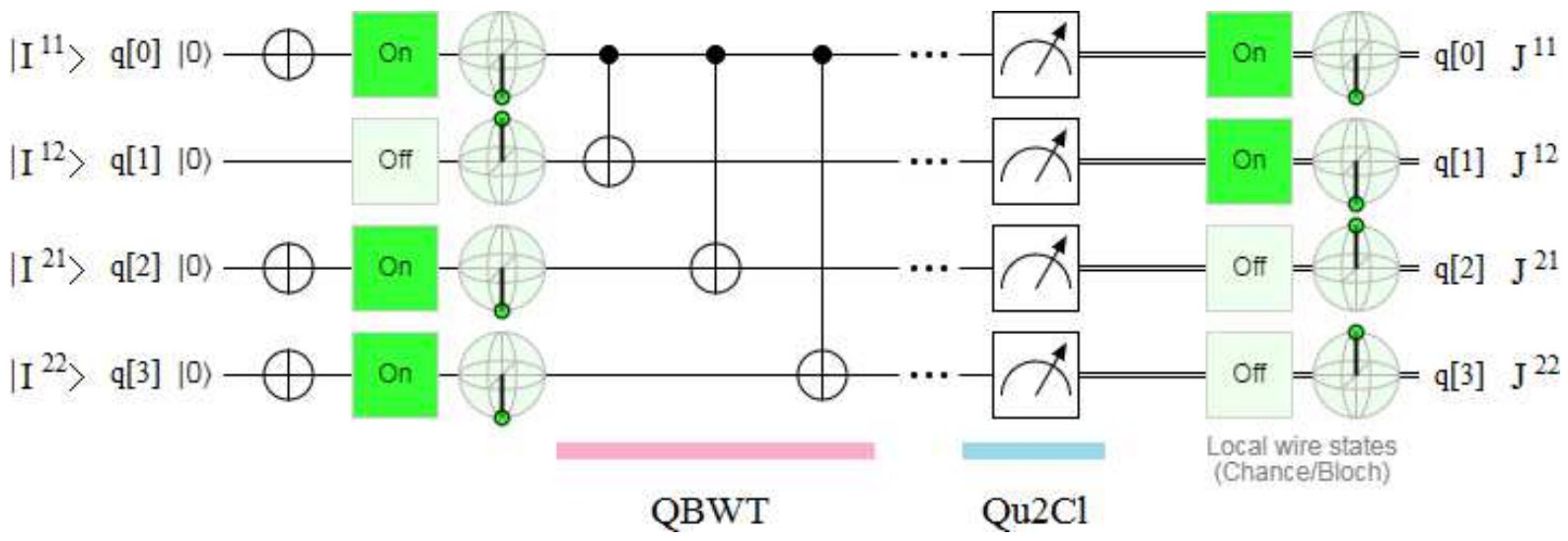

**Figure 23.** Outcomes on Quirk [9] for an implementation of QBWT.

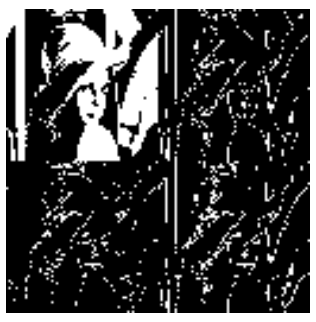

**Figure 24.** Result of the QBWT application, where we can see the following subbands: a) upper-left, low-frequency, b) upper-right, horizontal high-frequency, c) down-left, vertical high-frequency, and d) down-right, diagonal high-frequency.

As a consequence of what was said before and since:

a) Quirk [9] is a purely graphic *drag and drop* simulator (by the way, the most visual and pedagogical one), which does not allow an automatic load to happen and does not interact with the environment, and since

b) an image uploading by hand is unacceptable, then,

we settle for the example of the Figs. 21 and 23 as a proof of concept for QBWT algorithm.

Now, if we want to verify that we obtain the same bits as those on the left side of Eq.(47), when applying iQBWT, then, we must resort to Fig.25, where, we can see that the qubits on the left and the bits/qubits on the right of the figure are coincidental, closing the circle.

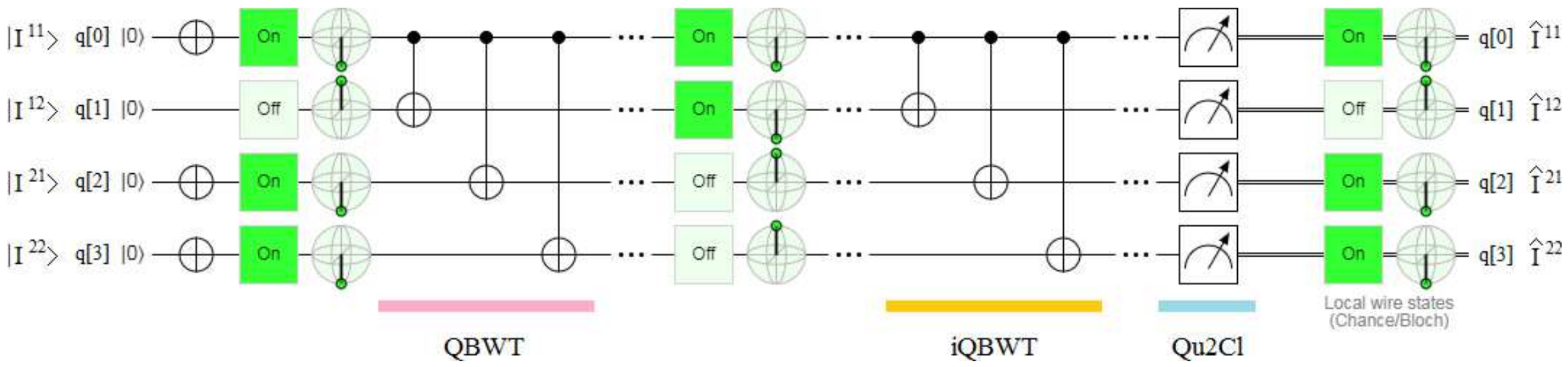

**Figure 25.** Outcomes on Quirk [9] for an implementation of QBWT and iQBWT.

**Outcomes on QPS [8]:** With the same order and criteria as in the previous case, we carried out the implementation of the QBWT on QPS [8]. The results are shown in Figs. 26 and 27. Figure 26 fully matches the results in Fig.23. Figure 27 shows us the probability, the angles and the Bloch's sphere. Unlike Quirk [9] that shows everything on a single screen, QPS does it on separate screens [8].

Although QPS [8] is a cloud service which can be connected to IBM Q [6] and Rigetti [7] to run the quantum algorithm on their QPUs, it is an extremely complicated procedure, apparently, to ingest it from the Cl2Qu interface. Then, we built a parser in C/C++/Python from the text plain, with the bits of the classical image, to qasm (Quantum Assembly Language) [6, 8]. This implies:

I. run the parser,
II. simulate or run the qasm on QPS [8], and
III. visually recover the results, as in the case of Quirk [9],

as many times as necessary, according to the size of the image.

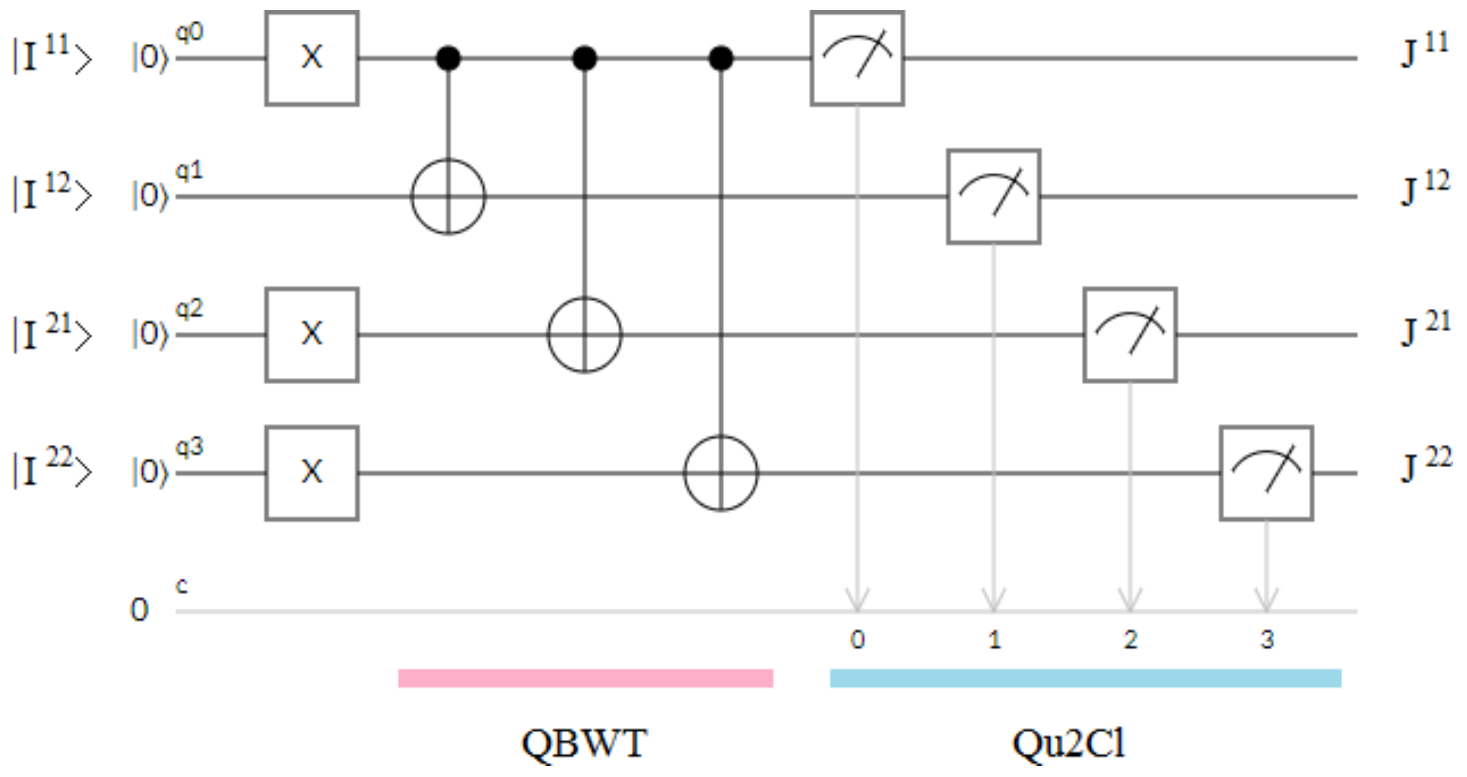

**Figure 26.** Graphic layout of QBWT on QPS [8].

Measure all

| Qubit | Measured | Probability of 1 | θ °rad | φ °rad | θ °deg | φ °deg | Bloch |
|---|---|---|---|---|---|---|---|
| q0 | 1 | 1 | 3.14159265358979 | 0 | 180 | 0 | |
| q1 | 1 | 1 | 3.14159265358979 | 0 | 180 | 0 | |
| q2 | 0 | 0 | 0 | 0 | 0 | 0 | |
| q3 | 0 | 0 | 0 | 0 | 0 | 0 | |

**Figure 27.** Outcomes on QPS [8] for an implementation of QBWT.

QPS [8] works in two modalities, i.e., simulates or executes quantum algorithms from:

a) qasm editor, or
b) drag and drop designer.

Similar to the previous case, if we want to verify the joint work of QBWT and iQBWT, then we must turn to Figures 28 and 29.

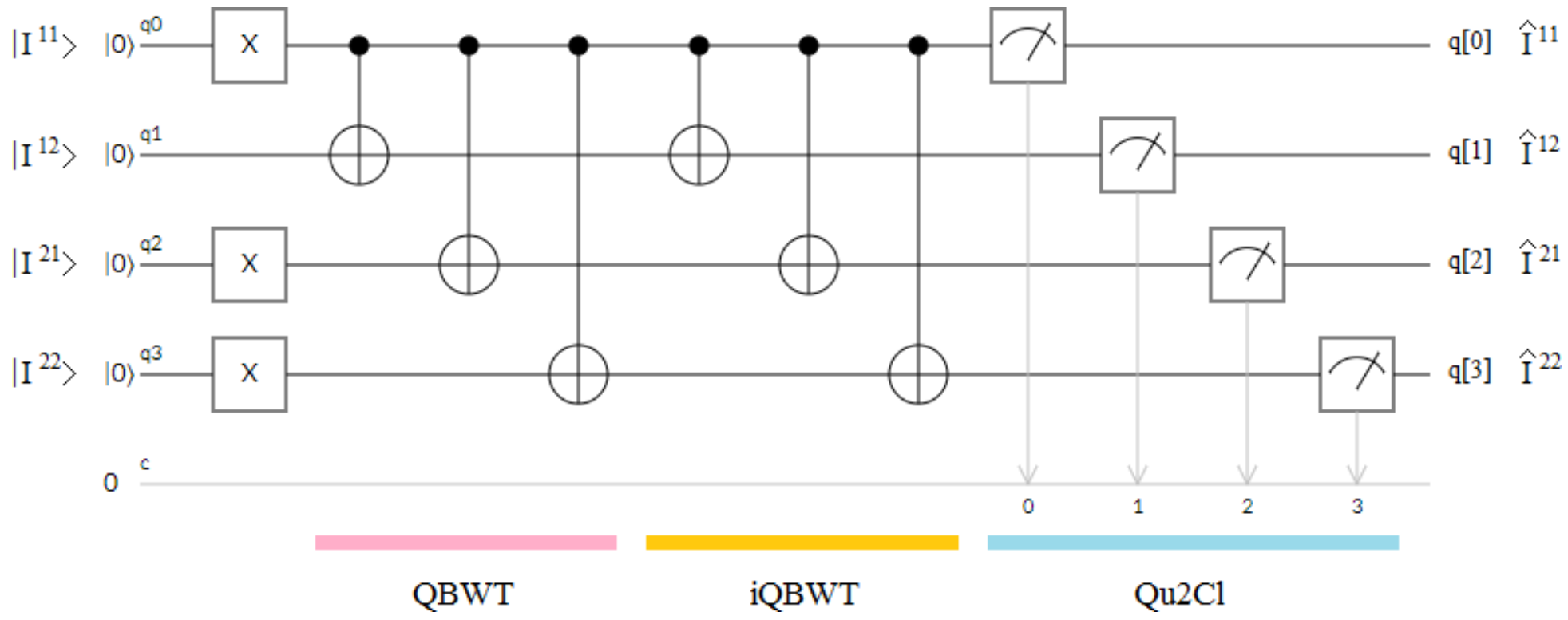

**Figure 28.** Graphic layout of QBWT and iQBWT on QPS [8].

The results obtained from Figs. 28 and 29 with QPS [8] are identical to those obtained with the Quirk platform [9] in Fig. 25.

For the two platforms that follow: IBM Q [6] and Rigetti [7], we can resort to the most recomended possibility presented below, i.e., to the architecture of Fig.30.

Measure all

| Qubit | Measured | Probability of 1 | θ °rad | φ °rad | θ °deg | φ °deg | Bloch |
|---|---|---|---|---|---|---|---|
| q0 | 1 | 1 | 3.14159265358979 | 0 | 180 | 0 | |
| q1 | 0 | 0 | 0 | 0 | 0 | 0 | |
| q2 | 1 | 1 | 3.14159265358979 | 0 | 180 | 0 | |
| q3 | 1 | 1 | 3.14159265358979 | 0 | 180 | 0 | |

**Figure 29.** Outcomes on QPS [8] for an implementation of QBWT and iQBWT.

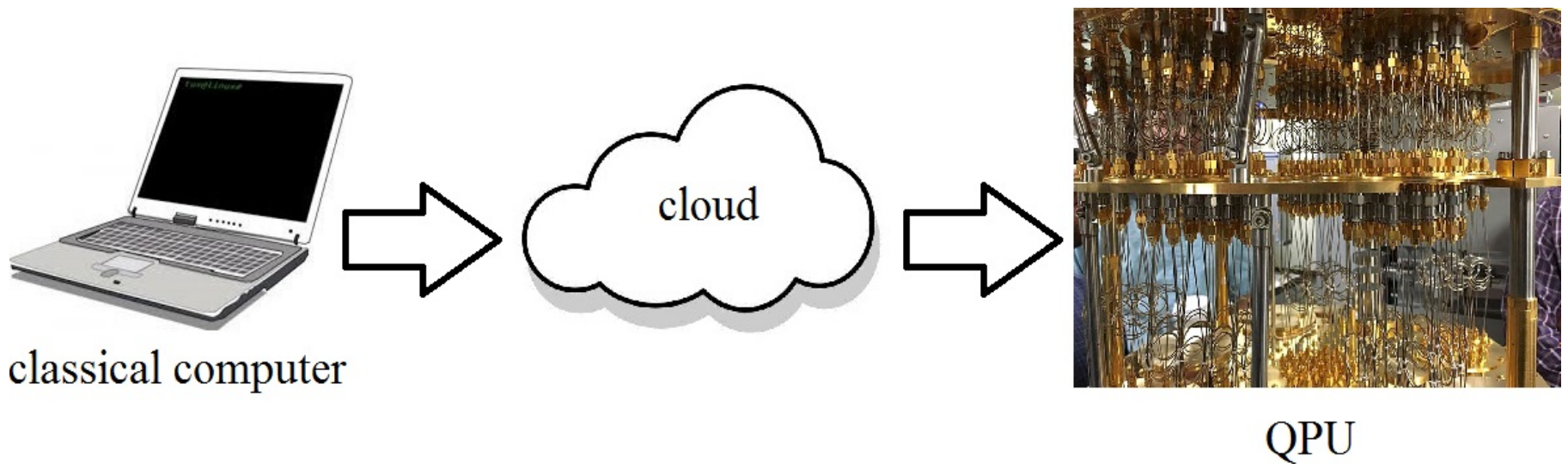

**Figure 30.** Architecture: classical computer + cloud + QPU of IBM Q [6] or Rigetti [7].

This architecture, Fig.30, involves only one classical computer, which is the host of the digital image (original and classical), the cloud accessed via Internet, and the corresponding QPU. It is clear that for this architecture it is not necessary that the classical computer and the QPU are in the same location. The key lies inside the classical computer. In this case, it will have:

a) a plain text, or eventually, a graphic file format with the digital image,

b) the C++ program with the bit-slacer procedure,

c) a connection software between the classical computer and the Quantum Machine Image (QMI) of the QPU via Putty [38] (a free SSH and telnet client for Windows), from the use of the cloud, in the case of Forest (development environment) by Rigetti [7], or a much more direct form in the case of IBM Q [6] from the use of qiskit (Quantum Information Science kit),

d) the Cl2Qu interface emulated in C++, and

e) the different parsers, i.e.:
- from plain text to qasm/qiskit for IBM Q [6], or
- from plain text to pyquil/quil (instruction set architecture) for Rigetti [7].

Next, we will use this last architecture, because it is available to everyone in a free and simple way (i.e., it is not necessary to access the facilities where the QPU is hosted) so that anyone can reproduce the experiments of this subsection without any restrictions.

**Outcomes on IBM Q [6]:** QBWT is implemented on IBM Q as is shown in Fig.31. Figure 32 shows the probability of the outcomes for this experiment. In the base of the blue bar, we will have the following qubits sequence, in order: q[3]q[2]q[1]q[0] = 0011. Figure 33 shows the characteristics of the used topology, it is about the IBM Q 5 Yorktown [ibmqx2] topology, and the implementation was carried out under the following parameters:

a) shots = 8192
b) seed = 7 and random

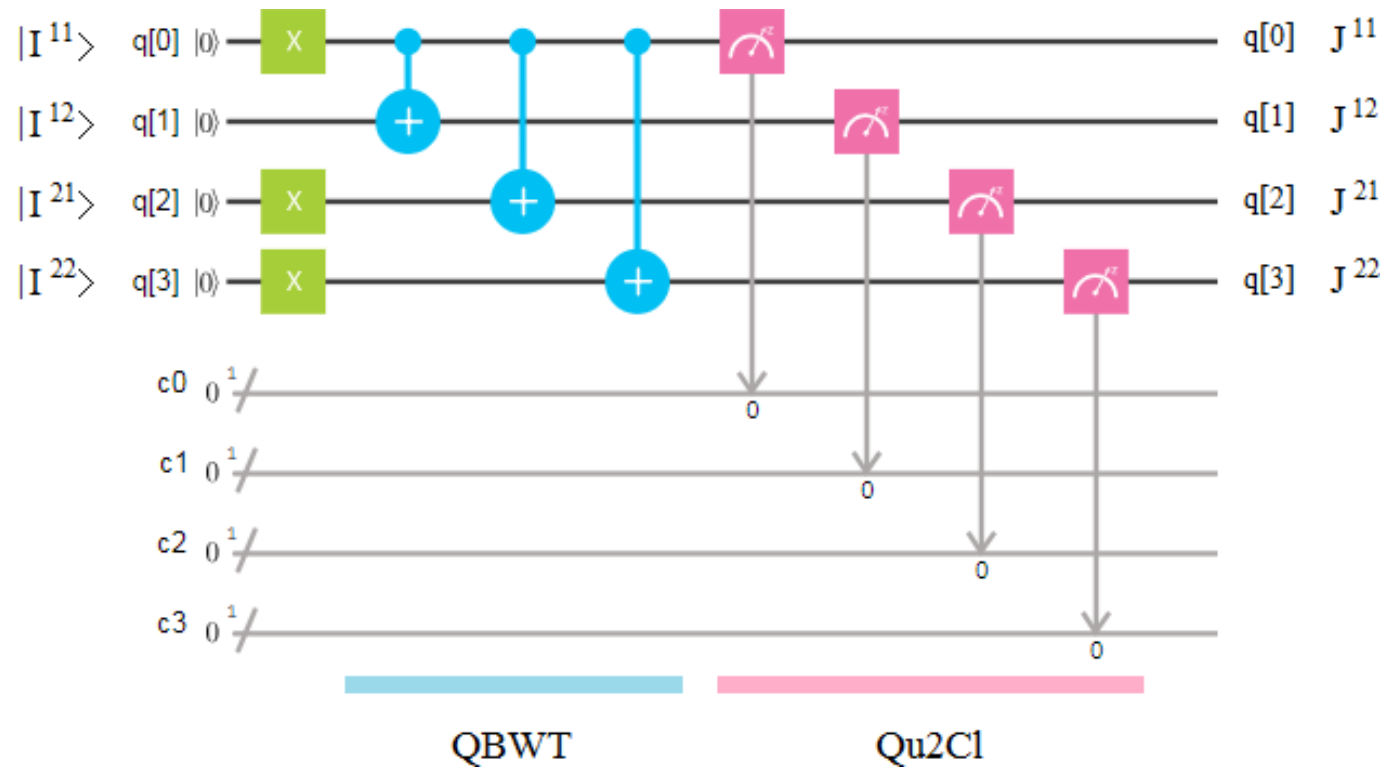

**Figure 31.** Graphic layout of QBWT on IBM Q [6].

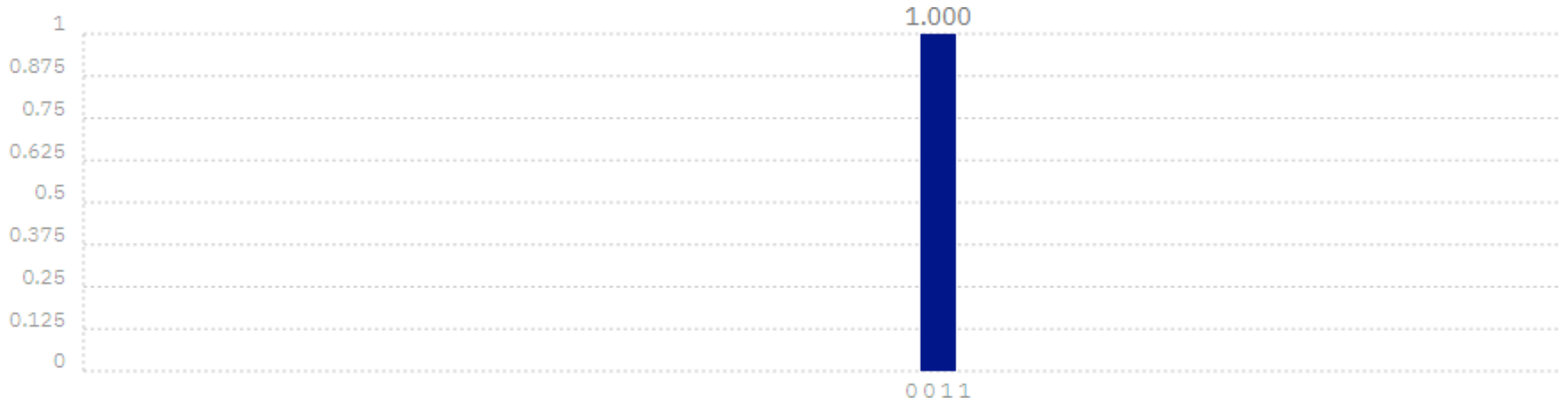

**Figure 32.** Probability of the outcomes for QBWT.

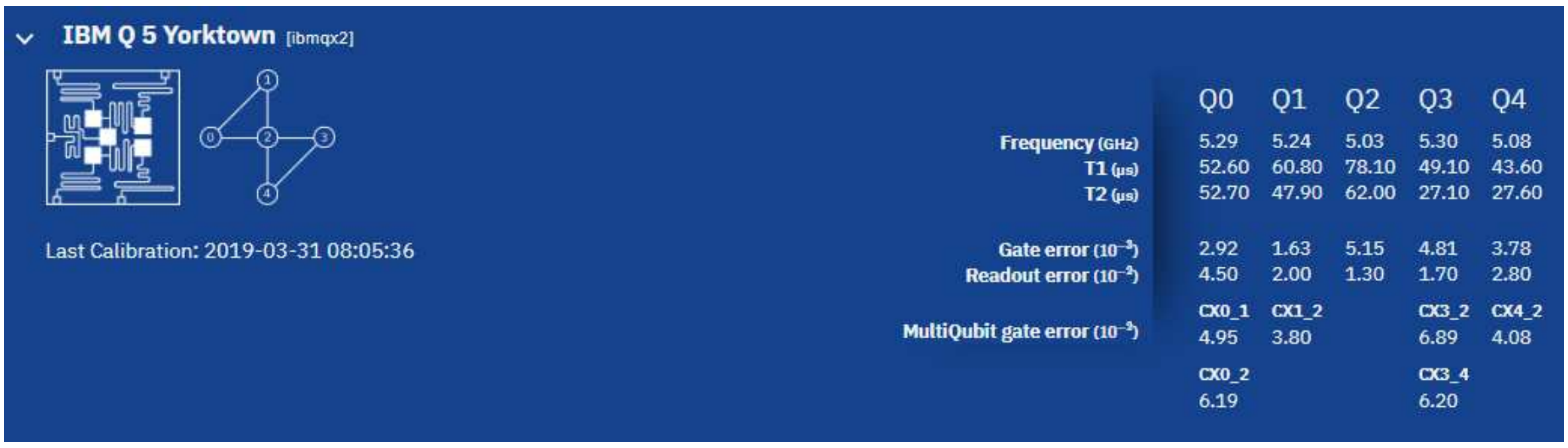

**Figure 33.** Characteristics of the used topology.

Now, we are going to verify the complete recovery of the original values by configuring Fig.34 and by applying QBWT, and iQBWT to the QBWT output.

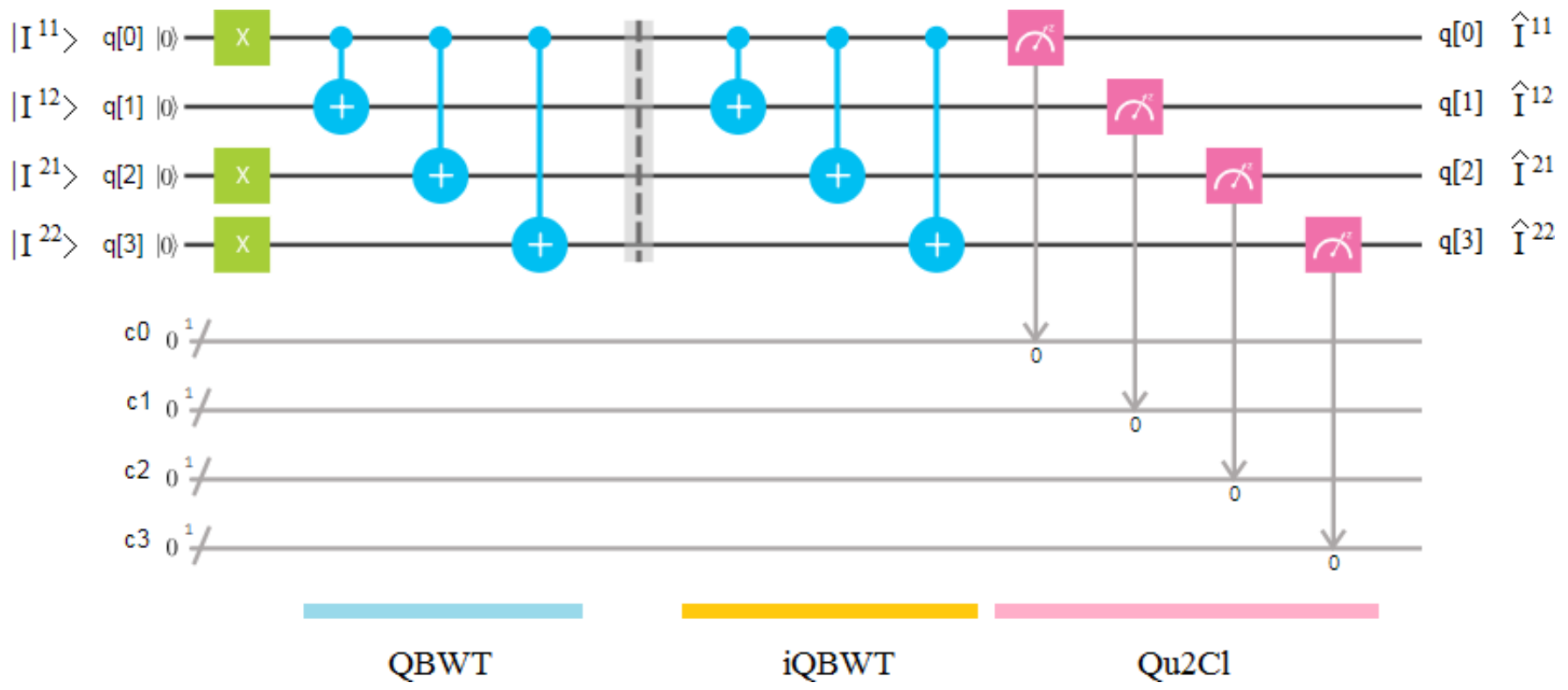

**Figure 34.** Graphic layout of QBWT and iQBWT on IBM Q [6].

Figure 35 shows the probability of the outcomes for this experiment. In the base of the blue bar, we will have the following qubits sequence, in order: q[3]q[2]q[1]q[0] = 1101. These results coincide in their totality with those obtained with the previous platforms.

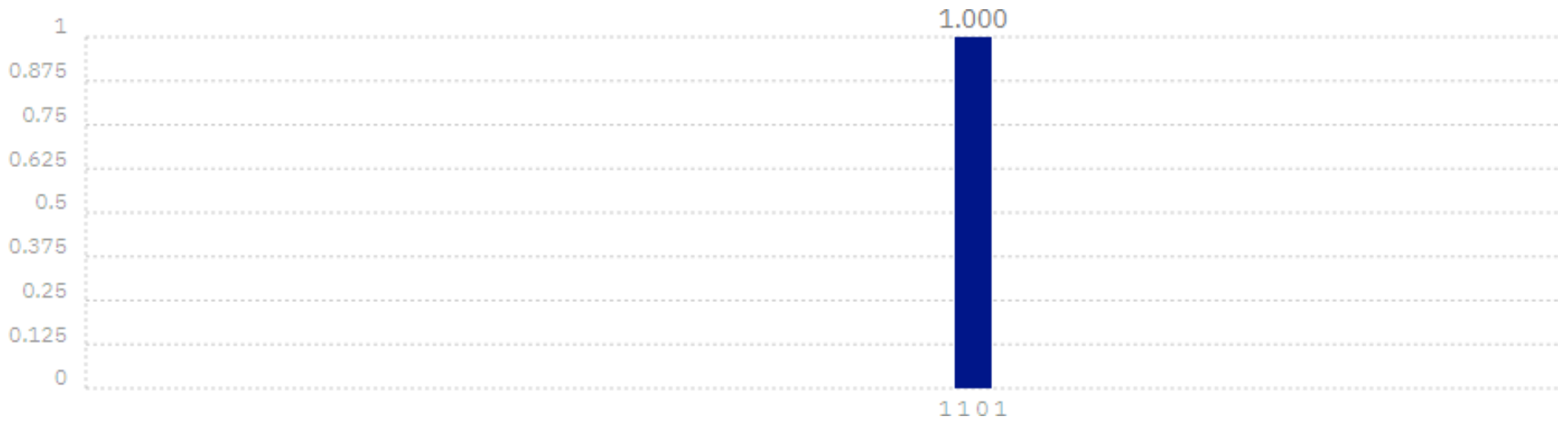

**Figure 35.** Probability of the outcomes for QBWT and iQBWT.

**Outcomes on Rigetti [7]:** The results on both the QVM and the QPU of Rigetti [7] in pyquil/quil, were identical to those of IBM Q [6] in qasm/qiskit. The lattice used was Aspen-3-4Q-F with the topology and characteristics shown in Fig.36.

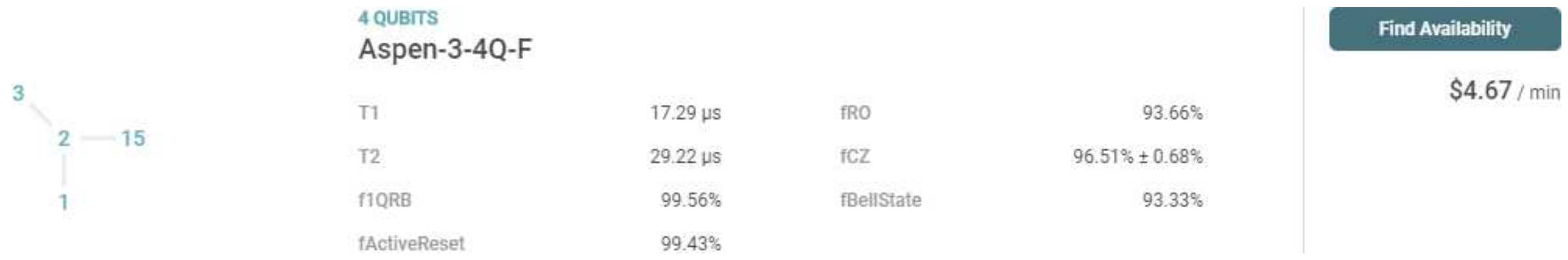

**Figure 36.** Characteristics of the used Rigetti's lattice.

It is not justified to continue implementing QBWT and iQBWT on the remaining platforms of the market: Xanadu [39], D-Wave [40], Google [41], Microsoft [42], and Quantum Inspire [43], since the performance of QBWT, iQBWT and the Cl2Qu interface (both in its implementation on a classic computer and in its dedicated physical version) has been fully demonstrated by the strength of the results obtained so far. Therefore, we will continue with the implementation of the next quantum algorithm.

### *6.2.2 Quantum Boolean Orthogonalization Process (QBOP)*

For this experiment, we will again decompose the image of Fig.13 in its 24 bitplanes. Figure 14 shows a simplified detail of this procedure for a sector of 4-by-4 pixels of Fig. 13, which is carried out by the same *bit-slicer*() function used in the previous quantum algorithm. Based on the work modality proposed in Fig.30, we will use the Cl2Qu interface of Fig.37 on the first 4 most significant bitplanes of Figs. 14 and 15.

Let us suppose the first 4 most significant bits of Fig.14 for a pixel in the position (r, c), being *r* the row and *c* the column in the corresponding tile. Equation (49) shows the conversion to quantum via a Cl2Qu interface.

$$\begin{bmatrix} I^0_{r,c} \\ I^1_{r,c} \\ I^2_{r,c} \\ I^3_{r,c} \end{bmatrix} = \begin{bmatrix} 1 \\ 0 \\ 1 \\ 0 \end{bmatrix} \rightarrow Cl2Qu \rightarrow \begin{bmatrix} \left| I^0_{r,c} \right\rangle \\ \left| I^1_{r,c} \right\rangle \\ \left| I^2_{r,c} \right\rangle \\ \left| I^3_{r,c} \right\rangle \end{bmatrix} = \begin{bmatrix} |1\rangle \\ |0\rangle \\ |1\rangle \\ |0\rangle \end{bmatrix} \tag{49}$$

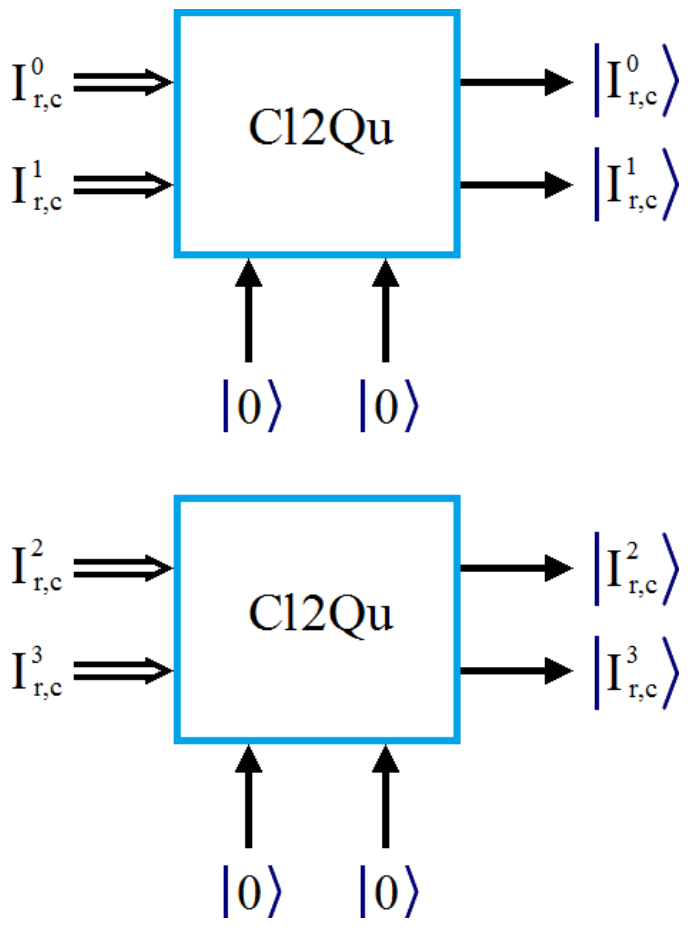

**Figure 37.** Cl2Qu interface for the intake of the first 4 bits of the Figures 14 and 15.

In its generic form, the QBOP algorithm represents a mapping between a non-orthogonal set $\left|I_{r,c}^{d}\right\rangle$, and an orthogonal set $\left|J_{r,c}^{d}\right\rangle$, as we can see in Eq.(50)

$$\begin{aligned}&\left|I_{r,c}^{d}\right\rangle \quad / \, I \in \{|0\rangle,|1\rangle\}, d\,(bitplane\ index) \in [1,8]\,(for\ every\ color), r \in [1, ROW], c \in [1, COL] \\ &\downarrow \\ &\left|J_{r,c}^{d}\right\rangle \quad / \, J \in \{|0\rangle,|1\rangle\}, d\,(bitplane\ index) \in [1,8]\,(for\ every\ color), r \in [1, ROW], c \in [1, COL]\end{aligned} \tag{50}$$

In its general form, QBOP is represented according to Eq.(51)

$$\begin{aligned}&\left|J_{r,c}^{0}\right\rangle = \left|I_{r,c}^{0}\right\rangle \\ &\left|J_{r,c}^{d}\right\rangle = \left|I_{r,c}^{d}\right\rangle \bigoplus_{k=0}^{d-1} \left|I_{r,c}^{d}\right\rangle \wedge \left|J_{r,c}^{k}\right\rangle \quad / \, d \in [1, D-1]\end{aligned} \tag{51}$$

where $\oplus$ is the *XOR* gate, $\wedge$ the *AND* gate, and $D$ is the bit depth for each color channel (generally $D$ = 8). In particular, for the example of Eq.(49), we will have,

$$\begin{aligned}&\left|J_{r,c}^{0}\right\rangle = \left|I_{r,c}^{0}\right\rangle \\ &\left|J_{r,c}^{1}\right\rangle = \left|I_{r,c}^{1}\right\rangle \oplus \left|I_{r,c}^{1}\right\rangle \wedge \left|J_{r,c}^{0}\right\rangle \\ &\left|J_{r,c}^{2}\right\rangle = \left|I_{r,c}^{2}\right\rangle \oplus \left|I_{r,c}^{2}\right\rangle \wedge \left|J_{r,c}^{0}\right\rangle \oplus \left|I_{r,c}^{2}\right\rangle \wedge \left|J_{r,c}^{1}\right\rangle \\ &\left|J_{r,c}^{3}\right\rangle = \left|I_{r,c}^{3}\right\rangle \oplus \left|I_{r,c}^{3}\right\rangle \wedge \left|J_{r,c}^{0}\right\rangle \oplus \left|I_{r,c}^{3}\right\rangle \wedge \left|J_{r,c}^{1}\right\rangle \oplus \left|I_{r,c}^{3}\right\rangle \wedge \left|J_{r,c}^{2}\right\rangle\end{aligned} \tag{52}$$

In a tensor way, we must see something like the following,

$$\left|J_{r,c}\right\rangle = \left|I_{r,c}\right\rangle \oplus \left|K_{r,c}\right\rangle \tag{53}$$

where,

$$\left|J_{r,c}\right\rangle = \begin{bmatrix} \left|J_{r,c}^{0}\right\rangle \\ \left|J_{r,c}^{1}\right\rangle \\ \left|J_{r,c}^{2}\right\rangle \\ \left|J_{r,c}^{3}\right\rangle \end{bmatrix} \tag{54}$$

$$\left|I_{r,c}\right\rangle = \begin{bmatrix} \left|I_{r,c}^{0}\right\rangle \\ \left|I_{r,c}^{1}\right\rangle \\ \left|I_{r,c}^{2}\right\rangle \\ \left|I_{r,c}^{3}\right\rangle \end{bmatrix} \tag{55}$$

$$\left|K_{r,c}\right\rangle = \begin{bmatrix} I_d \\ \left|I_{r,c}^{1}\right\rangle \wedge \left|J_{r,c}^{0}\right\rangle \\ \left|I_{r,c}^{2}\right\rangle \wedge \left|J_{r,c}^{0}\right\rangle \oplus \left|I_{r,c}^{2}\right\rangle \wedge \left|J_{r,c}^{1}\right\rangle \\ \left|I_{r,c}^{3}\right\rangle \wedge \left|J_{r,c}^{0}\right\rangle \oplus \left|I_{r,c}^{3}\right\rangle \wedge \left|J_{r,c}^{1}\right\rangle \oplus \left|I_{r,c}^{3}\right\rangle \wedge \left|J_{r,c}^{2}\right\rangle \end{bmatrix} \tag{56}$$

Being $I_d$ the identity matrix. Therefore, if we want to recover $\left|I_{r,c}\right\rangle$ from $\left|J_{r,c}\right\rangle$ and $\left|K_{r,c}\right\rangle$,

$$\left|I_{r,c}\right\rangle = \left|J_{r,c}\right\rangle \vee \left|K_{r,c}\right\rangle \tag{57}$$

where $\vee$ is the *OR* gate. Figure 38 represents Eq.(57) (taking into account all bits) for the red channel of Fig.13, while, Fig.39 shows us the complete set of bitplanes of these three components (*I*, *J*, and *K*). Figure 38 shows that the *J* image represents the dithering [29, 36] version of the original image *I*.

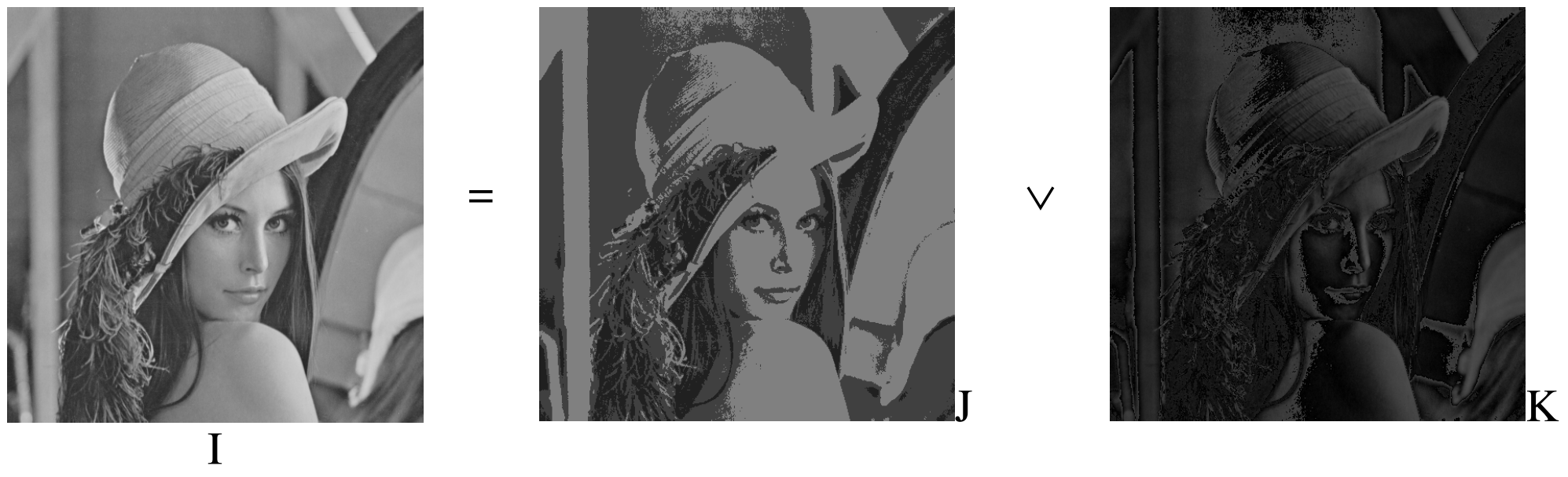

**Figure 38.** Detail of the three components of Eq.(57).

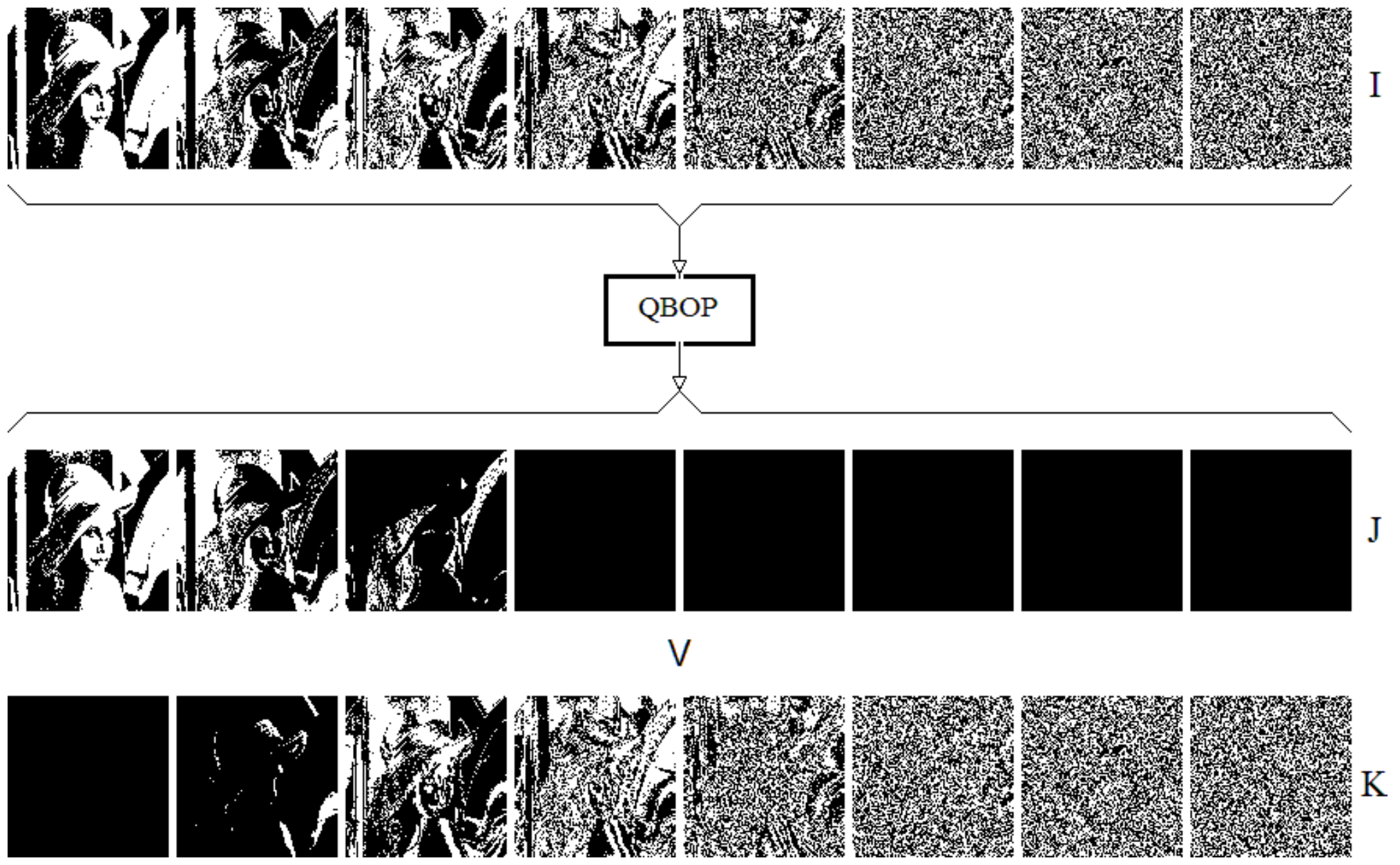

**Figure 39.** Slicing process with the three components of Eq.(57)

What we have explained up to now in theoretical form about the QBOP algorithm is what should be verified in the four quantum platforms [6-9] on which the implementations will be carried out next.

**Outcomes on Quirk [9]:** For the example of Eq.(49), we must obtain the results on the right side of Eq.(58). These results are perfectly verified in the implementation on Quirk of Fig.40, with an input sequence: $q[0]q[1]q[3]q[4]=|1\rangle|0\rangle|1\rangle|0\rangle$, and an output sequence: $q[0]q[2]q[4]q[5]=|1\rangle|0\rangle|0\rangle|0\rangle$.

$$\begin{bmatrix}|I^0_{r,c}\rangle\\|I^1_{r,c}\rangle\\|I^2_{r,c}\rangle\\|I^3_{r,c}\rangle\end{bmatrix}=\begin{bmatrix}|1\rangle\\|0\rangle\\|1\rangle\\|0\rangle\end{bmatrix}\to QBOP\to\begin{bmatrix}|J^0_{r,c}\rangle\\|J^1_{r,c}\rangle\\|J^2_{r,c}\rangle\\|J^3_{r,c}\rangle\end{bmatrix}=\begin{bmatrix}|1\rangle\\|0\rangle\\|0\rangle\\|0\rangle\end{bmatrix} \tag{58}$$

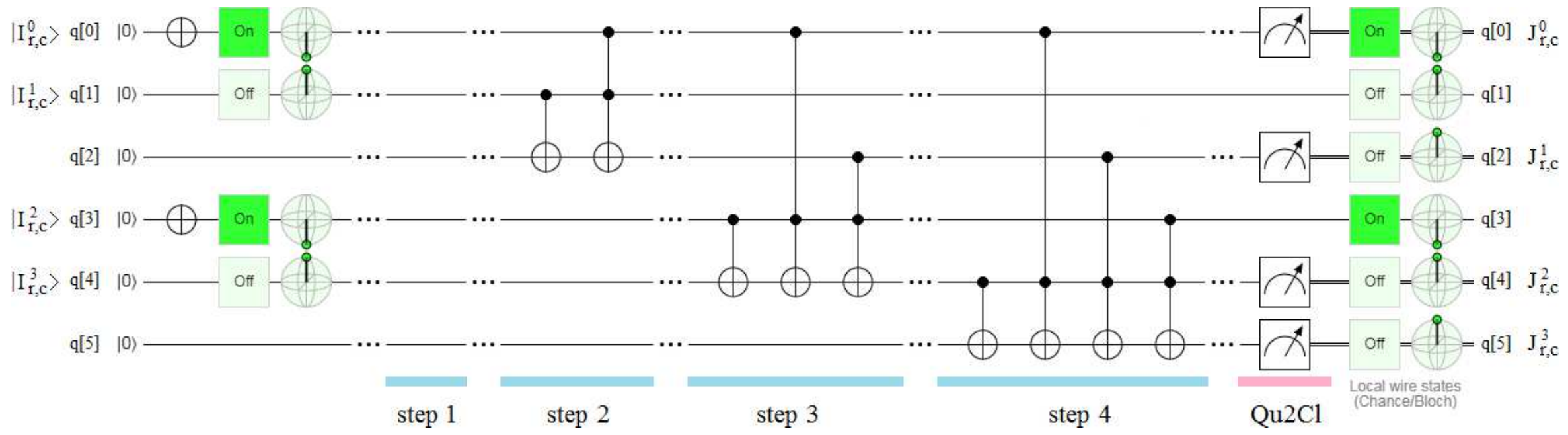

**Figure 40.** Outcomes on Quirk [9] for an implementation of QBOP.

The four rows of Eq.(52) are corresponding with the four steps of Fig.40. The output sequence automatically verifies the orthogonality, since:

$$\begin{aligned}|J^0_{r,c}\rangle\wedge|J^1_{r,c}\rangle&=|1\rangle\wedge|0\rangle=|0\rangle\\|J^0_{r,c}\rangle\wedge|J^2_{r,c}\rangle&=|1\rangle\wedge|0\rangle=|0\rangle\\|J^0_{r,c}\rangle\wedge|J^3_{r,c}\rangle&=|1\rangle\wedge|0\rangle=|0\rangle\\|J^1_{r,c}\rangle\wedge|J^2_{r,c}\rangle&=|0\rangle\wedge|0\rangle=|0\rangle\\|J^1_{r,c}\rangle\wedge|J^3_{r,c}\rangle&=|0\rangle\wedge|0\rangle=|0\rangle\\|J^2_{r,c}\rangle\wedge|J^3_{r,c}\rangle&=|0\rangle\wedge|0\rangle=|0\rangle\end{aligned} \tag{59}$$

**Outcomes on QPS [8]:** This hybrid platform (since it combines drag and drop scheme with qasm coding) allows us to access results identical to those obtained with Quirk [9]. Figure 41 shows the drag and drop scheme with the four steps of Eq.(52), where we have the following equivalences in relation to the inputs of Eq.(60):

$$|I^0_{r,c}\rangle|I^1_{r,c}\rangle|I^2_{r,c}\rangle|I^3_{r,c}\rangle\to q[0]q[1]q[3]q[4]=|1\rangle|0\rangle|1\rangle|0\rangle, \tag{60}$$

and the output of Eq.(61):

$$|J^0_{r,c}\rangle|J^1_{r,c}\rangle|J^2_{r,c}\rangle|J^3_{r,c}\rangle\to q[0]q[2]q[4]q[5]=|1\rangle|0\rangle|0\rangle|0\rangle. \tag{61}$$

The inputs q[2] and q[5] are ancillas, and the outpus q[1] and q[3] are trash [1].

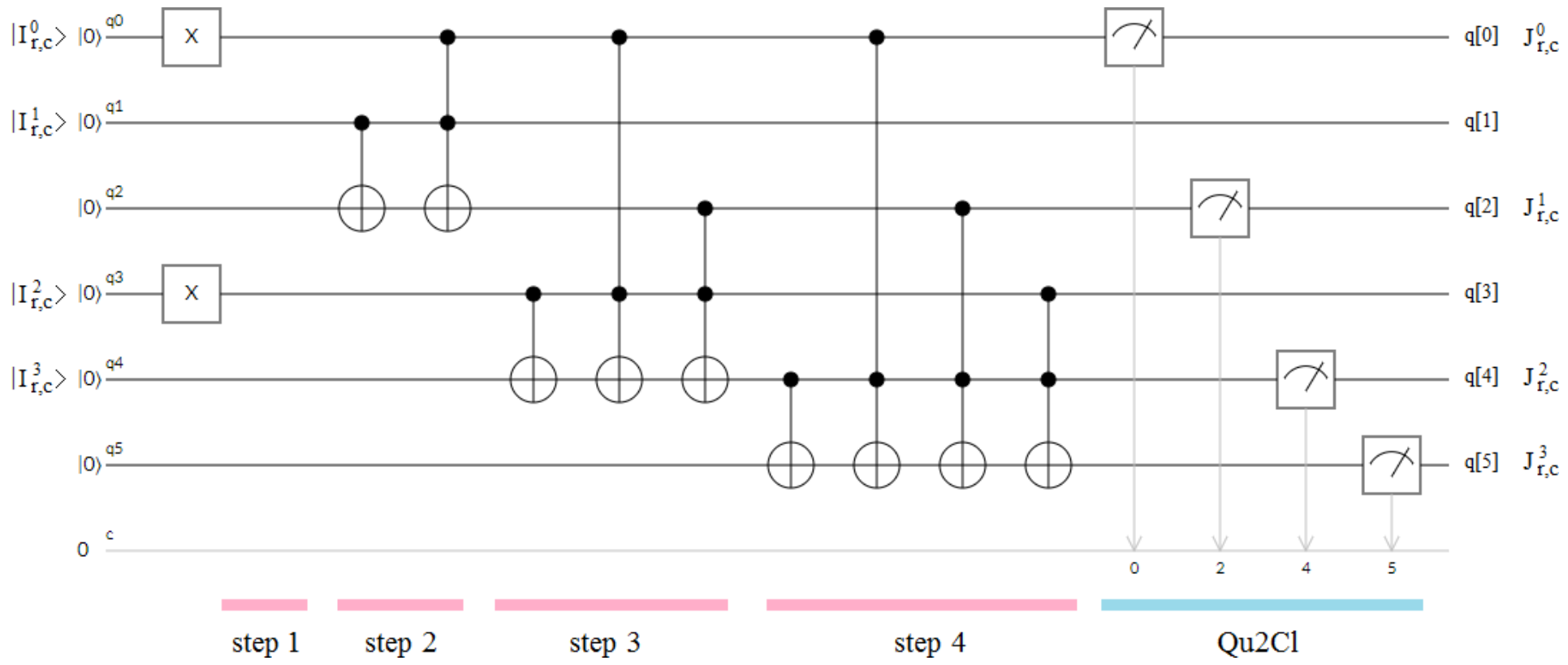

**Figure 41.** Graphic layout of QBOP on QPS [8].

The parser was used for the conversion between plaintext and qasm code. We must remember that QSP [8] has the option to run programs remotely on Rigetti's [7] and IBM Q [6] physical machines via a *ssh* connection to both QMIs. This option was used with total success. Finally, Fig.42 shows us the probabilities, angles and Bloch's spheres of the outcomes.

Measure all

| Qubit | Measured | Probability of 1 | θ °rad | φ °rad | θ °deg | φ °deg | Bloch |
|---|---|---|---|---|---|---|---|
| q0 | 1 | 1 | 3.14159265358979 | 0 | 180 | 0 | |
| q1 | 0 | 0 | 0 | 0 | 0 | 0 | |
| q2 | 0 | 0 | 0 | 0 | 0 | 0 | |
| q3 | 1 | 1 | 3.14159265358979 | 0 | 180 | 0 | |
| q4 | 0 | 0 | 0 | 0 | 0 | 0 | |
| q5 | 0 | 0 | 0 | 0 | 0 | 0 | |

**Figure 42.** Outcomes on QPS [8] for an implementation of QBOP.

**Outcomes on IBM Q [6]:** Figure 43 represents the implementation of the QBOP on this platform. The steps are separated thanks to three barriers for the purpose of eliminating optimizations *per se* by the platform that tends to group gates in an inappropriate way. The outcomes are identical to the previous platforms and their probabilities are shown in Fig.44.

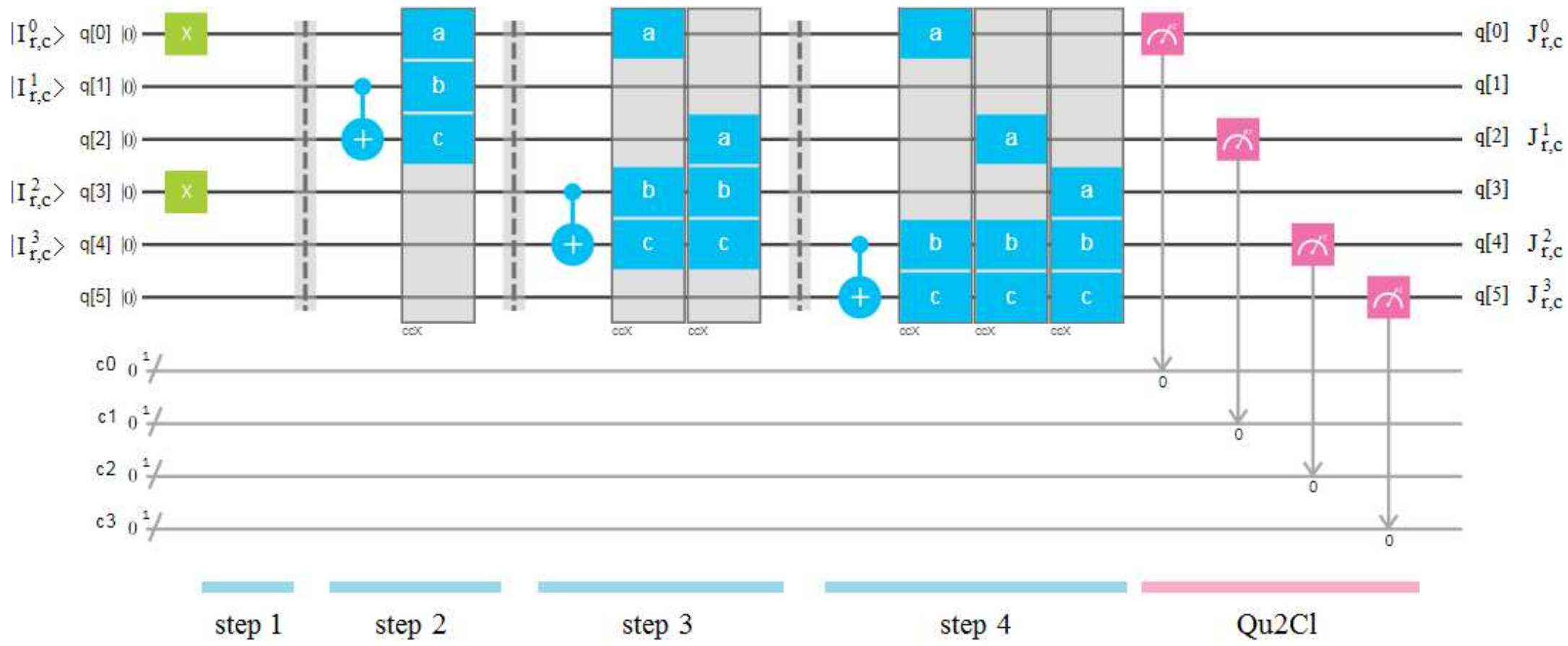

**Figure 43.** Graphic layout of QBOP on IBM Q [6].

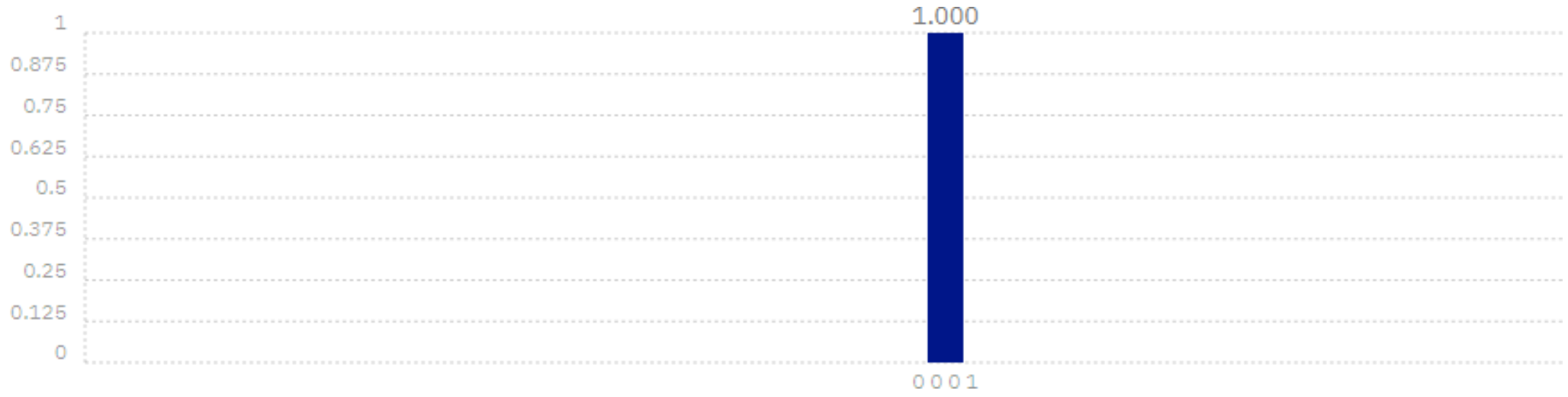

**Figure 44.** Probability of the outcomes for QBOP.

Figure 45 shows the characteristics of the used topology for this case, an IBM Q 14 Melbourne topology, which was coded in qiskit (for which, it is necessary a sign-up as a Beta user). Finally, the QBOP implementation was carried out under the following parameters:

a) shots = 8192
b) seed = 7 and random

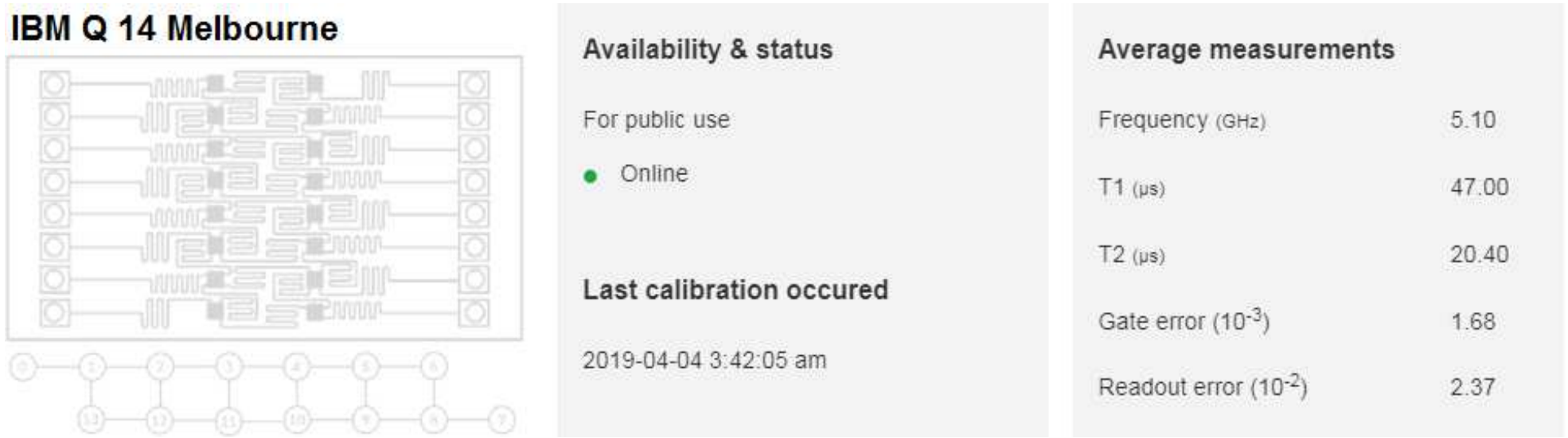

**Figure 45.** Characteristics of the used topology.

**Outcomes on Rigetti [7]:** The results in pyquil/quil both on the QVM and on the QPU of Rigetti [7] were identical to those of IBM Q [6]. The lattice used in this case was Aspen-3-6Q-B with the topology and characteristics shown in Fig.46.

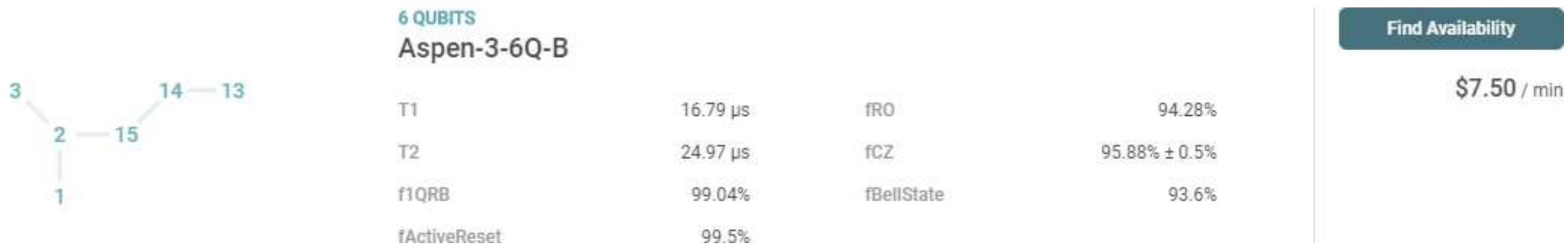

**Figure 46.** Characteristics of the used Rigetti's lattice.

### *6.3 Final comparison*

In Table III, we have a final comparison between FRQI, NEQR and QBIP, where the superscripts mean:

1. The fidelity of FRQI is reduced due to the immense amount of gates used. Given the circuit complexity of FQRI, it is practically impossible to do a mirror evaluation as in the example of Subsection 2.6 to calculate fidelity,
2. NEQR presents the same problem that we explained in the previous item for FRQI,
3. QBIP does not have gates because it is a criterion,
4. QBIP does not imply computational cost because it is a criterion. The only computational cost involved is that of the Cl2Qu interface and the associated quantum algorithm,
5. QBIP does not imply storage because it is a criterion. The only storage involved is that of the Cl2Qu interface and the associated quantum algorithm,
6. Since the qubits used are not orthogonal, then quantum measurement notably modifies the result,
7. NEQR presents the same problem that we explained in the previous item for FRQI,
8. PQC means physical quantum computer (QPU)

**TABLE III:** Final comparison between FRQI, NEQR and QBIP.

| Main features | Technique for the internal representation of the image | | |
|---|---|---|---|
| | FRQI | NEQR | QBIP |
| Does it have a Cl2Qu interface? | No | No | Yes |
| Data intake | By hand | By hand | Automatic via Cl2Qu |
| Are the qubits used orthogonal? | No (see Eq.20) | No (see Eq.33) | Yes |
| Impact of gate's noise and decoherence | High | High | Practically null |
| Fidelity | -[1] | -[2] | Practically equal to 1 |
| Number of gates | High | High | 0[3] |
| Computational cost | High | High | 0[4] |
| Number, dimensionality and size of the host registers inside QPU | High | High | Low |
| Required storage | High | High | 0[5] |
| Impact of quantum measurement | High[6] | High[7] | Practically null |
| Implementation on a PQC[8] | Unviable to date | Unviable to date | Successful on 8 PQC |

Table III synthesizes what has been demonstrated in the previous subsections regarding the performance of the three techniques.

## 7. Discussions

Nothing that was done with QBIP [4] on the nine mentioned platforms [6-9, 39-43] can be reproduced successfully with FRQI [1] and NEQR [3]. The experimental evidence obtained from the previous sections and synthesized in the comparative analysis expressed in Table III speaks by itself, beyond any possible doubt, about which technique works and which does not work in QImP.

## 8. Conclusions

As we saw in Sections 3 and 4, FRQI and NEQR do not work as a Cl2Qu interface, in fact, and as we saw in Section 6, they do not work at all. In fact, all the demonstrations made in this paper about the lack of performance of FRQI and NEQR are extensive to all their variants. On the other hand, the experiments carried out in Section 6 [44], are only a couple among many others carried out in laboratories around the world with similar results regarding the poor performance of FRQI and NEQR. In those laboratories, optical implementations of the three techniques were performed with results that are absolutely similar to those reported here. These results show us that the defects or virtues of each technique do not have anything to do with the platform of implementation selected by the researchers, but with the technique itself. Appendices A, B, C, D, and E show the entanglement coupling problem of NEQR for Quirk [9], QPS [8], IBM Q[6], Rigetti [7], and Quantum Inspire of QUTech [43]. The easy online access to simulators and QPUs of platforms such as IBM Q [6] and Rigetti [7] make the personal simulations of the QImP's authors on an HLI [2] absolutely incomprehensible.

Finally, and despite everything demonstrated in this work, if we use the correct techniques for the internal representation of the image, that is, those that really work, as is the case of Quantum Boolean Image Processing (QBIP), there is not the least doubt that Quantum Image Processing (QImP) will be one of the most successful tools within the Quantum Technology toolbox.

**Conflicts of Interest:** The author declares that there are no competing interests.

**Funding Statement:** The author acknowledges funding by Qubit Reset Labs, and American startup, under contract QImP-01#5/2/2019.

**Acknowledgments:** M. Mastriani thanks to the board of directors of Qubit Reset Labs by all its support. Besides, author thanks to Tushar Mittal, Amy Brown, and Eric C. Peterson from Rigetti for their complete description of the possible physical implementation of the qubit reset gate. Also to the Slack community of Rigetti for all its interesting observations. A very special thanks to Will J. Zeng, ex-Rigetti and ex-contributor of Quantum Programming Studio, and currently responsible for the Unitary Fund for all his kindness. I cannot fail to thank or forget the interesting exchange of ideas with Craig Gidney, head of Quirk, currently at Google, in particular about the practical feasibility of the qubit reset gate. A special acknowledgment to Antonio D. Corcoles- Gonzalez, and Jay M. Gambetta of IBM Q, as well as to their community for their important clarifications about the practical implementation of the qubit reset gate and the *if-then-else* statement on their Premium QPU, and Francisco Galvez, ex-IBM Spain and currently at ABDProf for teaching me the way to them. I also want to extend my thanks to Greg Gabrenya of D-Wave and Josh Izaac of Xanadu for answering all my questions so professionally. A very special recognition to Martín PiPuig of LIDI laboratory at National University of La Plata for having created all the configuration and programming for access to the Rigetti's QMI from our terminals. I also want to thank all staff of Qubit Reset Labs, for tolerating me in our presentations in Las Vegas, Santa Clara, Palo Alto, Stanford and New York in 2019. Finally, a special recognition to all for their patience, dedication and professionalism.

## Appendix A: Entanglement coupling on Quirk [9] for the alternative version of NEQR of Fig. 9

In this appendix, we will verify the presence of entanglement among the outcomes of the alternative version of NEQR in Fig. 9 on Quirk [9]. As in the case of Fig. 12 with the first version of NEQR in Fig. 8, if we use a pair of outcomes of this technique as an entangled pair, we will manage to teleport an arbitrary qubit state. Said teleportation will confirm the aforementioned entanglement, given that teleportation can only exist if it is previously vehiculized by the creation and distribution of an entangled pair. As in the previous case, the qubit to be teleported will be,

$$|\psi\rangle = HTHT|0\rangle \rightarrow \left(85.35534\,\%\,|0\rangle, 14.64466\,\%\,|1\rangle\right).$$

Figure A1 represents this experiment, where, $|\psi\rangle$ is introduced in q[10], while the entangled pair is taken from q[0] and q[9].

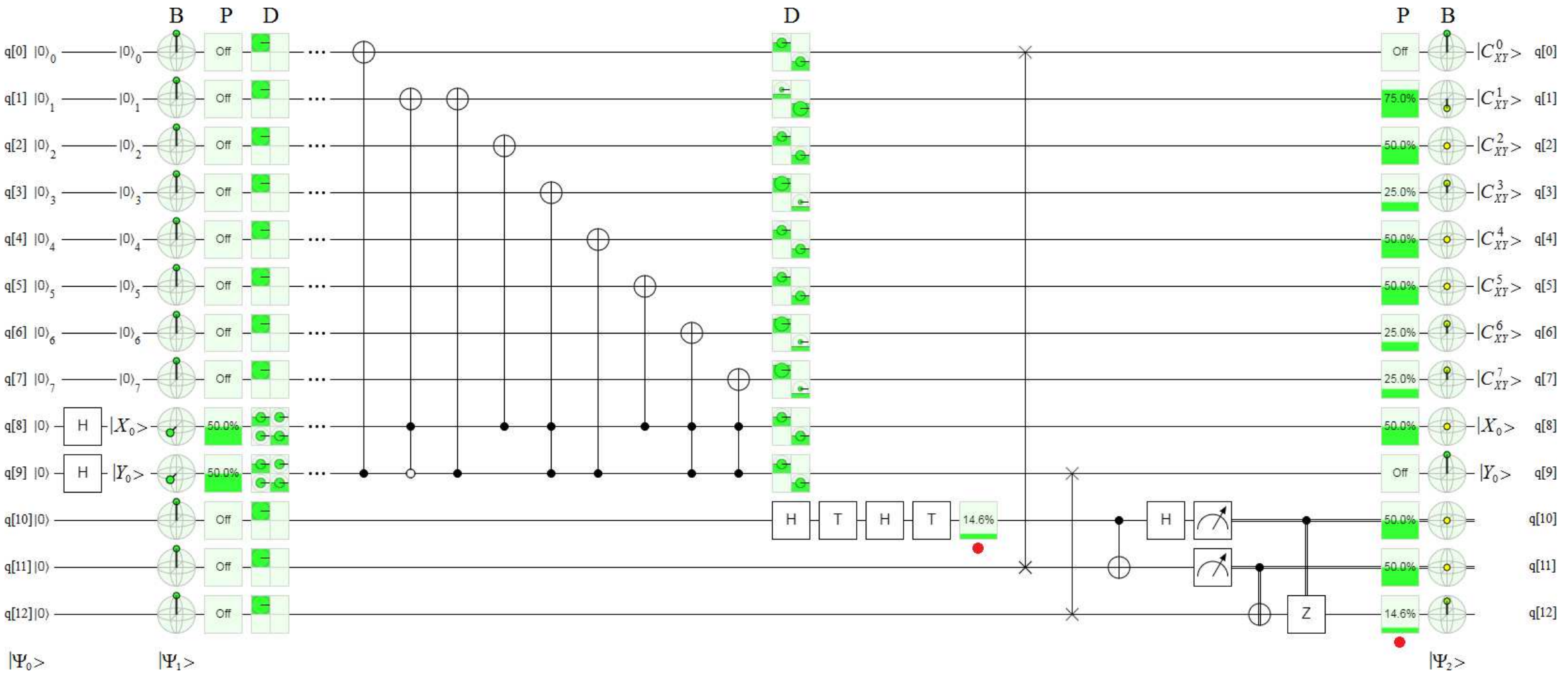

**Figure A1.** Implementation of the alternative version of NEQR of Fig. 9 on Quirk [9] with a teleportation module at its output. As in the case of Fig. 12 for the first version of NEQR, the red points under the *Probability of* |1> in qubits q[10] and q[12] shows the success of the teleportation.

The red points under the *Probability of* |1> in qubits q[10] and q[12] shows the success of the teleportation, where the teleported state $|\psi\rangle$ is recovered in q[12].

The fact that there is entanglement at the exit of both versions of NEQR is not a minor problem. In fact, it is a spurious effect that causes both elements that encode the pixel values $\left|C^i_{YX}\right\rangle$ and their locations in the tile $\left|Y\,X\right\rangle$ to be maximally entangled sharing the same values. Besides, both the $\left|C^i_{YX}\right\rangle$ and their locations in the tile $\left|Y\,X\right\rangle$ are not CBS $\{|0\rangle,|1\rangle\}$ as NEQR [3] states by definition. This spurious phenomenon between the values that represent the components of the pixel and their positions produces that both have the same state causing the morphological collapse of the image, which is, a complete nonsense. This was seen in detail in Subsection 6.2. Moreover, at the exit of both versions of NEQR there will always be the same spurious entanglement regardless of the image to be encoded, which constitutes even greater nonsense.

## Appendix B: Entanglement coupling on QPS [8] for the first version of NEQR of Fig. 8

In this appendix, we present two experiments:

1. the implementation of the first version of NEQR of Fig. 8 on QPS [8], and
2. the implementation of the first version of NEQR of Fig. 8 on QPS [8] with a teleportation module at its output.

***First version of NEQR:*** Figure B1 shows this implementation on QPS [8] thanks to a drag and drop editor very similar to Quirk [9].

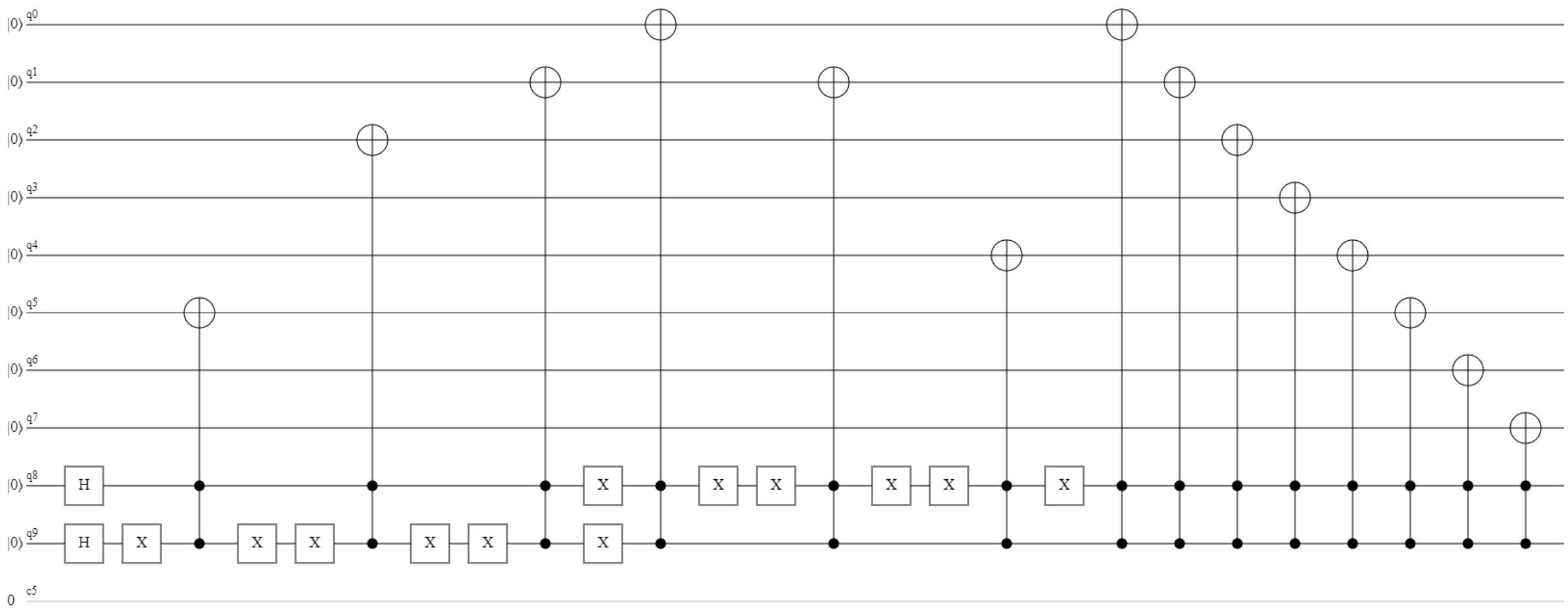

**Figure B1.** Implementation of the first version of NEQR of Fig. 8 on QPS [8].

Figure B2 shows the state vector for this experiment only showing the outcomes different to zero, where, the first column represents the sequence of qubits q[9]q[8]q[7]q[6]q[5]q[4]q[3]q[2]q[1]q[0], the second one means the obtained wavefunction, the third one means probability of |1>, while the forth column also represents probability of |1> but with a blue segment, in order to give it a more visual appearance. The outcomes clearly show the presence of maximally entanglement between some of them. On the other hand, Fig. B3 shows the detail of all involved qubits on the Bloch's sphere, while, Fig. B4 shows the mentioned entanglement between outcomes, where, the equal percentage of blue and yellow colors means maximally entangled states, while lower percentages of yellow color means non-maximally entangled states. As can be seen, in very few cases there is no degree of entanglement, although these cases fail to save the situation.

**State vector**

| | | |
|---|---|---|
| 0000000000 | 0.50000000+0.00000000i | 25.00000% |
| 0100100110 | 0.50000000+0.00000000i | 25.00000% |
| 1000010011 | 0.50000000+0.00000000i | 25.00000% |
| 1111111111 | 0.50000000+0.00000000i | 25.00000% |

**Figure B2.** State vector for this experiment only showing the outcomes different to zero, where, the first column represents the sequence of qubits q[9]q[8]q[7]q[6]q[5]q[4]q[3]q[2]q[1]q[0], the second one means wavefunction, the third one means probability of |1>, while the forth column also represents probability of |1> but with a blue segment. The outcomes clearly show the presence of entanglement.

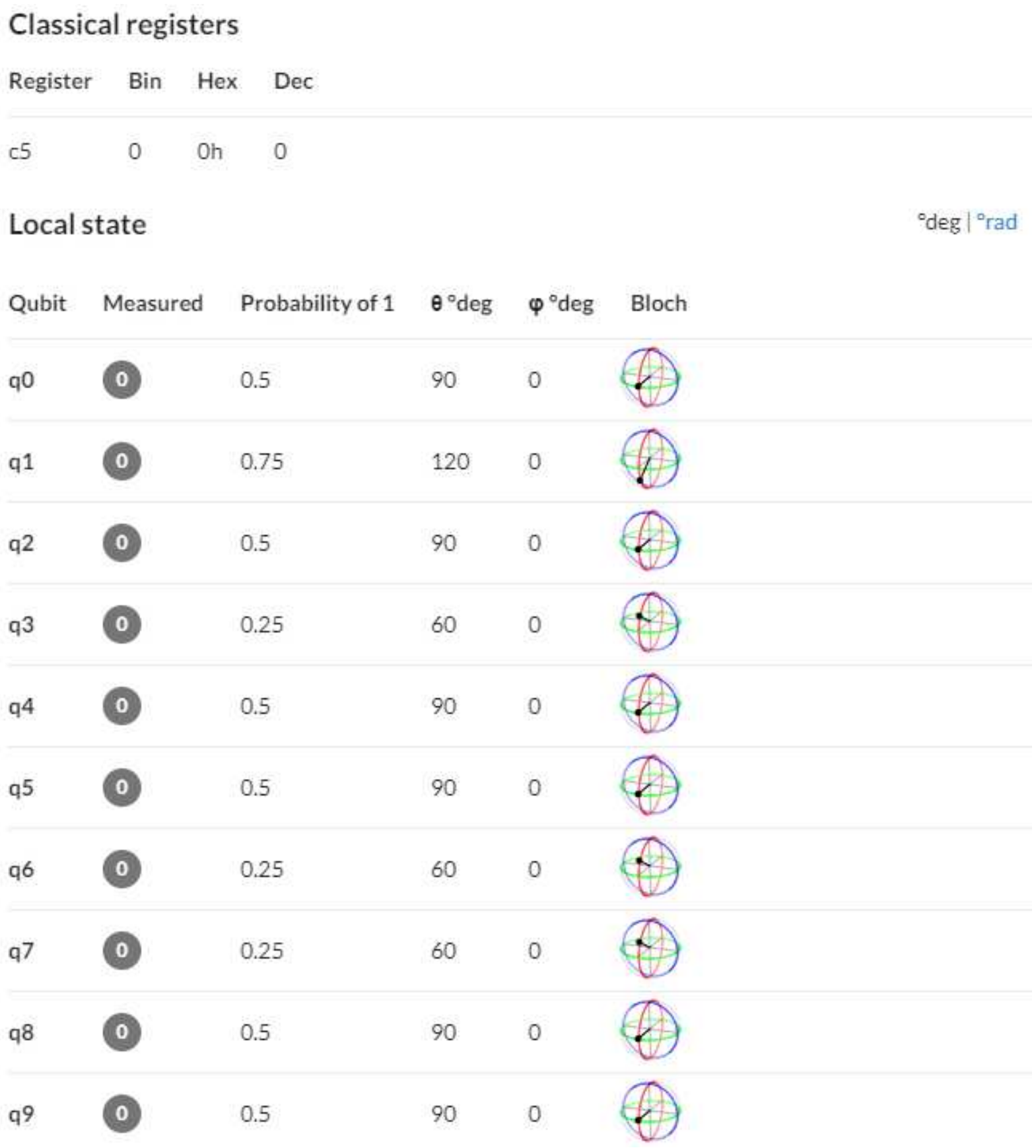

**Figure B3.** Detail of all involved qubits on the Bloch's sphere.

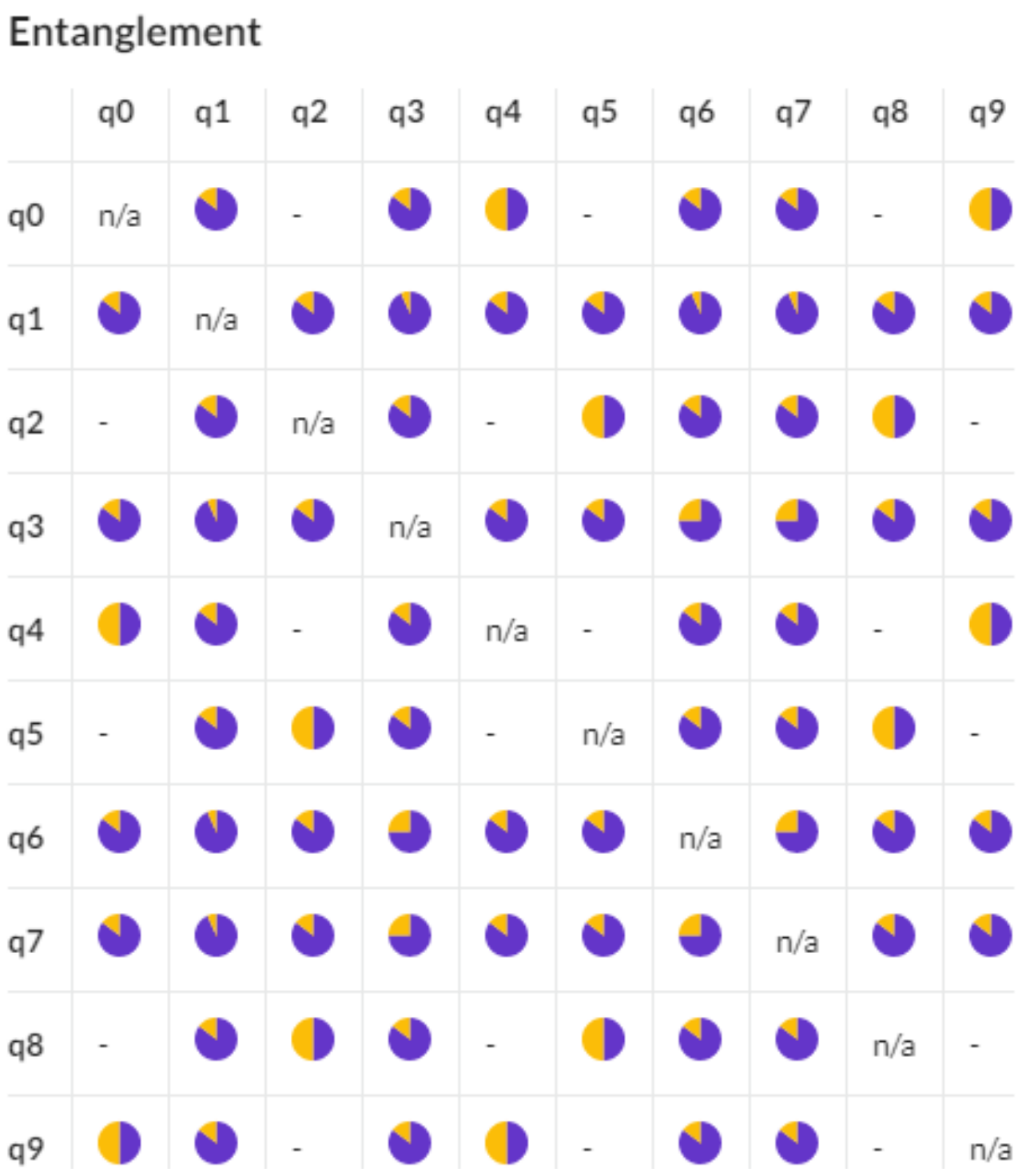

**Figure B4.** Entanglement between outcomes, where, equal percentage of blue and yellow colors means maximally entangled states, while lower percentages of yellow color means non-maximally entangled states. As can be seen, in very few cases there is no degree of entanglement.

***First version of NEQR with teleportation and its output:*** Figure B5 shows this implementation on QPS [8] where the teleportation module involved to the qubits q[0], q[9], q[10], q[11], and q[12].

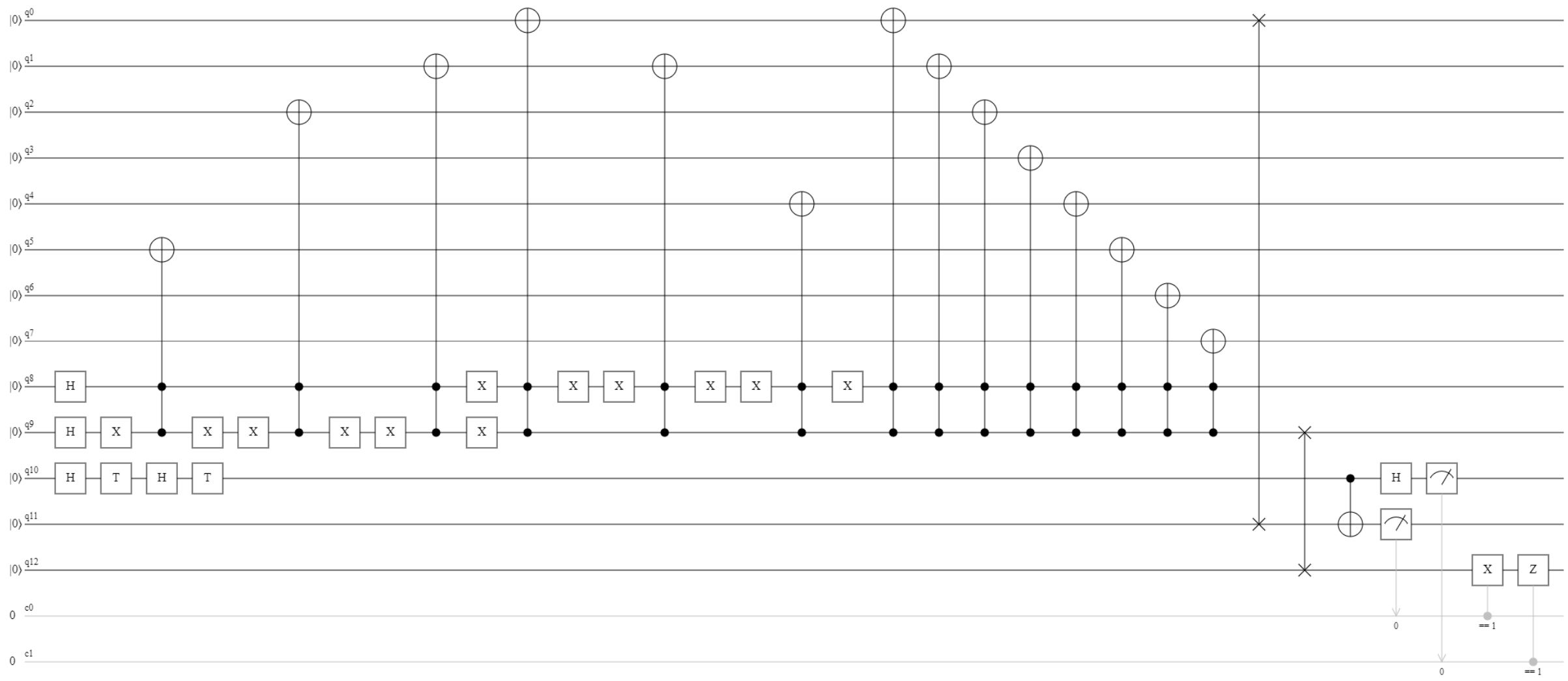

**Figure B5.** Implementation of the first version of NEQR of Fig. 8 on QPS [8] with a teleportation module at its output.

Figure B6 shows the state vector of the outcomes of Fig. B5, showing the outcomes different to zero, where, the first column represents the sequence of qubits q[9]q[8]q[7]q[6]q[5]q[4]q[3]q[2]q[1]q[0], the second one represents the wavefunction, the third one means probability of |1>, while the forth column also represents probability of |1> but with a blue segment. The outcomes clearly show a successful teleportation given that 42.67767 % + 42.67767 % = 85.35534 % which is exactly the probability of |0> of the qubit to be teleported, while, 7.32233 % + 7.32233 % = 14.64466 % which is exactly the probability of |1> of the qubit to be teleported.

**State vector**

| | | |
|---|---|---|
| 0010000000000 | 0.60355339+0.25000000i | 42.67767% |
| 0010100100110 | 0.60355339+0.25000000i | 42.67767% |
| 1010000010010 | 0.25000000-0.10355339i | 7.32233% |
| 1010111111110 | 0.25000000-0.10355339i | 7.32233% |

**Figure B6.** State vector for this experiment only showing the outcomes different to zero, where, the first column represents the sequence of qubits q[9]q[8]q[7]q[6]q[5]q[4]q[3]q[2]q[1]q[0], the second one means wavefunction, the third one means probability of |1>, while the forth column also represents probability of |1> but with a blue segment. The outcomes clearly show a successful teleportation given that 42.67767 % + 42.67767 % = 85.35534 % which is exactly the probability of |0> of the qubit to be teleported, while, 7.32233 % + 7.32233 % = 14.64466 % which is exactly the probability of |1> of the qubit to be teleported.

Figure B7 represents the detail of all involved qubits on the Bloch's sphere, while, Fig. B8 represents entanglement between outcomes, where, equal percentage of blue and yellow colors means maximally entangled states, while lower percentages of yellow color means non-maximally entangled states. As can be seen, the qubit to be teleported, originally in q[10], has disappeared from q[10] in order not to violate the No-Cloning Theorem [25] and has appeared in q[12], i.e., it has been successfully teleported, unequivocally verifying the existence of entanglement between outcomes of NEQR.

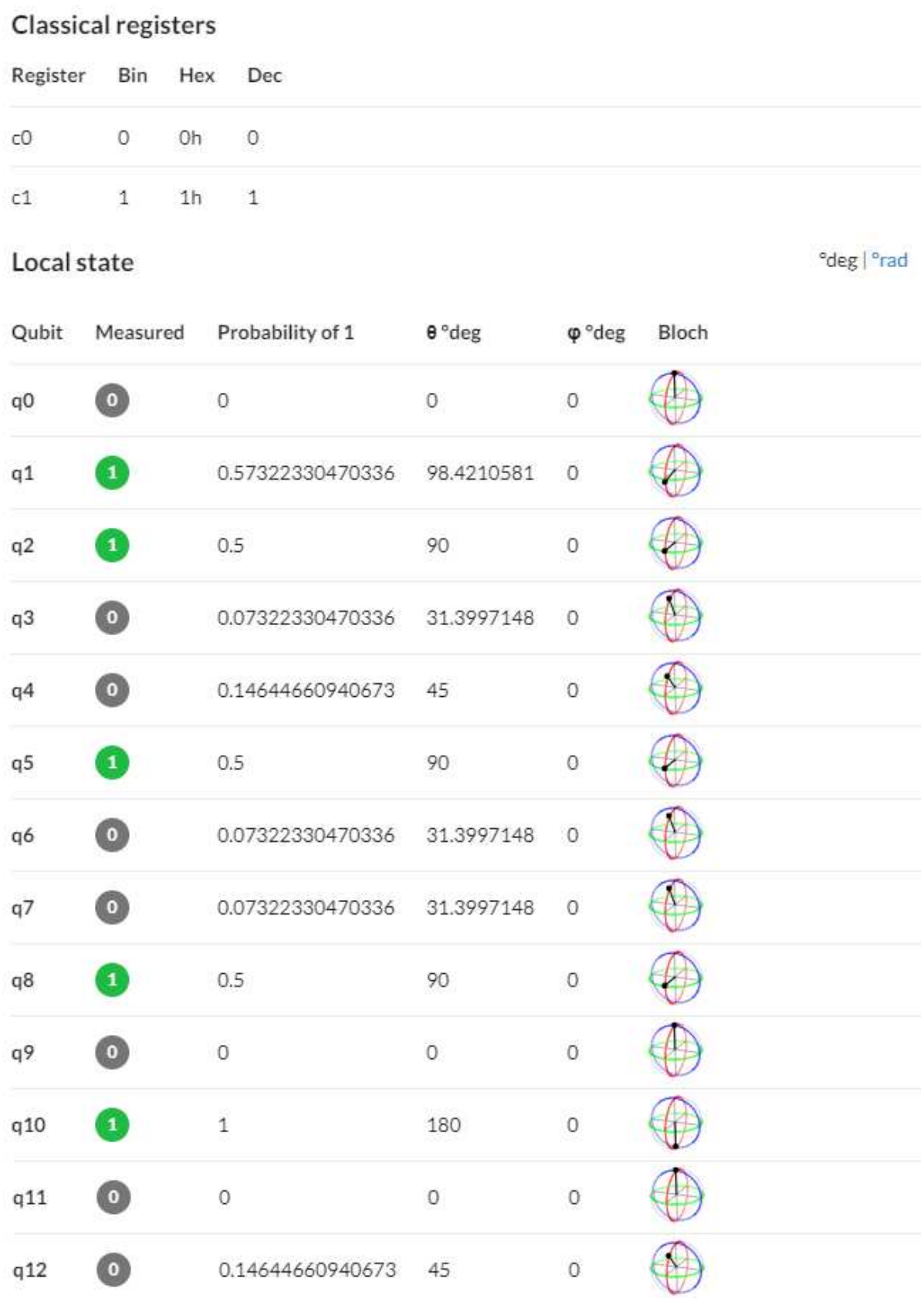

Classical registers

| Register | Bin | Hex | Dec |
|---|---|---|---|
| c0 | 0 | 0h | 0 |
| c1 | 1 | 1h | 1 |

Local state °deg | °rad

| Qubit | Measured | Probability of 1 | θ °deg | φ °deg | Bloch |
|---|---|---|---|---|---|
| q0 | 0 | 0 | 0 | 0 | |
| q1 | 1 | 0.57322330470336 | 98.4210581 | 0 | |
| q2 | 1 | 0.5 | 90 | 0 | |
| q3 | 0 | 0.07322330470336 | 31.3997148 | 0 | |
| q4 | 0 | 0.14644660940673 | 45 | 0 | |
| q5 | 1 | 0.5 | 90 | 0 | |
| q6 | 0 | 0.07322330470336 | 31.3997148 | 0 | |
| q7 | 0 | 0.07322330470336 | 31.3997148 | 0 | |
| q8 | 1 | 0.5 | 90 | 0 | |
| q9 | 0 | 0 | 0 | 0 | |
| q10 | 1 | 1 | 180 | 0 | |
| q11 | 0 | 0 | 0 | 0 | |
| q12 | 0 | 0.14644660940673 | 45 | 0 | |

**Figure B7.** Detail of all involved qubits on the Bloch's sphere.

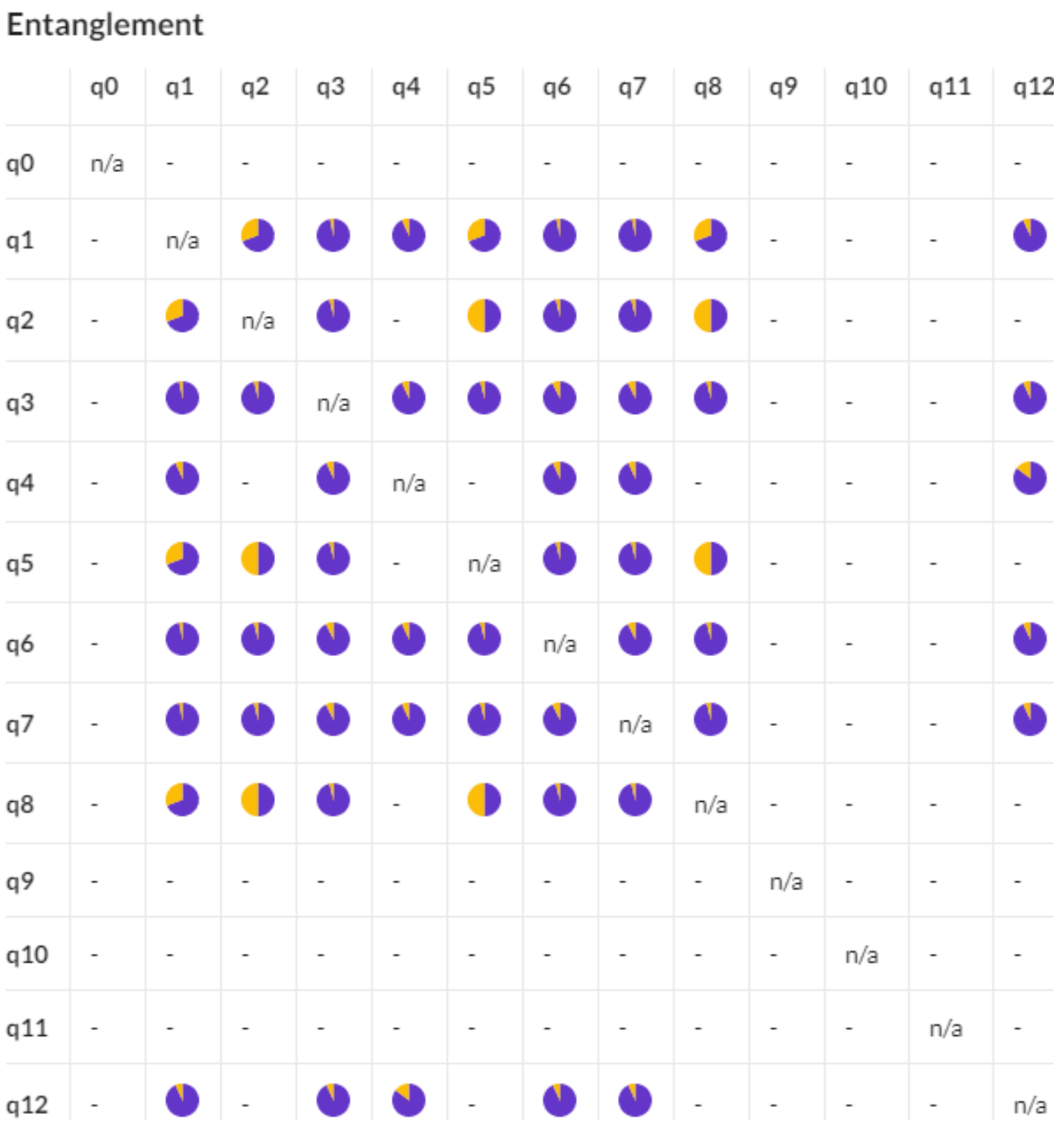

**Figure B8.** Entanglement between outcomes, where, equal percentage of blue and yellow colors means maximally entangled states, while lower percentages of yellow color means non-maximally entangled states. As can be seen, the qubit to be teleported, originally in q[10], has disappeared from q[10] in order not to violate the No-Cloning Theorem [25] and has appeared in q[12], i.e., it has been successfully teleported.

## Appendix C: Entanglement coupling on IBM Q [6] for the first version of NEQR of Fig. 8

As in the previous case, we will have both closely linked experiments here.

***First version of NEQR:*** Figure C1 shows this implementation on IBM Q [6] thanks to a drag and drop editor very similar to Quirk [9], and QPS [8].

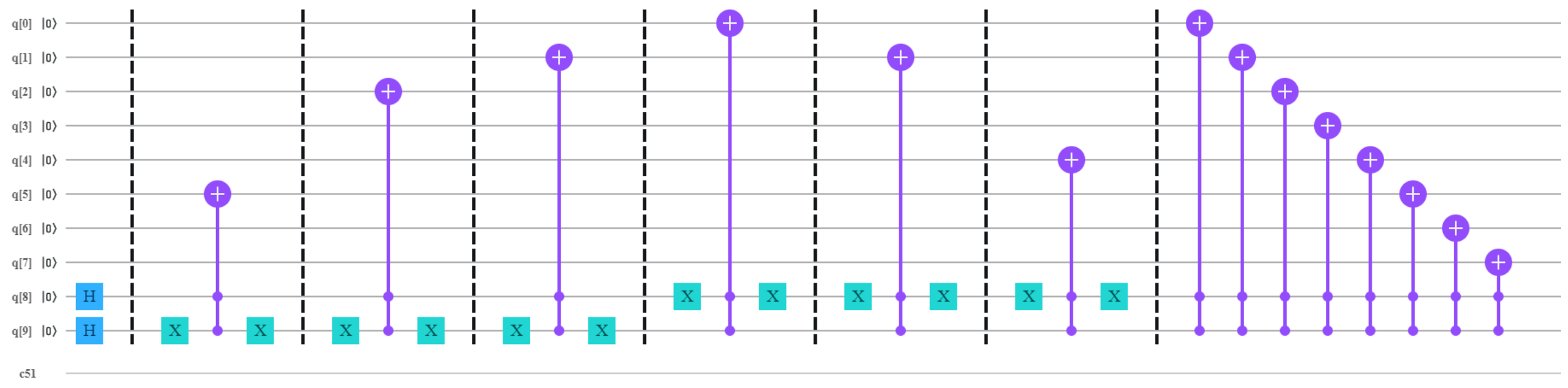

**Figure C1.** Implementation of the first version of NEQR of Fig. 8 on IBM Q [6].

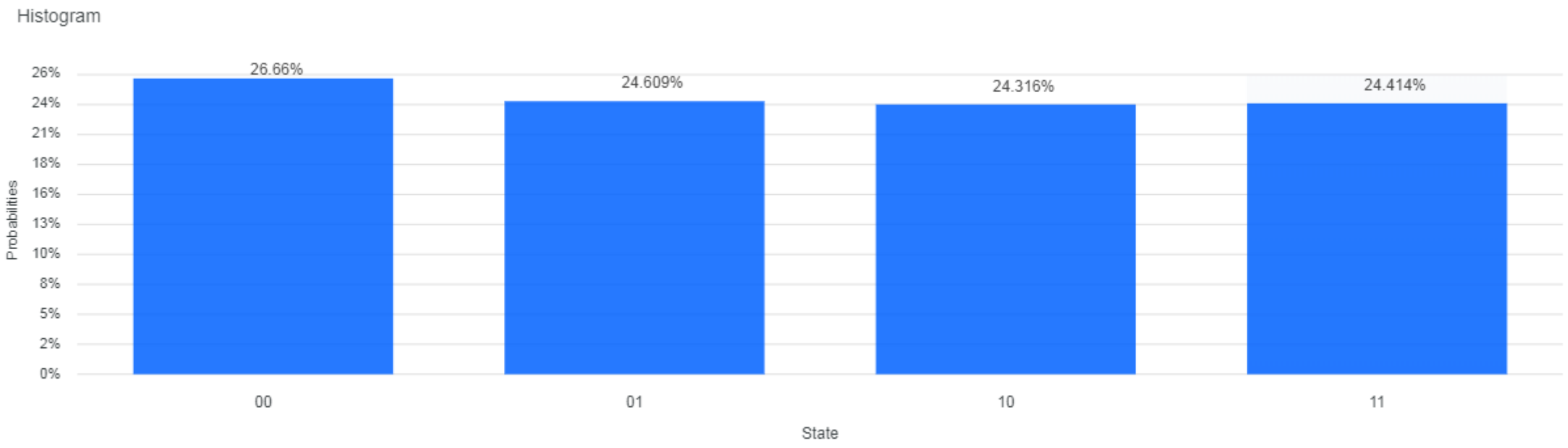

**Figure C2.** Probability of |1> for the only nonzero outcomes, exclusively. These outcomes clearly show the presence of entanglement.

Figure C2 shows the probability of |1> for the only nonzero outcomes, exclusively, with similar results to those of Fig. B2 for the case of QPS [8], i.e., it is a maximally entangled case with the following outcomes:

WAVEFUNCTION
(0.5+0j) |0000000000>
(0.5+0j) |0100100110>
(0.5+0j) |1000010011>
(0.5+0j) |1111111111>

AMPLITUDES
(0.5+0j) |0000000000>
(0.5+0j) |0100100110>
(0.5+0j) |1000010011>
(0.5+0j) |1111111111>

PROBABILITIES
0.25 |0000000000>
0.25 |0100100110>
0.25 |1000010011>
0.25 |1111111111>

***First version of NEQR with teleportation and its output:*** Figure C3 shows this implementation on IBM Q [6] where the teleportation module involved to the qubits q[0], q[9], q[10], q[11], and q[12].

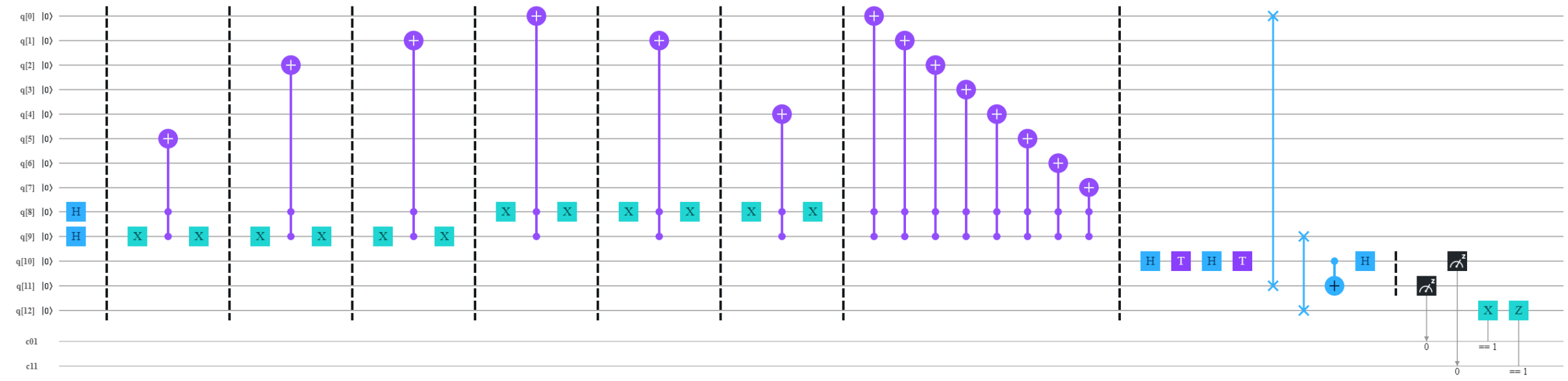

**Figure C3.** Implementation of the first version of NEQR of Fig. 8 on IBM Q [6] with a teleportation module at its output.

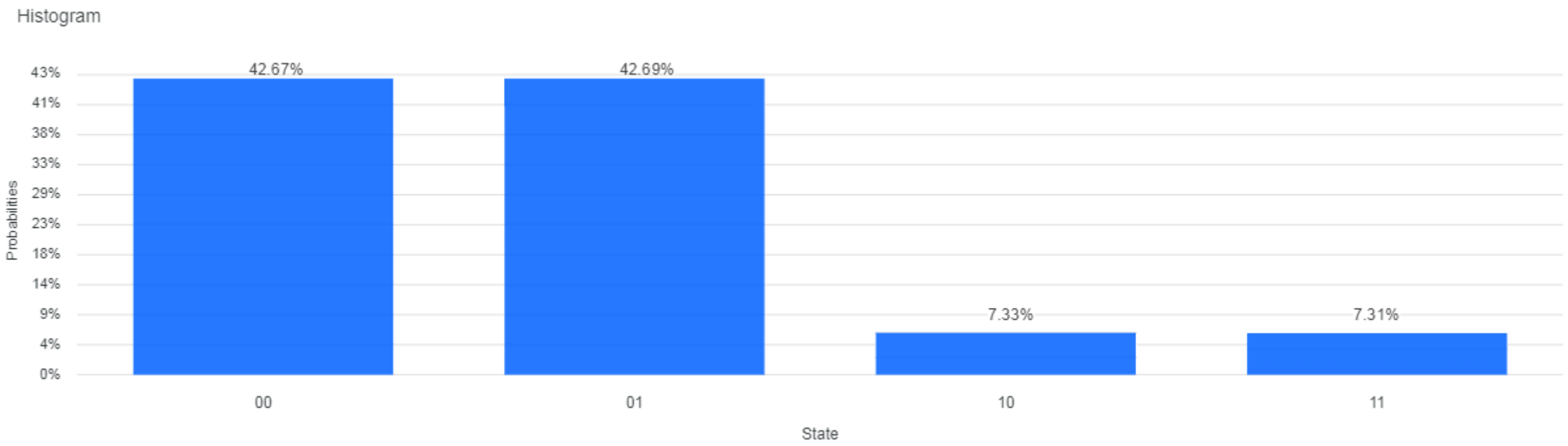

**Figure C4.** Probability of |1> for the nonzero outcomes, exclusively. The teleportation has been successful, given that 42.67 % + 42.69 % = 85.36 % which is exactly the probability of |0> of the qubit to be teleported, while, 7.33 % + 7.31 % = 14.64 % which is exactly the probability of |1> of the qubit to be teleported.

Figure C4 shows the probability of |1> for the nonzero outcomes, exclusively, with results similar to those of Fig. B6 for the case of QPS [8], i.e., it is a completely successful teleportation, demonstrating in an indirect way the existence of entanglement among the involved outcomes of NEQR, with the following results:

WAVEFUNCTION
(0.6035533906+0.25J) |0000000000000>
(0.6035533906+0.25J) |0000100100110>
(0.25-0.1035533906J) |1000000010010>
(0.25-0.1035533906J) |1000111111110>

AMPLITUDES
(0.6035533906+0.25J) |0000000000000>
(0.6035533906+0.25J) |0000100100110>
(0.25-0.1035533906J) |1000000010010>
(0.25-0.1035533906J) |1000111111110>

PROBABILITIES
0.4267766953 |0000000000000>
0.4267766953 |0000100100110>
0.0732233047 |1000000010010>
0.0732233047 |1000111111110>

### Appendix D: Entanglement coupling on Rigetti [7] for the first version of NEQR of Fig. 8

Rigetti [7] has not a graphic interface, therefore, the only way to show the results is by capturing the screens at the end of each running of QVM for both experiments.

***First version of NEQR:*** Figure D1 shows the captured screen of Rigetti's results, for the first version of NEQR of Fig. 8, showing the quil code, wavefunctions (only those that are nonzero), and the amplitudes. The entanglement between the outcomes is clearly evident.

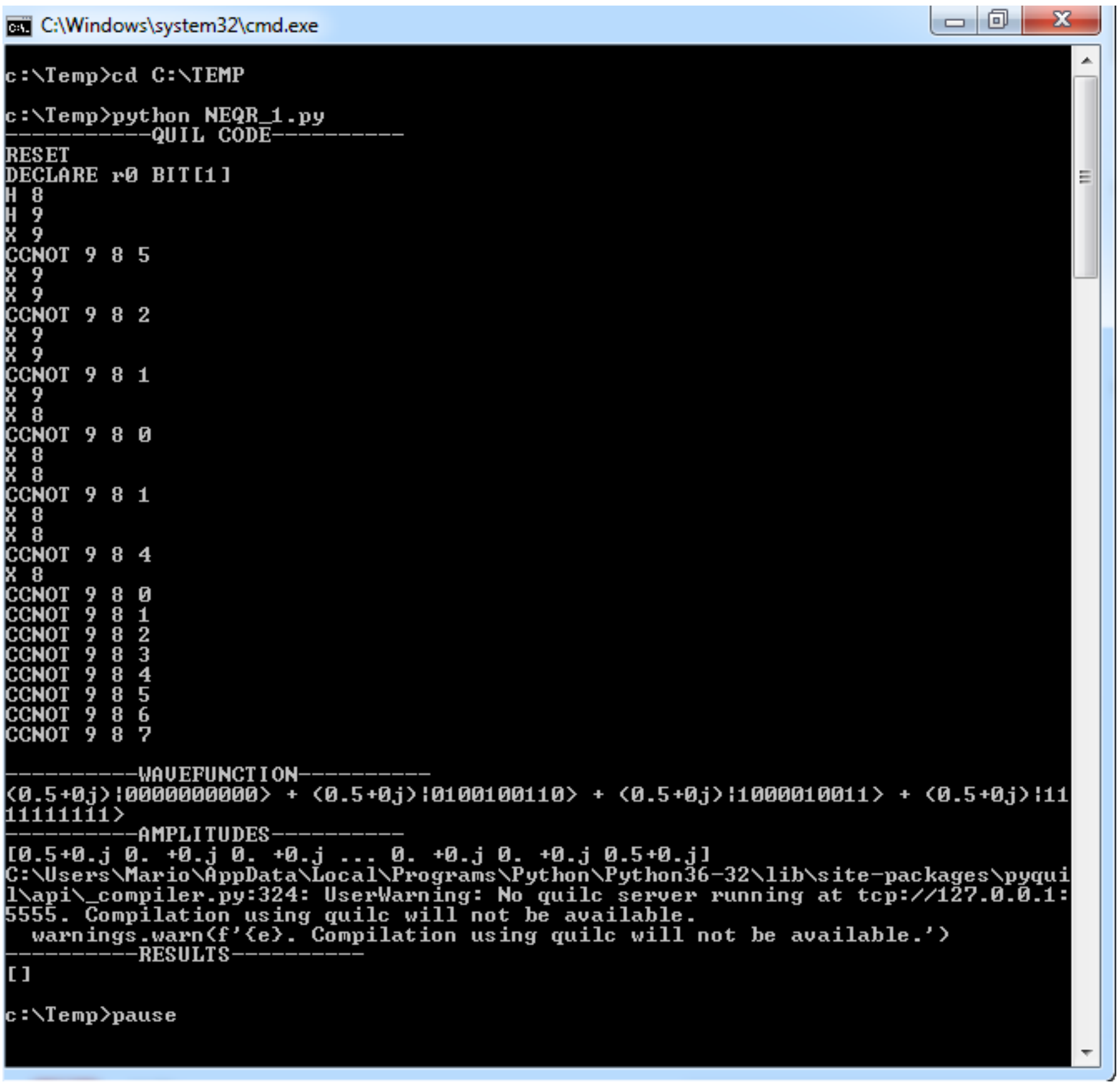

**Figure D1.** Captured screen of Rigetti's results, for the first version of NEQR of Fig. 8, showing the quil code, wavefunctions (nonzero), and amplitudes. The entanglement between the outcomes is so explicit as in the previous cases.

The specific results of this experiment are:

WAVEFUNCTION
(0.5+0j) |0000000000>
(0.5+0j) |0100100110>
(0.5+0j) |1000010011>
(0.5+0j) |1111111111>

AMPLITUDES
(0.5+0j) |0000000000>
(0.5+0j) |0100100110>
(0.5+0j) |1000010011>
(0.5+0j) |1111111111>

PROBABILITIES
0.25 |0000000000>
0.25 |0100100110>
0.25 |1000010011>
0.25 |1111111111>

***First version of NEQR with teleportation and its output:*** Figure D2 shows the captured screen of Rigetti's results, for the first version of NEQR of Fig. 8 with a teleportation module involved to the qubits q[0], q[9], q[10], q[11], and q[12], showing the quil code, wavefunctions (only those that are nonzero), and the amplitudes. The teleportation was a complete success due to:

0.6035533906 + 0.25 = 0.8535533906, which is the exact probability of |0>, and
0.25 - 0.1035533906 = 0.1464466094, which is the exact probability of |1>.

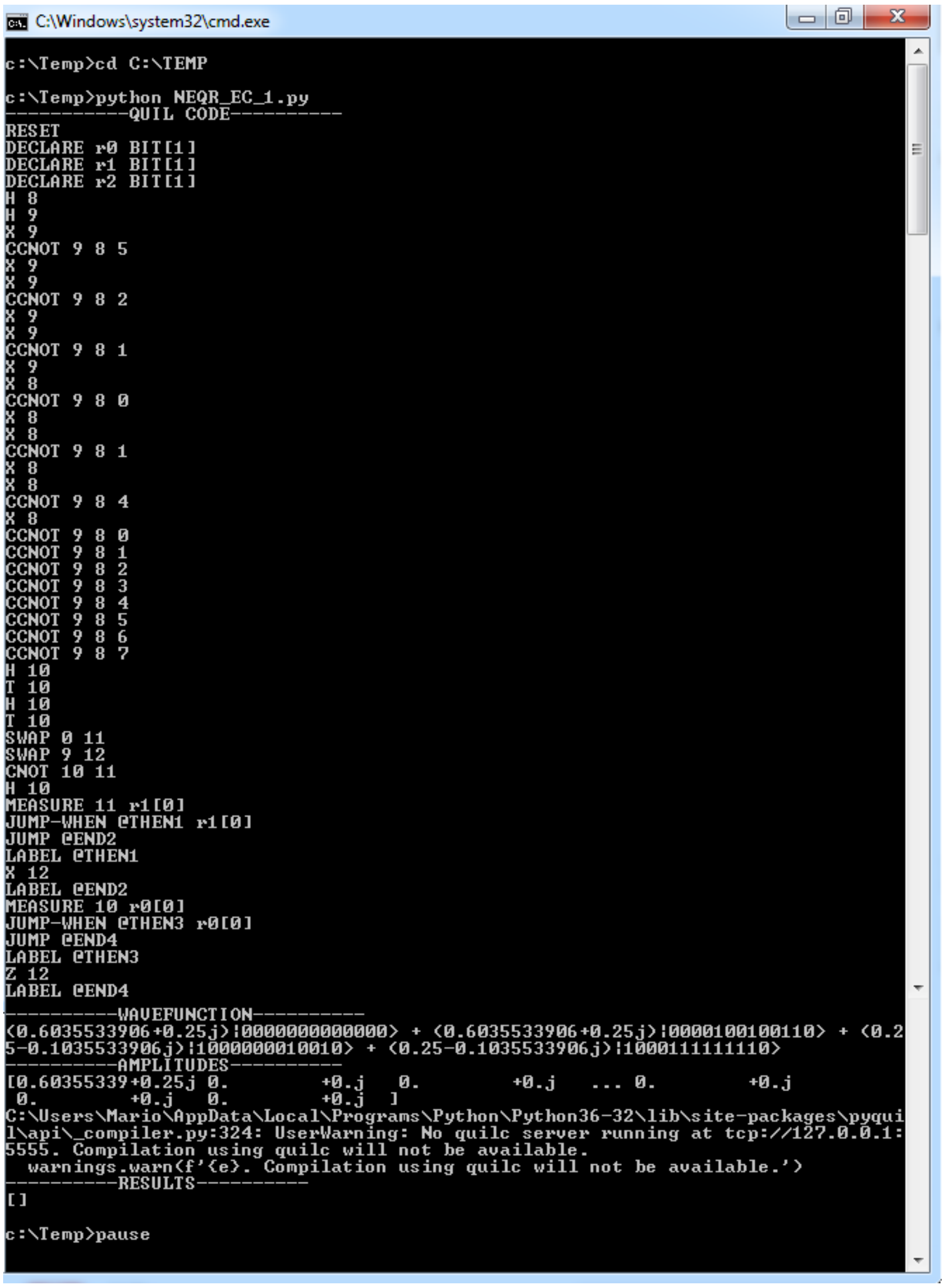

```
C:\Windows\system32\cmd.exe
c:\Temp>cd C:\TEMP

c:\Temp>python NEQR_EC_1.py
-----------QUIL CODE----------
RESET
DECLARE r0 BIT[1]
DECLARE r1 BIT[1]
DECLARE r2 BIT[1]
H 8
H 9
X 9
CCNOT 9 8 5
X 9
X 9
CCNOT 9 8 2
X 9
X 9
CCNOT 9 8 1
X 9
X 8
CCNOT 9 8 0
X 8
X 8
CCNOT 9 8 1
X 8
X 8
CCNOT 9 8 4
X 8
CCNOT 9 8 0
CCNOT 9 8 1
CCNOT 9 8 2
CCNOT 9 8 3
CCNOT 9 8 4
CCNOT 9 8 5
CCNOT 9 8 6
CCNOT 9 8 7
H 10
T 10
H 10
T 10
SWAP 0 11
SWAP 9 12
CNOT 10 11
H 10
MEASURE 11 r1[0]
JUMP-WHEN @THEN1 r1[0]
JUMP @END2
LABEL @THEN1
X 12
LABEL @END2
MEASURE 10 r0[0]
JUMP-WHEN @THEN3 r0[0]
JUMP @END4
LABEL @THEN3
Z 12
LABEL @END4
----------WAVEFUNCTION----------
(0.6035533906+0.25j)|0000000000000> + (0.6035533906+0.25j)|0000100100110> + (0.2
5-0.1035533906j)|1000000010010> + (0.25-0.1035533906j)|1000111111110>
----------AMPLITUDES----------
[0.60355339+0.25j 0.         +0.j   0.         +0.j   ... 0.         +0.j
 0.         +0.j   0.         +0.j   ]
C:\Users\Mario\AppData\Local\Programs\Python\Python36-32\lib\site-packages\pyqui
l\api\_compiler.py:324: UserWarning: No quilc server running at tcp://127.0.0.1:
5555. Compilation using quilc will not be available.
  warnings.warn(f'{e}. Compilation using quilc will not be available.')
----------RESULTS----------
[]

c:\Temp>pause
```

**Figure D2.** Captured screen of Rigetti's results, for the first version of NEQR with a teleportation module and its output, which is similar to the case of Fig. 12 on Quirk [9], showing the quil code, wavefunctions (nonzero), and amplitudes. The success of teleportation is irrefutable.

The entanglement between the outcomes is clearly evident, since if such entanglement had not existed, such a perfect teleportation would have never been possible.

The specific results of this experiment are:

WAVEFUNCTION
(0.6035533906+0.25J) |0000000000000>
(0.6035533906+0.25J) |0000100100110>
(0.25-0.1035533906J) |1000000010010>
(0.25-0.1035533906J) |1000111111110>

AMPLITUDES
(0.6035533906+0.25J) |0000000000000>
(0.6035533906+0.25J) |0000100100110>
(0.25-0.1035533906J) |1000000010010>
(0.25-0.1035533906J) |1000111111110>

PROBABILITIES
0.4267766953 |0000000000000>
0.4267766953 |0000100100110>
0.0732233047 |1000000010010>
0.0732233047 |1000111111110>

The code in pyquil, as well as the complete results in a non-format file are found in [44], along with the rest of the implementations of this paper.

## Appendix E: Implementation of NEQR on the Quantum Inspire platform of QUTech [43]

Since QUTech [43] does not have the *if-then-else* statement, we can only implement both versions of NEQR, therefore, teleportation is excluded because it requires such a statement.

***First versión of NEQR:*** Figure E1 shows this implementation on Quantum Inspire of QUTech [43] thanks to a drag and drop editor very similar to Quirk [9], QPS [8], and IBM Q [6].

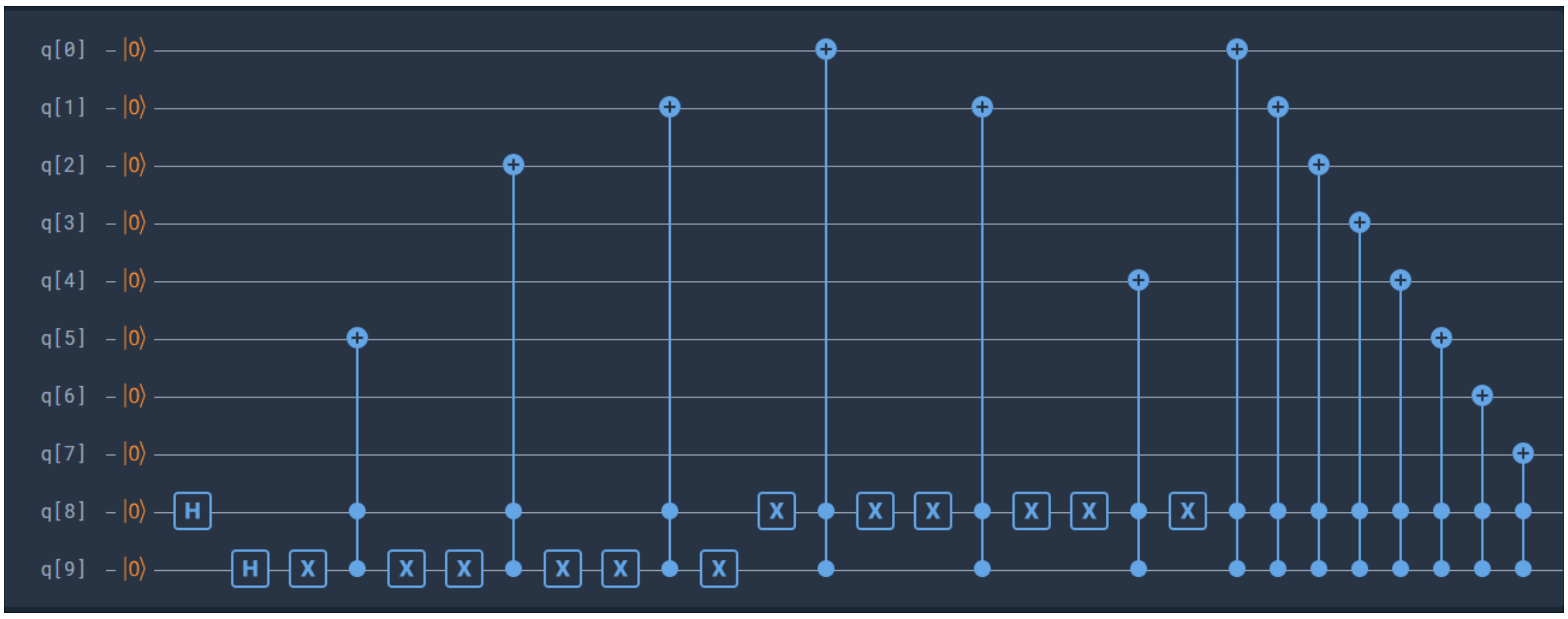

**Figure E1.** Implementation of the first version of NEQR of Fig. 8 on Quantum Inspire [43].

Figure E2 shows the probability of |1> of the outcomes for the first version of NEQR of Fig. 8, only in those cases where the outcomes are nonzero. These outcomes clearly show the presence of entanglement and they are similar to those of Fig. C2 for IBM Q. As in the case of IBM Q [6], the number of shots was 1024. The total coincidence with the results of the previous platforms does not leave any doubt about the spurious presence of an entanglement with a different degree of participation between the outcomes of the first version of NEQR, which, as we have already mentioned, it is responsible for the disastrous results of Subsection 6.2.

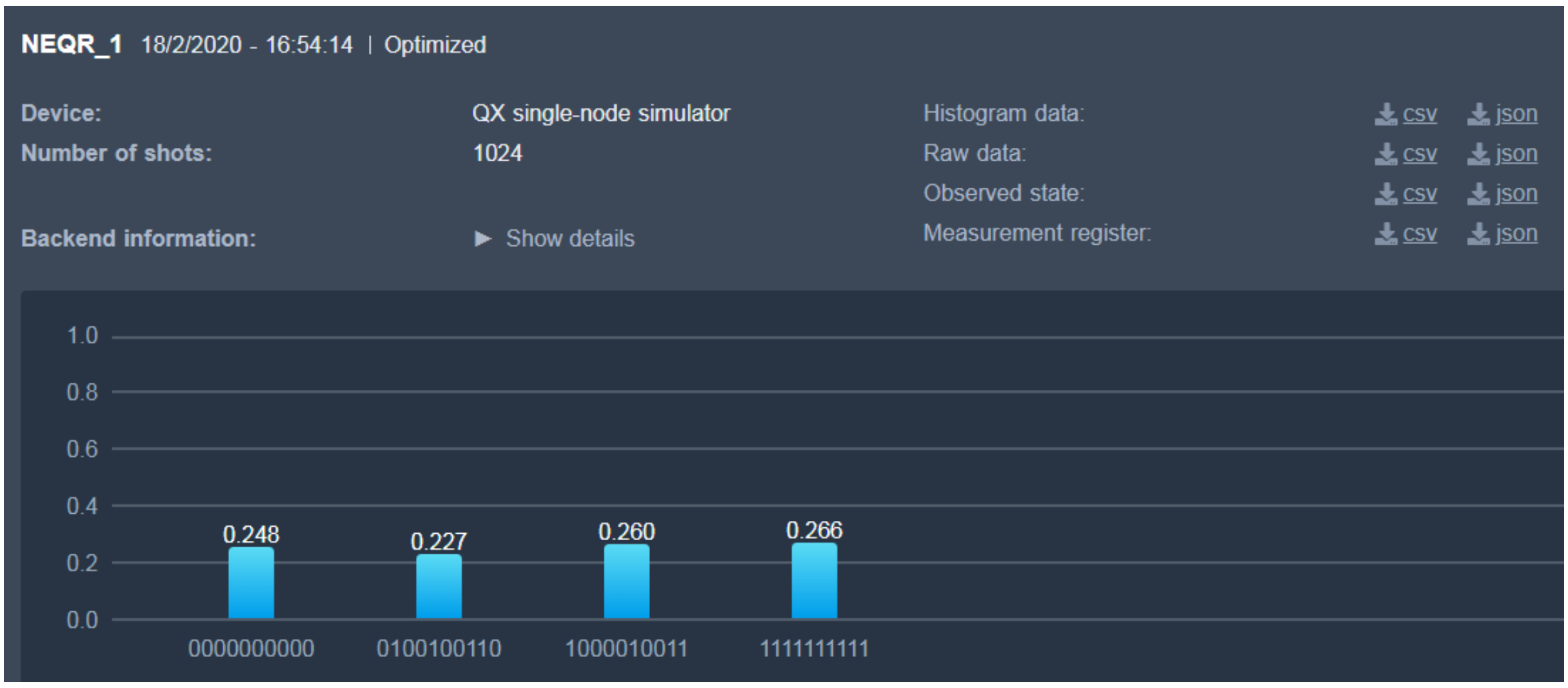

**Figure E2.** Probability of |1> of the outcomes for the first version of NEQR of Fig. 8, exclusively for the nonzero cases. These outcomes clearly show the presence of entanglement and they are similar to those of Fig. C2 for IBM Q.

***Alternative versión of NEQR:*** Figure E3 shows this implementation on Quantum Inspire of QUTech [43]. Although this version implies a certain gate economy compared to the first version, its devastating results are similar to those of the previous version, as we can see in Fig. E4. Therefore, NEQR, in any of its forms, cannot escape this complex problem, which clearly and irremediably compromises its use in Quantum Image Processing (QImP).

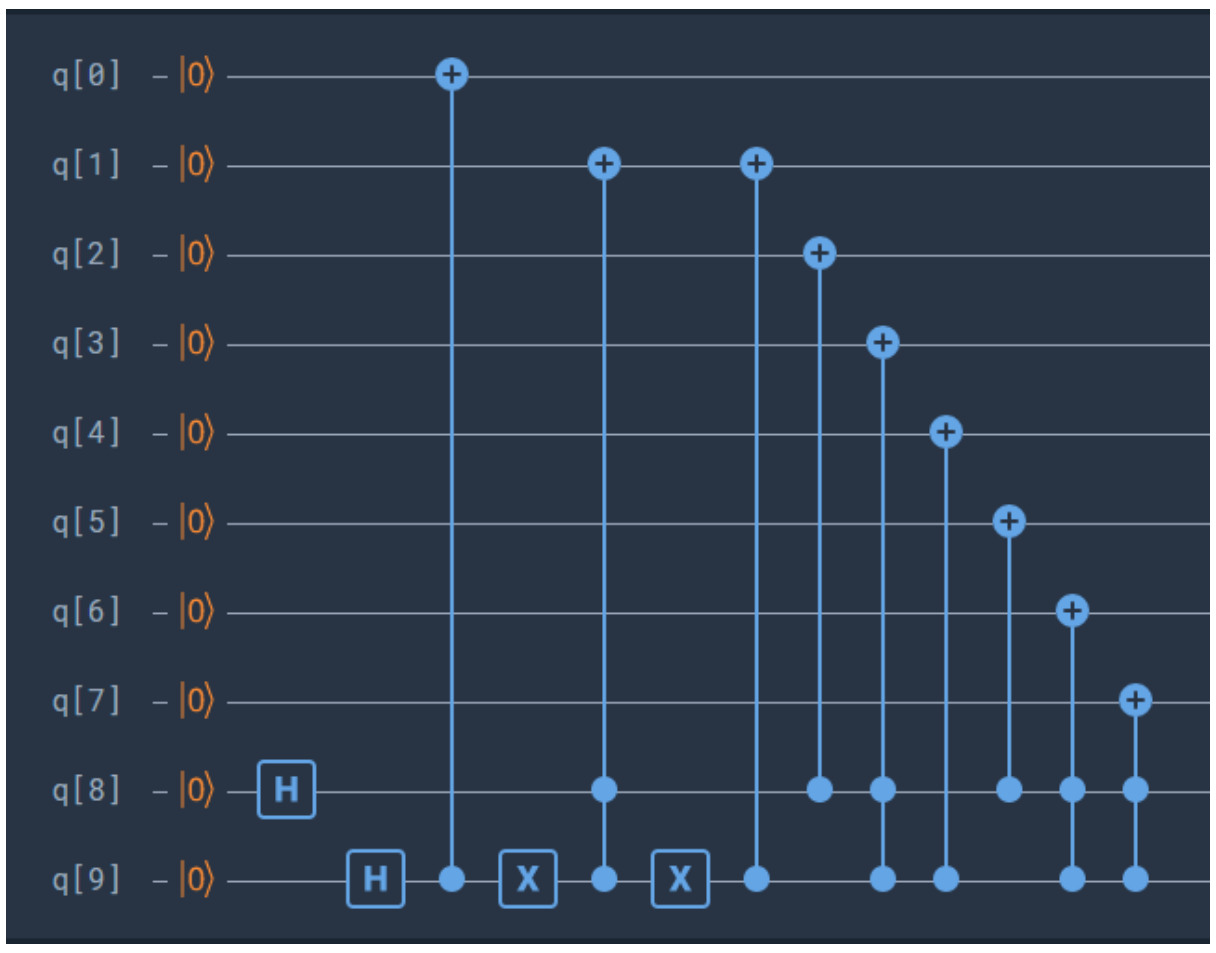

**Figure E3.** Implementation of the alternative version of NEQR of Fig. 9 on Quantum Inspire [43].

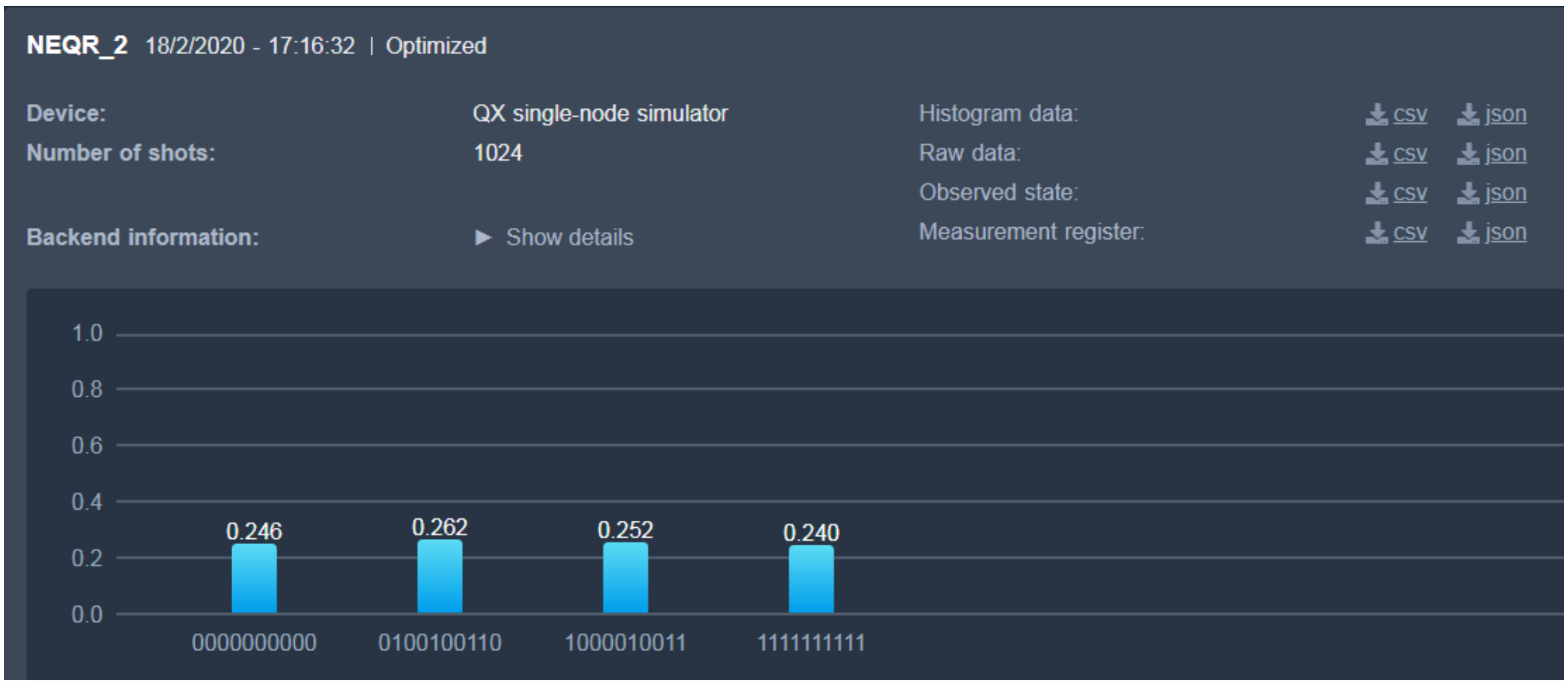

**Figure E4.** Probability of |1> of the outcomes for the alternative version of NEQR of Fig. 9, exclusively for the nonzero cases. In this case too, these outcomes clearly show the presence of entanglement being similar to those of Fig. C2 for IBM Q.